\documentclass[12pt]{article}
\usepackage{mathrsfs}
\makeatletter \@addtoreset{equation}{section} \makeatother
\renewcommand{\theequation}{\thesection.\arabic{equation}}
\newcommand{\ba}{\begin{array}}
\newcommand{\ea}{\end{array}}
\newcommand{\beq}{\begin{equation}}
\newcommand{\eeq}{\end{equation}}
\newcommand{\bea}{\begin{eqnarray}}
\newcommand{\eea}{\end{eqnarray}}

\def\bce{\begin{center}}
\def\ece{\end{center}}

\def\nonu{\nonumber}

\def\pa{\partial}
\def\al{\alpha}
\def\be{\beta}
\def\ga{\gamma}
\def\Ga{\Gamma}
\def\de{\delta}

\def\la{\lambda}
\def\La{\Lambda}

\def\eps6{{\displaystyle \mathop{\epsilon}^{6}}{}}
\def\g6{{\displaystyle \mathop{g}^{6}}{}}

\def\nab6{{\displaystyle \mathop{\nabla}^{6}}{}}

\def\0{{\sst{(0)}}}
\def\1{{\sst{(1)}}}
\def\2{{\sst{(2)}}}
\def\3{{\sst{(3)}}}
\def\4{{\sst{(4)}}}
\def\5{{\sst{(5)}}}
\def\6{{\sst{(6)}}}
\def\7{{\sst{(7)}}}
\def\8{{\sst{(8)}}}

\def\ba{\begin{array}}
\def\ea{\end{array}}
\def\beq{\begin{equation}}
\def\eeq{\end{equation}}
\def\be{\begin{equation}}
\def\ee{\end{equation}}

\def\la{\lambda}
\def\eps{\epsilon}

\def\ba{\begin{array}}
\def\ea{\end{array}}
\def\beq{\begin{equation}}
\def\eeq{\end{equation}}
\def\be{\begin{equation}}
\def\ee{\end{equation}}

\def\la{\lambda}
\def\eps{\epsilon}

\def\eps6{{\displaystyle \mathop{\epsilon}^{6}}{}}

\def\nab6{{\displaystyle \mathop{\nabla}^{6}}{}}

\newcommand{\bean}{\begin{eqnarray*}}
\newcommand{\eean}{\end{eqnarray*}}

\begin{document}
\thispagestyle{empty} \addtocounter{page}{-1}
   \begin{flushright}
\end{flushright}

\vspace*{1.3cm}
  
\centerline{ \Large \bf
The ${\cal N}=4$ Supersymmetric Linear $W_{\infty}[\la]$ Algebra
}
\vspace*{1.5cm}
\centerline{ {\bf  Changhyun Ahn}
} 
\vspace*{1.0cm} 
\centerline{\it 
 Department of Physics, Kyungpook National University, Taegu
41566, Korea} 
\vspace*{0.5cm}
\centerline{\tt ahn@knu.ac.kr
} 
\vskip2cm

\centerline{\bf Abstract}
\vspace*{0.5cm}

From the recently known ${\cal N}=2$ supersymmetric linear
$W_{\infty}^{K,K}[\la]$ algebra where $K$ is the dimension of
fundamental (or antifundamental) representation
of bifundamental $\beta \, \ga$ and $b \, c$
ghost system,
we determine its ${\cal N}=4$ supersymmetric enhancement at $K=2$.
We construct the ${\cal N}=4$ stress energy tensor, 
the first ${\cal N}=4$ multiplet and their operator product expansions
(OPEs) in terms of above bifundamentals.
We show that the OPEs between the first ${\cal N}=4$ multiplet and itself
are the same as the corresponding ones in the
${\cal N}=4$ coset $\frac{SU(N+2)}{SU(N)}$ model under the large $(N,k)$
't Hooft-like limit with fixed $\la_{co} \equiv
\frac{(N+1)}{(k+N+2)}$, up to two central terms.
The two parameters are related to
each other $\la =\frac{1}{2}\, \la_{co}$.
We also provide other  OPEs by considering the
second, the third and the fourth ${\cal N}=4$ multiplets
in the ${\cal N}=4$ supersymmetric linear $W_{\infty}[\la]$ algebra.


\baselineskip=18pt
\newpage
\renewcommand{\theequation}
{\arabic{section}\mbox{.}\arabic{equation}}

\tableofcontents

\section{ Introduction}

The free field construction in two dimensional conformal field theory
is useful to study the extension of the conformal symmetries in string
theory. Because their operator product expansions (OPEs)
take the simple form in the sense that the right hand side
of OPEs does not contain the fields, contrary to
the affine Kac-Moody algebra, it is straightforward to
determine the conserved currents of any (conformal) weights
in terms of the quadratic free fields
with multiple derivatives. In order to describe the supersymmetric
theory, the fermionic free field is necessary
to describe the symmetries as well as the bosonic
free field. Depending on the weights of the bosonic and fermionic
fields, the weights of the currents we can make by using them
are determined naturally from a simple counting of weights.
The central charge of the Virasoro algebra consisting of the
stress energy tensor of weight-$2$ is fixed by the number of
(bosonic and fermionic) free fields.
Usually, the bosonic field has the weight-$1$ (or zero) while the
fermionic field has the weight-$\frac{1}{2}$.

More generally,
the above weights of the bosonic and fermionic fields
can be deformed by a parameter
$\la$ \cite{FMS}.
Although the weights of
each bosonic and fermionic field depend on
this $\la$ parameter explicitly, due to the plus and minus signs
in the coefficients in front of $\la$ of the weights,
the weights of particular composite
combinations of these free fields do not contain
the parameter $\la$.
Then we can construct the currents of
integer (or half integer)
weights in terms of
free fields as mentioned before by considering
that the weights of each term should not depend on the $\la$.
Of course, the $\la$-dependence in
the coefficients in front of free fields in the
expression of the currents occurs in very nontrivial way
\cite{BVd1,BVd2}. 
This is a new feature because
the structure constants of the resulting algebra
contain the $\la$ dependence explicitly,
compared to the ones in the previous paragraph.

So far, we have two bosonic and two fermionic
free fields.
There exist two fundamental OPEs between them.
We can introduce the multiple bosonic and fermionic
fields which transform as bifundamentals.
Because they are independent fields and the multiple defining OPEs
satisfy independently,
all the previous analysis can be generalized to describe
the symmetries easily.
For example, the central charge of
the Virasoro algebra is simply a sum over each contribution
from bosonic and fermionic free fields.
For each current of weight-$h$, there exist
multicomponent generators.
The corresponding $W_{\infty}$ algebras (without any deformation
parameter $\la$)
are obtained in \cite{BK,OS,Odake}.
By construction, because there are many fermionic currents,
there is more room for the supersymmetric theory we would like to
obtain.

Then it is natural to
consider the multiple bosonic and fermionic free fields
together with the deformation of $\la$.
Recently, in \cite{Ahn2203}, 
the ${\cal N}=2$ supersymmetric linear $W_{\infty}^{K,K}[\la]$
algebra is obtained by analyzing the currents
from the multiple bosonic and
fermionic free fields with derivatives. 
Here $K$ is the dimension of fundamental representation of
above bifundamentals.
The number of bosonic currents of each weight
($h=1,2, \cdots$) is given by $2 K^2$
which is equal to the number of fermionic currents of each weight
($h=\frac{3}{2}, \frac{5}{2}, \cdots$).
The factor $2$ appears because we are considering the complex
free fields.
Among $2K^2$-fermionic currents,
two of them play the role of ${\cal N}=2$ supersymmetry generators.
There are also $K^2$-fermionic currents of weight-$\frac{1}{2}$
and this fact will affect the structure of the ${\cal N}=4$
superconformal algebra
\footnote{There are also similar constructions in \cite{CHR,EGR}.}.

In this paper,
we would like to construct
the ${\cal N}=4$ supersymmetric linear
$W_{\infty}^{2,2}[\la]$ algebra by focusing on the $K=2$ case
which is very special in the sense that only this $K=2$
will provide the supersymmetric theory we want to obtain.
Then among eight fermionic currents of weight-$\frac{3}{2}$,
the half of them play the role of ${\cal N}=4$ supersymmetry
generators. The remaining half of them belong to
the first ${\cal N}=4$ multiplet.
Moreover, the lowest fermionic currents of weight-$\frac{1}{2}$
can join the generators of the ${\cal N}=4$ superconformal algebra.
For the weight-$1$ currents, the seven of them
play the role of the bosonic generators of the
${\cal N}=4$ superconformal algebra and
the remaining one will appear in the lowest operator
in the first ${\cal N}=4$ multiplet.
For the weight-$2$, one of them is given by the stress energy tensor,
six of them will appear in the generators of
the first ${\cal N}=4$ multiplet  
and the remaining one will arise in the lowest operator
in the second ${\cal N}=4$ multiplet.
For the weights greater than $2$, we can analyze similarly
and they can be placed into the corresponding
${\cal N}=4$ multiplets appropriately, according to $SO(4)$ indices
of ${\cal N}=4$ superspace.

We would like to determine the explicit algebra how the above
analysis on the weight contents fits in the ${\cal N}=4$ supersymmetric
linear $W_{\infty}[\la]$ algebra \footnote{
  The terminology of
  $W_{\infty}[\la]$ algebra is used here instead of using the
  previous terminology
  of $W_{\infty}^{2,2}[\la]$ algebra for simplicity. }.

  In this paper, we determine
  the ${\cal N}=4$ stress energy tensor,
  the first, the second, the third and the fourth
  ${\cal N}=4$ multiplets in the ${\cal N}=4$
  supersymmetric $W_{\infty}[\la]$ algebra, by using
  the $\beta \, \ga$ and $b \, c$ ghost systems.
  We calculate the various OPEs
  between them, where the sum of
  two (super) weights appearing on the left hand side
  is less than or equal to $4$ ($h_1+h_2 \leq 4$),
  in ${\cal N}=4$ superspace.
  As in the abstract, 
  the case of $h_1=1=h_2$ reproduces the corresponding one
  \cite{AKK1910,AK1509} 
  in the ${\cal N}=4$ coset model under the large $(N,k)$
  't Hooft-like limit.
  
In section $2$,
we review the $\beta \, \ga$ and $b \, c$ ghost systems
and 
the bosonic and fermionic currents
can be written in terms of these fields.

In section $3$, we determine
the ${\cal N}=4$ stress energy tensor, the first
${\cal N}=4$ multiplet and their OPEs explicitly.

In section $4$,
we summarize what we have obtained in this paper and future
directions are described.

In Appendices, some of the details in section
$3$ are presented explicitly.

We are heavily using the Thielemans package \cite{Thielemans}
with the help of
a mathematica \cite{mathematica}.

\section{Review}

\subsection{The fundamental OPEs}

The bosonic $\beta \, \gamma$ and fermionic
$b \, c$ ghost systems
satisfy the following OPEs \cite{CHR,Ahn2203}
\bea
\ga^{i,\bar{a}}(z)\, \beta^{\bar{j},b}(w) =
\frac{1}{(z-w)}\, \de^{i \bar{j}}\, \de^{\bar{a} b} + \cdots,
\qquad
c^{i, \bar{a}}(z) \, b^{\bar{j},b}(w) =
\frac{1}{(z-w)}\, \de^{i \bar{j}}\, \de^{\bar{a} b} + \cdots.
\label{fundOPE}
\eea
The fundamental indices $a, b $ run over $a, b =1,2$
and the antifundamental indices $\bar{a}, \bar{b}$ run over
$\bar{a}, \bar{b}=1,2$.
The  fundamental indices $i, j $ of $SU(N)$
run over $i, j =1,2, \cdots, N$
and the antifundamental indices $\bar{i}, \bar{j}$ of
$SU(N)$ run over
$\bar{i}, \bar{j}=1,2, \cdots, N$.

\subsection{The quadratic bosonic and fermionic operators}

We can construct the bosonic and fermionic operators (or currents)
by taking the quadratic expressions of above
$\beta \, \gamma$ and $b \,c $ ghost systems in the presence
of various holomorphic derivatives
as follows: \cite{CHR,Ahn2203}
\bea
V_{\la,\bar{a} b}^{(s)+} & = & \sum_{i=0}^{s-1}\, a^i(s, \la)\,
\pa^{s-1-i}\,
(( \pa^i \, \beta^{\bar{l} b} ) \, \de_{l \bar{l}}  \,
\ga^{l \bar{a}}) +
 \sum_{i=0}^{s-1}\, a^i(s, \la+\frac{1}{2})\, \pa^{s-1-i}\,
 (( \pa^i \, b^{\bar{l} b} ) \,  \de_{l \bar{l}} \,
 c^{l \bar{a}} ),
 \nonu \\
 V_{\la,\bar{a} b}^{(s)-} & = & -\frac{(s-1+2\la)}{(2s-1)}\,
 \sum_{i=0}^{s-1}\, a^i(s, \la)\, \pa^{s-1-i}\,
 (( \pa^i \, \beta^{\bar{l} b} ) \, \de_{l \bar{l}} \, 
 \ga^{l \bar{a}}) \nonu \\
 & + &
 \frac{(s-2\la)}{(2s-1)}\,
 \sum_{i=0}^{s-1}\, a^i(s, \la+\frac{1}{2})\, \pa^{s-1-i}\,
 (( \pa^i \, b^{\bar{l} b} ) \,  \de_{l \bar{l}} \,
 c^{l \bar{a}} ),
\nonu \\
Q_{\la,\bar{a} b}^{(s)+} & = & \sum_{i=0}^{s-1}\, \al^i(s, \la)\,
\pa^{s-1-i}\,
(( \pa^i \, \beta^{\bar{l} b} ) \, \de_{l \bar{l}} \,
 c^{l \bar{a}}) -
 \sum_{i=0}^{s-2}\, \beta^i(s, \la)\, \pa^{s-2-i}\,
 (( \pa^i \, b^{\bar{l} b} ) \, \de_{l \bar{l}} \,
 \ga^{l \bar{a}} ),
\nonu \\
Q_{\la,\bar{a} b}^{(s)-} & = & \sum_{i=0}^{s-1}\, \al^i(s, \la)\,
\pa^{s-1-i}\,
(( \pa^i \, \beta^{\bar{l} b} ) \, \de_{l \bar{l}} \,
 c^{l \bar{a}}) +
 \sum_{i=0}^{s-2}\, \beta^i(s, \la)\, \pa^{s-2-i}\,
 (( \pa^i \, b^{\bar{l} b} ) \, \de_{l \bar{l}} \,
 \ga^{l \bar{a}} ).
\label{VVQQla}
\eea
The first two operators of weight $s$
are bosonic and the last two operators
of weight $(s-\frac{1}{2})$ are fermionic.
Each term on the right hand sides
has the summation over the indices $l$ and $\bar{l}$
of $SU(N)$.
Each operator has four components, $11,12,21$ and $22$ in the indices
$\bar{a}$ and $b$.
Each coefficient on the right hand sides
depends on the weight $s$ (or $(s-\frac{1}{2})$)
and the $\la$.
They can be summarized by \cite{BVd1,BVd2}
\bea 
 a^i(s, \la) \equiv \left(\begin{array}{c}
s-1 \\  i \\
 \end{array}\right) \, \frac{(-2\la-s+2)_{s-1-i}}{(s+i)_{s-1-i}},
 \qquad 0 \leq i \leq (s-1),
 \nonu \\
 \al^i(s, \la) \equiv \left(\begin{array}{c}
s-1 \\  i \\
 \end{array}\right) \, \frac{(-2\la-s+2)_{s-1-i}}{(s+i-1)_{s-1-i}},
 \qquad 0 \leq i \leq (s-1),
 \nonu \\
  \beta^i(s, \la) \equiv \left(\begin{array}{c}
s-2 \\  i \\
  \end{array}\right) \, \frac{(-2\la-s+2)_{s-2-i}}{(s+i)_{s-2-i}},
  \qquad 0 \leq i \leq (s-2).
  \label{coeff}
 \eea
 The parentheses in (\ref{coeff})
 stand for the binomial coefficients and
 the $(a)_n$ symbols stand for the rising Pochhammer symbol
 $(a)_n \equiv a(a+1)\cdots (a+n-1)$.
 There are nontrivial relations between these coefficients
 \cite{BVd1,BVd2}.
 
\subsection{ The $\la$-dependent currents}

We can split the above bosonic operators into the one
written in terms of bosonic fields and the other
written in terms of fermionic fields by simple linear combinations.
For the fermionic operators, we also split them in terms of
the one having only one kind of fermion fields and the other
having only the other kind of fermion fields.
Then we obtain the following operators with the explicit
bifundamental indices \cite{Ahn2203}
\bea
W^{\la,\bar{a} b}_{F,h} & = &
\frac{n_{W_{F,h}}}{q^{h-2}}\,
\frac{(-1)^h}{\sum_{i=0}^{h-1}\, a^i( h, \frac{1}{2})}\,
\Bigg[\frac{(h-1+2\la)}{(2h-1)}\, V_{\la,\bar{a} b}^{(h)+} + V_{\la,\bar{a} b}^{(h)-}
\Bigg],
\nonu \\
W^{\la,\bar{a} b}_{B,h} & = &
 \frac{n_{W_{B,h}}}{q^{h-2}}\,
 \frac{(-1)^h}{\sum_{i=0}^{h-1}\, a^i( h, 0)}
\,
\Bigg[\frac{(h-2\la)}{(2h-1)}\, V_{\la,\bar{a} b}^{(h)+} - V_{\la,\bar{a} b}^{(h)-}
  \Bigg],
\nonu \\
Q^{\la, \bar{a} b}_{h+\frac{1}{2}} & = & \frac{1}{2} \,\frac{n_{W_{Q,h+\frac{1}{2}}}}{q^{h-1}}
\, \frac{(-1)^{h+1}  \, h }{
   \sum_{i=0}^{h-1} \, \beta^i( h+1, 0)}\, \Bigg[
  Q_{\la,\bar{a} b}^{(h+1)-} - Q_{\la,\bar{a} b}^{(h+1)+}\Bigg],
\nonu \\
\bar{Q}^{\la, b \bar{a}}_{h+\frac{1}{2}} & = & \frac{1}{2} \,
\frac{n_{W_{Q,h+\frac{1}{2}}}}{q^{h-1}} \,
 \frac{(-1)^{h+1}  }{
   \sum_{i=0}^{h} \, \al^i( h+1, 0)} \,
\Bigg[ Q_{\la,\bar{a} b}^{(h+1)-} +
  Q_{\la,\bar{a} b}^{(h+1)+}\Bigg].
\label{WWQQnonzerola}
\eea
The first two operators of weight $h$ are bosonic
and the last two operators of weight $(h+\frac{1}{2})$ are fermionic
\footnote{
The normalizations are given by \cite{BPRSS}
  \bea
n_{W_{\mathrm{F},h}} =  \frac{2^{h-3}(h-1)!}{(2h-3)!!}\,q^{h-2},
n_{W_{\mathrm{B},h}}=\frac{2^{h-3}\,h!}{(2h-3)!!}\,q^{h-2},
n_{Q_{h+\frac{1}{2}}} \!& = \!&
\frac{2^{h-\frac{1}{2}}h!}{(2h-1)!!}\,q^{h-1}
= n_{\bar{Q}_{h+\frac{1}{2}}}.
\label{nor}
\eea
Then the $q$ dependence in (\ref{WWQQnonzerola}) disappears.}.
The overall coefficients do not depend on the $\la$.
We list the explicit expressions for low weights
by substituting (\ref{coeff}) and (\ref{nor})
into (\ref{WWQQnonzerola}) as follows \cite{Ahn2203}: 
\bea
W^{\la,\bar{a} b}_{F,1} & = & -\frac{1}{4}\,
\Big( V_{\la,\bar{a} b}^{(1)-} + 2\la \, V^{(1)+}_{\la,\bar{a} b}\Big),
\qquad
W^{\la,\bar{a} b}_{F,2}  = 
\Big( V_{\la,\bar{a} b}^{(2)-} + \frac{1}{3} \, (1+2\la) \,
V^{(2)+}_{\la,\bar{a} b}\Big),
\nonu \\
W^{\la,\bar{a} b}_{F,3}  & = & -4 \, 
\Big( V_{\la,\bar{a} b}^{(3)-} + \frac{1}{5} \, (2+2\la) \,
V^{(3)+}_{\la,\bar{a} b}\Big),
\qquad
W^{\la,\bar{a} b}_{F,4}  =16 \, 
\Big( V_{\la,\bar{a} b}^{(4)-} + \frac{1}{7} \, (3+2\la) \,
V^{(4)+}_{\la,\bar{a} b}\Big),
\nonu \\
W^{\la,\bar{a} b}_{B,1} & = & -\frac{1}{4}\,
\Big( -V_{\la,\bar{a} b}^{(1)-} + (1-2\la) \, V^{(1)+}_{\la,\bar{a} b}\Big),
\qquad
W^{\la,\bar{a} b}_{B,2}  = 
\Big( -V_{\la,\bar{a} b}^{(2)-} + \frac{1}{3} \,
(2-2\la) \, V^{(2)+}_{\la,\bar{a} b}\Big),
\nonu \\
W^{\la,\bar{a} b}_{B,3}  & = & -4 \, 
\Big(- V_{\la,\bar{a} b}^{(3)-} +
\frac{1}{5} \, (3-2\la) \, V^{(3)+}_{\la,\bar{a} b}\Big),
\qquad
W^{\la,\bar{a} b}_{B,4}  =16 \, 
\Big( -V_{\la,\bar{a} b}^{(4)-} + \frac{1}{7} \, (4-2\la) \,
V^{(4)+}_{\la,\bar{a} b}\Big),
\nonu \\
Q^{\la,\bar{a} b}_{ \frac{3}{2}} & = & \frac{1}{\sqrt{2}} \,
\Big( Q_{\la,\bar{a} b}^{ (2) -} -  Q_{\la,\bar{a} b}^{ (2) +}\Big),
\qquad
Q^{\la,\bar{a} b}_{ \frac{5}{2}}  =  - 2 \sqrt{2} \,
\Big( Q_{\la,\bar{a} b}^{ (3) -} -  Q_{\la,\bar{a} b}^{ (3) +}\Big),
\nonu \\
Q^{\la,\bar{a} b}_{ \frac{7}{2}}  & = &  8 \sqrt{2} \,
\Big( Q_{\la,\bar{a} b}^{ (4) -} -  Q_{\la,\bar{a} b}^{ (4) +}\Big),
\qquad Q^{\la,\bar{a} b}_{ \frac{9}{2}}   =   -32 \sqrt{2} \,
\Big( Q_{\la,\bar{a} b}^{ (5) -} -  Q_{\la,\bar{a} b}^{ (5) +}\Big),
\nonu \\
\bar{Q}^{\la, b \bar{a}}_{ \frac{1}{2}} & = & -\frac{1}{2\sqrt{2}} \,
\Big( \bar{Q}_{\la,\bar{a} b}^{ (1) -} +  \bar{Q}_{\la,\bar{a} b}^{ (1) +}\Big),
\qquad
\bar{Q}^{\la, b \bar{a}}_{ \frac{3}{2}}  =  \frac{1}{\sqrt{2}} \,
\Big( \bar{Q}_{\la,\bar{a} b}^{ (2) -} +  \bar{Q}_{\la,\bar{a} b}^{ (2) +}\Big),
\nonu \\
\bar{Q}^{\la, b \bar{a}}_{ \frac{5}{2}} & = & -2\sqrt{2} \,
\Big( \bar{Q}_{\la,\bar{a} b}^{ (3) -} +  \bar{Q}_{\la,\bar{a} b}^{ (3) +}\Big),
\qquad
\bar{Q}^{\la, b \bar{a}}_{ \frac{7}{2}}  =  8 \sqrt{2} \,
\Big( \bar{Q}_{\la,\bar{a} b}^{ (4) -} +  \bar{Q}_{\la,\bar{a} b}^{ (4) +}\Big),
\nonu \\
\bar{Q}^{\la, b \bar{a}}_{ \frac{9}{2}} & = & -32 \sqrt{2} \,
\Big( \bar{Q}_{\la,\bar{a} b}^{ (5) -} +  \bar{Q}_{\la,\bar{a} b}^{ (5) +}\Big),
\qquad \cdots \qquad.
\label{lowspincurrents}
\eea
We can easily see that the normalization for the overall factor
is increased by $-4$ when we increase the weight.
Note that the lowest weight for the bosonic operators
is given by $1$ and the one for the fermionic
operators is given by $\frac{1}{2}$.
The $Q^{\la,\bar{a} b}_{ \frac{1}{2}}$ is identically zero. 

Then we have eight bosonic currents for the weight
$h=1,2, \cdots $ and eight
fermionic currents
for the weight
$h+\frac{1}{2}=\frac{3}{2}, \frac{5}{2}, \cdots$ in
(\ref{WWQQnonzerola}). In next section, we
will construct the ${\cal N}=4$ multiplets as well as
the ${\cal N}=4$ stress energy tensor by using
(\ref{WWQQnonzerola}).

\section{ The ${\cal N}=4$ supersymmetric linear
  $W_{\infty}[\la]$ algebra}

\subsection{The construction of  the ${\cal N}=4$ stress energy tensor}

\subsubsection{The $\la$-dependent quasiprimary stress energy tensor}

The stress energy tensor of weight-$2$
from (\ref{lowspincurrents}) is given by \cite{BVd1,BVd2}
\bea
L & = &
\Big(W^{\la,11}_{\mathrm{B},2}+W^{\la,22}_{\mathrm{B},2}+W^{\la,11}_{\mathrm{F},2}+
W^{\la,22}_{\mathrm{F},2} \Big),
\label{Lterm}
\eea
which is equal to $V^{(2)+}_{\la,\bar{a} b} \, \de_{b \bar{a}}$.
The central charge, which is the same as the
fourth order pole of the OPE
$L(z) \, L(w)$ times two, is 
\bea
c= 6\,N\, (1-4\la).
\label{central}
\eea
At $\la=0$, the central charge becomes $6N$.
We will observe that the remaining seven weight-$2$
operators appear in the first and the second
${\cal N}=4$ multiplets.

\subsubsection{The
 $\la$-dependent
  weight-$\frac{3}{2}$ primary supersymmetry currents}

It is natural to consider that the weight-$\frac{3}{2}$ operators,
which depend on the $\la$, can be obtained from
the corresponding ones at $\la=0$.
Our starting point is the following ansatz for the
 weight-$\frac{3}{2}$ operators
\bea
G^1
&
=&
-\frac{1}{2}\,\Big(
Q^{\la,11}_{\frac{3}{2}}
+i\sqrt{2}\,Q^{\la,12}_{\frac{3}{2}}
+2i \sqrt{2}\,Q^{\la,21}_{\frac{3}{2}}
-2\,Q^{\la,22}_{\frac{3}{2}}
\nonu \\
&- & 2\,\bar{Q}^{\la,11}_{\frac{3}{2}}
-2i \sqrt{2}\,\bar{Q}^{\la,12}_{\frac{3}{2}}
-i\sqrt{2}\,\bar{Q}^{\la,21}_{\frac{3}{2}}
+\bar{Q}^{\la,22}_{\frac{3}{2}}
\Big)\,,
\nonu\\
G^2
&
=&
\frac{i}{2}\,\Big(
Q^{\la,11}_{\frac{3}{2}}
+2i\sqrt{2} \,Q^{\la,21}_{\frac{3}{2}}
-2 \,Q^{\la,22}_{\frac{3}{2}}
-2\, \bar{Q}^{\la,11}_{\frac{3}{2}}
-2i\sqrt{2} \, \bar{Q}^{\la,12}_{\frac{3}{2}}
+\bar{Q}^{\la,22}_{\frac{3}{2}}
\Big)\,,
\nonu\\
G^3
&
=&
\frac{i}{2}\,\Big(
Q^{\la,11}_{\frac{3}{2}}
+i\sqrt{2} \,Q^{\la,12}_{\frac{3}{2}}
-2\,Q^{\la,22}_{\frac{3}{2}}
-2\, \bar{Q}^{\la,11}_{\frac{3}{2}}
-i \sqrt{2} \, \bar{Q}^{\la,21}_{\frac{3}{2}}
+\bar{Q}^{\la,22}_{\frac{3}{2}}
\Big)\,,
\nonu\\
G^4
&
=&
\frac{1}{2}\,Q^{\la,11}_{\frac{3}{2}}
+Q^{\la,22}_{\frac{3}{2}}
+\bar{Q}^{\la,11}_{\frac{3}{2}}
+\frac{1}{2} \, \bar{Q}^{\la,22}_{\frac{3}{2}}
\,.
\label{fourG}
\eea
We can check that the third order pole
of $G^i(z)\, G^j(w)$ is given by
$\frac{2}{3}\, c \, \de^{ij}$ with (\ref{central}).
We have seen the half of the weight-$\frac{3}{2}$ operators
in (\ref{fourG}) and the remaining ones will be given in next
subsection of the first ${\cal N}=4$ multiplet.

From now on, we construct the remaining operators
in the ${\cal N}=4$ superconformal algebra based on the
explicit expressions of (\ref{fourG}).

\subsubsection{The
 $\la$-independent
  weight-$1$ primary operators}

From the defining equations of the second order pole 
in the OPE $G^i(z)\, G^{j}(w)$,
which are given by $- 2 i \, \Big(T^{i j} +
\frac{1}{2}\, (1-4\la)\, \varepsilon^{i j k l}\, T^{k l} \Big)(w)$,
we can determine the following six
weight-$1$ operators which do not
depend on the $\la$ \footnote{We can easily see that
  the parameter $\al=\frac{1}{2}\, \frac{(k^+-k^-)}{(k^++k^-)}$
  with $k^+=k+1$ and $k^-=N+1$
  in the ${\cal N}=4$ coset model becomes
$\al =\frac{1}{2} \, (1-2 \la_{co})= \frac{1}{2} \, (1-4 \la)$
  under the large $(N,k)$ limit with fixed $\la_{co} \equiv
  \frac{(N+1)}{(k+N+2)}$ \cite{Schoutens,AK1509,AKK1910}.
  See also \cite{GG1305}.
\label{laco}}
\bea
T^{12}
&
=&
-i \,  \Big(2i\,W^{\la,11}_{\mathrm{B},1}
-\sqrt{2}\,W^{\la,12}_{\mathrm{B},1}
-2i\,\,W^{\la,22}_{\mathrm{B},1}
+2i\,W^{\la,11}_{\mathrm{F},1}
-2\sqrt{2}\,W^{\la,12}_{\mathrm{F},1}
-2i\,W^{\la,22}_{\mathrm{F},1} \Big)\,,
\nonu\\
T^{13}
&
=&
-i \, \Big(-2i\,W^{\la,11}_{\mathrm{B},1}
+4\sqrt{2}\,W^{\la,21}_{\mathrm{B},1}
+2i\,\,W^{\la,22}_{\mathrm{B},1}
-2i\,W^{\la,11}_{\mathrm{F},1}
+2\sqrt{2}\,W^{\la,21}_{\mathrm{F},1}
+2i\,W^{\la,22}_{\mathrm{F},1}\Big)\,,
\nonu\\
T^{14}
&
=& - i\,  \Big(
2\,W^{\la,11}_{\mathrm{B},1}
+i\sqrt{2}\,W^{\la,12}_{\mathrm{B},1}
+4i\sqrt{2}\,\,W^{\la,21}_{\mathrm{B},1}
-2\,W^{\la,22}_{\mathrm{B},1}
-2\,W^{\la,11}_{\mathrm{F},1}
-2i\sqrt{2}\,W^{\la,12}_{\mathrm{F},1}
\nonu \\
& - & 2i\sqrt{2}\,W^{\la,21}_{\mathrm{F},1}
+2\,W^{\la,22}_{\mathrm{F},1}\Big)\,,
\nonu\\
T^{23}
&
=& - i  \,\Big(
-2\,W^{\la,11}_{\mathrm{B},1}
-i\sqrt{2}\,W^{\la,12}_{\mathrm{B},1}
-4i\sqrt{2}\,\,W^{\la,21}_{\mathrm{B},1}
+2\,W^{\la,22}_{\mathrm{B},1}
-2\,W^{\la,11}_{\mathrm{F},1}
-2i\sqrt{2}\,W^{\la,12}_{\mathrm{F},1}
\nonu \\
& - & 2i\sqrt{2}\,W^{\la,21}_{\mathrm{F},1}
+2\,W^{\la,22}_{\mathrm{F},1} \Big)\,,
\nonu\\
T^{24}
&
=& - i  \,\Big(
-2i\,W^{\la,11}_{\mathrm{B},1}
+4\sqrt{2}\,W^{\la,21}_{\mathrm{B},1}
+2i\,\,W^{\la,22}_{\mathrm{B},1}
+2i\,W^{\la,11}_{\mathrm{F},1}
-2\sqrt{2}\,W^{\la,21}_{\mathrm{F},1}
-2i\,W^{\la,22}_{\mathrm{F},1} \Big)\,,
\nonu\\
T^{34}
&
=&
-i  \,  \Big(-2i\,W^{\la,11}_{\mathrm{B},1}
+\sqrt{2}\,W^{\la,12}_{\mathrm{B},1}
+2i\,\,W^{\la,22}_{\mathrm{B},1}
+2i\,W^{\la,11}_{\mathrm{F},1}
-2\sqrt{2}\,W^{\la,12}_{\mathrm{F},1}
-2i\,W^{\la,22}_{\mathrm{F},1}\Big)\,.
\label{spin1}
\eea
Note that the right hand sides of (\ref{spin1})
are proportional to the expressions of $\Phi_1^{(1),ij}$
at $\la=0$
when we replace the weight-$2$ in the
$W^{\bar{a} b}_{\mathrm{B},2} $ and $W^{\bar{a} b}_{\mathrm{F},2}$
\cite{AKK1910}
with the weight-$1$ by generalizing to the $\la$ dependent ones.
We can check that the first order pole of
the  OPE $G^i(z) \, G^{j}
(w)$ provides the correct quasiprimary operator $L(w)$
and the corresponding descendants of (\ref{spin1}).

\subsubsection{The
 $\la$-independent
  weight-$\frac{1}{2}$ primary operators}

Again, the defining equation of the second order pole
of the OPE $G^i(z)\, T^{j k}(w)$,
$-\Big(\varepsilon^{i j k l}\, \Ga^l + (1-4\la)\, (\de^{i k}\, \Ga^j-
\de^{i j}\, \Ga^k)\Big)(w)$ allows us to
determine the following
weight-$\frac{1}{2}$ operators
\bea
\Gamma^1
&
=&
-\frac{i}{2}\,\Big(-2\,\bar{Q}^{\la,11}_{\frac{1}{2}}
-2i \sqrt{2}\,\bar{Q}^{\la,12}_{\frac{1}{2}}
-i\sqrt{2}\,\bar{Q}^{\la,21}_{\frac{1}{2}}
+\bar{Q}^{\la,22}_{\frac{1}{2}}
\Big)\,,
\nonu\\
\Gamma^2
&
=&
-\frac{1}{2}\,\Big(
-2\, \bar{Q}^{\la,11}_{\frac{1}{2}}
-2i\sqrt{2} \, \bar{Q}^{\la,12}_{\frac{1}{2}}
+\bar{Q}^{\la,22}_{\frac{1}{2}}
\Big)\,,
\nonu\\
\Gamma^3
&
=&
-\frac{1}{2}\,\Big(
-2\, \bar{Q}^{\la,11}_{\frac{1}{2}}
-i \sqrt{2} \, \bar{Q}^{\la,21}_{\frac{1}{2}}
+\bar{Q}^{\la,22}_{\frac{1}{2}}
\Big)\,,
\nonu\\
\Gamma^4
&
=&
i \Big( \bar{Q}^{\la,11}_{\frac{1}{2}}
+\frac{1}{2} \, \bar{Q}^{\la,22}_{\frac{1}{2}} \Big)
\,.
\label{fourGa}
\eea
Note that the right hand sides of
(\ref{fourGa}) can be obtained
from (\ref{fourG}) by replacing the weight-$\frac{3}{2}$
with the weight-$\frac{1}{2}$ with an overall factor $i$.
Further analysis for the first order pole of
the OPE $G^i(z)\, T^{j k}(w)$ provides
the correct descendants of (\ref{fourGa}) and the
primary operators of weight-$\frac{3}{2}$ in (\ref{fourG}).
These are not dependent of the $\la$ because
the lowest fermionic operators $\bar{Q}^{\la,a \bar{b}}_{\frac{1}{2}}$
do not contain the $\la$ from (\ref{VVQQla}) and (\ref{WWQQnonzerola}).

\subsubsection{The
$\la$-independent
  weight-$1$ quasiprimary operator}

For the final weight-$1$ operator, we can use
the defining equation for the first order pole of the OPE
$G^i(z)\, \Ga^j(w)$
which is equal to
$\Big(-\varepsilon^{i j k l}\, T^{k l} + i \, \de^{i j}\, U
\Big)(w)$. It turns out that
\bea
U
=
2 \, \Big[ (W^{\la,11}_{\mathrm{F},1}+W^{\la,22}_{\mathrm{F},1})+
  (W^{\la,11}_{\mathrm{B},1}+W^{\la,22}_{\mathrm{B},1}) \Big]\,,
\label{U}
\eea
which does not contain the $\la$.
Therefore, the eight independent
weight-$1$ operators from $W_{B,2}^{\bar{a} b}$ and
$W_{F,2}^{\bar{a} b}$ are given by (\ref{spin1}) and (\ref{U}).
The remaining one will be given in next subsection of
the first ${\cal N}=4$ multiplet.

In next subsection, we will describe 
whether the above five kinds of operators will produce the known
${\cal N}=4$ superconformal algebra or not.

\subsection{ The OPEs between the ${\cal N}=4$ stress energy tensor
  and itself}

We calculate the OPEs between the
five operators in
the ${\cal N}=4$ stress energy tensor
and the weight-$1$ operator
in  that  multiplet.

\subsubsection{The OPE between the weight-$1,\frac{1}{2},1$
operators
  and the weight-$1$ operator}

From the explicit expressions (\ref{U}), (\ref{fourGa}), (\ref{spin1}),
(\ref{lowspincurrents}) and (\ref{fundOPE}),
we can check the following OPEs
\bea
U(z) \, U(w) & = &  + \cdots,
\nonu \\
\Gamma^i(z) \, U(w) & = & + \cdots,
\nonu \\
T^{i j}(z) \, U(w) & = &  + \cdots.
\label{trivialope}
\eea
The last two are the standard results \cite{Schoutens,AK1509}
while the first one
is rather trivial result. This is due to the fact that
the expression for the $U$ in (\ref{U})has the
same relative coefficients. The standard result for the first one
leads to the nontrivial second order pole, which is given by
a central term.
We
expect that the OPE between $L(z) \, U(w)$ contains the third
order pole because the OPE $\pa \, U(z) \, U(w)$ has no singular
term. 

\subsubsection{The OPE between the weight-$\frac{3}{2}$ operators
  and the weight-$1$ operator}

Similarly, by using (\ref{fourG}), (\ref{U})
and previous defining equations, we obtain
the following OPE
\bea
G^{i}(z) \, U(w) & = & 
-\frac{1}{(z-w)^{2}} \, \Bigg[i\:\Gamma^{i}\Bigg](w)-
\frac{1}{(z-w)} \, \Bigg[i\:\partial\Gamma^{i}\Bigg](w) +\cdots.
\label{GiU}
\eea
The weight-$1$ operator $U$ plays the role of keeping
the structure of the weight-$\frac{3}{2}$ operator on the left hand side
with the weight reduced to $\frac{1}{2}$. The relative coefficient
for the descendant can be seen from the standard conformal field theory
analysis \footnote{
  There is 
  \bea
\Big(G^{i} - i (1-4\la)\, \pa \, \Ga^i\Big)(z) \, U(w) & = & 
-\frac{1}{(z-w)^{2}} \, \Bigg[i\:\Gamma^{i}\Bigg](w)-
\frac{1}{(z-w)} \, \Bigg[i\:\partial\Gamma^{i}\Bigg](w) +\cdots,
\label{GitildeU}
\eea
due to one of the relations in (\ref{trivialope}).
This will be used in the ${\cal N}=4$ superspace description.}.

\subsubsection{The OPE between the weight-$2$ operator
  and the weight-$1$ operator}

Let us consider the final nontrivial OPE
with (\ref{Lterm}) and (\ref{U}).
It turns out that there exists 
\bea
L(z) \, U(w) & = &-\frac{1}{(z-w)^3}\, \Bigg[N\Bigg]+
\frac{1}{(z-w)^{2}} \, \Bigg[U\Bigg](w)+
\frac{1}{(z-w)} \, \Bigg[\pa U\Bigg](w) +\cdots.
\label{LU}
\eea
Compared to the standard result \cite{Schoutens,AK1509},
the above OPE contains the central term, as mentioned before.
This is due to the fact that the expression of (\ref{U})
has particular relative coefficients
\footnote{We have
  \bea
  \Big( L -\frac{1}{2}\, (1-4\la)\, \pa \, U\Big)(z) \,
  U(w) & = &-\frac{1}{(z-w)^3}\, N+
\frac{1}{(z-w)^{2}} \, U(w)+
\frac{1}{(z-w)} \, \pa U(w) +\cdots,
\label{LtildeU}
\eea
which will be used later.
}.

\subsubsection{The ${\cal N}=4$ supersymmetric OPE in
the ${\cal N}=4$ superspace}

We can put the operators of
${\cal N}=4$ superconformal algebra found in previous section
into a super field in the ${\cal N}=4$ superspace
\footnote{The coordinates of ${\cal N}=4$  superspace 
can be described as 
$(Z, \overline{Z})$ where 
$Z=(z, \theta^i)$, $\overline{Z} =(\bar{z}, \bar{\theta}^i)$
and the $SO(4)$-vector index $i$ runs over $i=1, \cdots, 4$. 
The left covariant spinor derivative 
is given by 
$ D^i = \theta^i \frac{\pa}{\pa z } +  \frac{\pa}{\pa {\theta^i}}$
with nontrivial anticommutators 
$
\{ D^i, D^j \} = 2 \delta^{ij} \frac{\pa}{\pa z}$.
The simplified notation
$\theta^{4-0}$ stands for 
$\theta^1 \,\theta^2 \,\theta^3 \,\theta^4$. 
The complement $4-i$ is defined such that
$\theta^1 \,\theta^2 \,\theta^3 \, \theta^{4} = \theta^{4-i}\,
\theta^{i}$ \cite{Schoutens,AK1509}.
}.
Then we have the ${\cal N}=4$ stress energy tensor
\cite{Schoutens,AK1509}
\bea
    {\bf J} & = &
- \Delta + i \, \theta^{j} \, \Gamma^{j}-
i \, \theta^{4-jk} \, T^{jk}-\theta^{4-j} \, (G^{j}-
i \, (1-4\la)  \, \partial \, \Gamma^{j})+\theta^{4-0}  
\, (2 \, L-\frac{1}{2}\, (1-4\la)\, \partial^2 \, \Delta)
\nonu \\
    & \equiv & \Bigg(-\Delta, \, i \Gamma^i,
\, -i \, T^{ij}, \, - \Big(G^i-i \, (1-4\la)\, \pa \, \Gamma^i
\Big), \, 2 \Big(L-\frac{1}{2}\,
(1-4\la)\,
\pa^2 \, \Delta \Big)\Bigg),
\label{BigJ}
\eea
where the lowest component has the following
relation $ - \pa\, \Delta \equiv U$.
We will use the operator $U$ rather than
the operator $\Delta$.
The precise relations between the components and its
superfields in (\ref{BigJ})
at vanishing fermionic coordinates can be summarized by
\cite{Schoutens,AK1509}
\bea
U   &
\leftrightarrow &   \pa \, {\bf J},
\qquad
\Gamma^i  \leftrightarrow  -i \, D^i {\bf J},
\qquad
T^{ij}  \leftrightarrow  -\frac{i}{2!} \, \varepsilon^{ijkl} \,
D^k D^l {\bf J},
\nonu \\
G^i
& \leftrightarrow &  -\frac{1}{3!} \,
\varepsilon^{ijkl} \, D^j D^k D^l {\bf J}
-(1-4\la)\, \pa \, D^i \, {\bf J},
\nonu \\
L  & \leftrightarrow & \frac{1}{2 \cdot 4!} \,
\varepsilon^{ijkl} \, D^{i} D^j D^k D^l {\bf J}-
 \frac{1}{2}\, (1-4\la)\,
\pa^2 \, {\bf J}.
\label{comptosuper1}
\eea
Due to the extra terms in the fourth and fifth elements
of the ${\cal N}=4$ stress energy tensor,
there are extra terms in the corresponding
expressions of (\ref{comptosuper1}).

Then we can write down the previous
equations (\ref{trivialope}), (\ref{GiU}) (or (\ref{GitildeU}))
and (\ref{LU}) (or (\ref{LtildeU}))
including other OPEs in the component approach
in terms of the following single OPE
in the ${\cal N}=4$ superspace
\bea
{\bf J}(Z_{1})\,{\bf J}(Z_{2})  & = & 
-\frac{\theta_{12}^{4-0}}{z_{12}^{2}}\,
N
+\frac{\theta_{12}^{4-i}}{z_{12}}\,
D^i\, {\bf J}
(Z_{2})
+\frac{\theta_{12}^{4-0}}{z_{12}}\,
2\, \pa\, {\bf J}
(Z_{2}) + \cdots.
\label{JJ}
\eea
Compared to the standard expression \cite{Schoutens,AK1509},
there is no $\mbox{log}(z_{12})$ term \footnote{
  We have the corresponding OPE 
  \bea
{\bf J}(Z_{1})\, \pa \, {\bf J}(Z_{2})  & = & 
-\frac{\theta_{12}^{4-0}}{z_{12}^{3}}\,
2 \, N
+\frac{\theta_{12}^{4-i}}{z^2_{12}}\,
D^i\, {\bf J}
(Z_{2})
+\frac{\theta_{12}^{4-0}}{z^2_{12}}\,
2\, \pa\, {\bf J}
(Z_{2}) +
\frac{\theta_{12}^{4-i}}{z_{12}}\,
\pa \, D^i\, {\bf J}
(Z_{2})
+\frac{\theta_{12}^{4-0}}{z_{12}}\,
2\, \pa^2 \, {\bf J}
(Z_{2})
\nonu \\
& + & \cdots.
\label{JJ'}
\eea
From (\ref{JJ'}) which is more relevant to
the previous three component results ((\ref{trivialope}),
(\ref{GitildeU})
and (\ref{LtildeU})), we obtain (\ref{JJ})
after the integrations.
}.

In Appendix $A$, all the component OPEs  are summarized
explicitly \footnote{Compared to the
construction in \cite{AKK1910}, the presence of
$W_{B,1}^{\la,\bar{a} b}$ and $\bar{Q}_{\frac{1}{2}}^{\la,b \bar{a}}$
in (\ref{spin1}), (\ref{fourGa}) and (\ref{U})
is new and these will change the structure of the algebra.}.
It is straightforward to obtain these OPEs
from (\ref{JJ}) by acting the various super derivatives
with the relations (\ref{comptosuper1}).

\subsection{The construction of  the first ${\cal N}=4$ multiplet}

\subsubsection{ The $\la$-dependent
  weight-$1$ primary operator}

Let us start with the final weight-$1$ primary operator
of the ${\cal N}=2$ superconformal algebra \cite{Ahn2203,Ahn2107}
\bea
\Phi^{(1)}_{0}
=
4 \, \Bigg[ (1-2\la)\, (W^{\la,11}_{\mathrm{F},1}+W^{\la,22}_{\mathrm{F},1})-
2\la \,(W^{\la,11}_{\mathrm{B},1}+W^{\la,22}_{\mathrm{B},1})\Bigg]\,.
\label{spinone}
\eea
The field contents of (\ref{spinone}) are the
same as the one in (\ref{U}).
At $\la=0$, the only first two terms in
(\ref{spinone}) contribute to the final expression
and reproduces the one in \cite{AKK1910}.
Compared to the previous construction on the weight-$1$
operator,
the $\la$-dependent coefficients appear in the above.

\subsubsection{ The $\la$-dependent weight-$\frac{3}{2}$
  primary operators}

From the defining equation \cite{Schoutens,BCG,AK1509} of
\bea
G^{i}(z) \, \Phi_{0}^{(1)}(w)\Bigg|_{\frac{1}{(z-w)}} & = & 
- \Phi_{\frac{1}{2}}^{(1),i}(w),
\label{Relone}
\eea
we can determine 
the following primary (under the stress energy tensor
(\ref{Lterm})) operators of weight-$\frac{3}{2}$
\bea
\Phi^{(1),1}_{\frac{1}{2}}
&
=&
\frac{1}{2}\,\Big(
Q^{\la,11}_{\frac{3}{2}}
+i\sqrt{2}\,Q^{\la,12}_{\frac{3}{2}}
+2i \sqrt{2}\,Q^{\la,21}_{\frac{3}{2}}
-2\,Q^{\la,22}_{\frac{3}{2}}
\nonu \\
& + & 2\,\bar{Q}^{\la,11}_{\frac{3}{2}}
+2i \sqrt{2}\, \bar{Q}^{\la,12}_{\frac{3}{2}}
+i\sqrt{2}\,\bar{Q}^{\la,21}_{\frac{3}{2}}
-\bar{Q}^{\la,22}_{\frac{3}{2}}
\Big)\,,
\nonu\\
\Phi^{(1),2}_{\frac{1}{2}}
&
=&
-\frac{i}{2}\,\Big(
Q^{\la,11}_{\frac{3}{2}}
+2i\sqrt{2} \,Q^{\la,21}_{\frac{3}{2}}
-2 \,Q^{\la,22}_{\frac{3}{2}}
+2\, \bar{Q}^{\la,11}_{\frac{3}{2}}
+2i\sqrt{2} \, \bar{Q}^{\la,12}_{\frac{3}{2}}
-\bar{Q}^{\la,22}_{\frac{3}{2}}
\Big)\,,
\nonu\\
\Phi^{(1),3}_{\frac{1}{2}}
&
=&
-\frac{i}{2}\,\Big(
Q^{\la,11}_{\frac{3}{2}}
+i\sqrt{2} \,Q^{\la,12}_{\frac{3}{2}}
-2\,Q^{\la,22}_{\frac{3}{2}}
+2\, \bar{Q}^{\la,11}_{\frac{3}{2}}
+i \sqrt{2} \, \bar{Q}^{\la,21}_{\frac{3}{2}}
-\bar{Q}^{\la,22}_{\frac{3}{2}}
\Big)\,,
\nonu\\
\Phi^{(1),4}_{\frac{1}{2}}
&
=&
-\frac{1}{2}\,Q^{\la,11}_{\frac{3}{2}}
-Q^{\la,22}_{\frac{3}{2}}
+\bar{Q}^{\la,11}_{\frac{3}{2}}
+\frac{1}{2}\,\bar{Q}^{\la,22}_{\frac{3}{2}}
\,.
\label{higherspin3half}
\eea
The field contents of (\ref{higherspin3half})
are the same as the ones in (\ref{fourG}).
The only difference appears in the minus signs
of  $Q^{\la, \bar{a} b}_{\frac{3}{2}}$.
Then we have the complete weight-$\frac{3}{2}$ operators
in (\ref{fourG}) and (\ref{higherspin3half}).
Compared to the $\la=0$ case in \cite{AKK1910},
the generalization of the fermionic operators
to the nonzero $\la$ case in
(\ref{WWQQnonzerola})
provides the exact relative coefficients in (\ref{higherspin3half}).
In other words, for the expressions in \cite{AKK1910}
at $\la=0$, a simple generalization of (\ref{WWQQnonzerola})
leads to the above result in (\ref{higherspin3half}).
This is also true for other remaining operators of
weights-$2, \frac{5}{2},3$.

\subsubsection{ The $\la$-dependent
  weight-$2$ primary operators}

By using the following defining
equation \cite{AK1509}
\bea
G^{i}(z) \, \Phi_{\frac{1}{2}}^{(1),j}(w)\Bigg|_{\frac{1}{(z-w)}} & = & 
-
 \Bigg[ \delta^{ij}\, \partial\Phi_{0}^{(1)}-
   \frac{1}{2} \, \varepsilon^{i j k l} \,
   \Phi_{1}^{(1),k l} \Bigg](w),
\label{Reltwo}
 \eea
we obtain the weight-$2$ primary operators, by taking
two different indices, as follows:
\bea
\Phi^{(1),12}_{1}
&
=&
2i\,W^{\la,11}_{\mathrm{B},2}
-\sqrt{2}\,W^{\la,12}_{\mathrm{B},2}
-2i\,\,W^{\la,22}_{\mathrm{B},2}
+2i\,W^{\la,11}_{\mathrm{F},2}
-2\sqrt{2}\,W^{\la,12}_{\mathrm{F},2}
-2i\,W^{\la,22}_{\mathrm{F},2}\,,
\nonu\\
\Phi^{(1),13}_{1}
&
=&
-2i\,W^{\la,11}_{\mathrm{B},2}
+4\sqrt{2}\,W^{\la,21}_{\mathrm{B},2}
+2i\,\,W^{\la,22}_{\mathrm{B},2}
-2i\,W^{\la,11}_{\mathrm{F},2}
+2\sqrt{2}\,W^{\la,21}_{\mathrm{F},2}
+2i\,W^{\la,22}_{\mathrm{F},2}\,,
\nonu\\
\Phi^{(1),14}_{1}
&
=&
2\,W^{\la,11}_{\mathrm{B},2}
+i\sqrt{2}\,W^{\la,12}_{\mathrm{B},2}
+4i\sqrt{2}\,\,W^{\la,21}_{\mathrm{B},2}
-2\,W^{\la,22}_{\mathrm{B},2}
-2\,W^{\la,11}_{\mathrm{F},2}
-2i\sqrt{2}\,W^{\la,12}_{\mathrm{F},2}
\nonu \\
& - & 2i\sqrt{2}\,W^{\la,21}_{\mathrm{F},2}
+2\,W^{\la,22}_{\mathrm{F},2}\,,
\nonu\\
\Phi^{(1),23}_{1}
&
=&
-2\,W^{\la,11}_{\mathrm{B},2}
-i\sqrt{2}\,W^{\la,12}_{\mathrm{B},2}
-4i\sqrt{2}\,\,W^{\la,21}_{\mathrm{B},2}
+2\,W^{\la,22}_{\mathrm{B},2}
-2\,W^{\la,11}_{\mathrm{F},2}
-2i\sqrt{2}\,W^{\la,12}_{\mathrm{F},2}
\nonu \\
& - & 2i\sqrt{2}\,W^{\la,21}_{\mathrm{F},2}
+2\,W^{\la,22}_{\mathrm{F},2}\,,
\nonu\\
\Phi^{(1),24}_{1}
&
=&
-2i\,W^{\la,11}_{\mathrm{B},2}
+4\sqrt{2}\,W^{\la,21}_{\mathrm{B},2}
+2i\,\,W^{\la,22}_{\mathrm{B},2}
+2i\,W^{\la,11}_{\mathrm{F},2}
-2\sqrt{2}\,W^{\la,21}_{\mathrm{F},2}
-2i\,W^{\la,22}_{\mathrm{F},2}\,,
\nonu\\
\Phi^{(1),34}_{1}
&
=&
-2i\,W^{\la,11}_{\mathrm{B},2}
+\sqrt{2}\,W^{\la,12}_{\mathrm{B},2}
+2i\,\,W^{\la,22}_{\mathrm{B},2}
+2i\,W^{\la,11}_{\mathrm{F},2}
-2\sqrt{2}\,W^{\la,12}_{\mathrm{F},2}
-2i\,W^{\la,22}_{\mathrm{F},2}\,.
\label{spintwo}
\eea
We observe, as described before, that
by taking the corresponding expressions
for the weight-$1$ operators at $\la =0$
and replacing them with the ones in (\ref{WWQQnonzerola}),
the above results can be obtained.
So far, we have obtained the seven weight-$2$ operators
consisting of (\ref{Lterm}) and (\ref{spintwo}) and the
the remaining one will appear in the lowest component of
the second ${\cal N}=4$ multiplet.

\subsubsection{ The
$\la$-dependent
  weight-$\frac{5}{2}$ quasiprimary operators}

From the defining equation \cite{AK1509} of
\bea
G^{i}(z) \, \Phi_{1}^{(1),jk}(w)\Bigg|_{\frac{1}{(z-w)}} & = & 
-  \Bigg[ 
\Big( \delta^{ij}\, \Phi_{\frac{3}{2}}^{(1),k}- \delta^{ik} \, 
\Phi_{\frac{3}{2}}^{(1),j}\Big)+
\varepsilon^{i j k l} \,
\partial\, \Phi_{\frac{1}{2}}^{(1),l} \Bigg](w),
\label{Relthree}
\eea
we can determine the following weight-$\frac{5}{2}$
quasiprimary operators
\bea
\tilde{\Phi}^{(1),1}_{\frac{3}{2}}
&
\equiv &
\Phi^{(1),1}_{\frac{3}{2}} -\frac{1}{3}\, (1-4\la)\,
\pa \,\Phi^{(1),1}_{\frac{1}{2}}
\nonu \\
&=& -\frac{1}{2}\,\Big(
Q^{\la,11}_{\frac{5}{2}}
+i\sqrt{2}\,Q^{\la,12}_{\frac{5}{2}}
+2i\sqrt{2}\,Q^{\la,21}_{\frac{5}{2}}
-2\,Q^{\la,22}_{\frac{5}{2}}
\nonu \\
& - & 2\,\bar{Q}^{\la,11}_{\frac{5}{2}}
-2i\sqrt{2}\,\bar{Q}^{\la,12}_{\frac{5}{2}}
-i\sqrt{2}\,\bar{Q}^{\la,21}_{\frac{5}{2}}
+\bar{Q}^{\la,22}_{\frac{5}{2}}
\Big)\,,
\nonu\\
\tilde{\Phi}^{(1),2}_{\frac{3}{2}}
&
\equiv &
\Phi^{(1),2}_{\frac{3}{2}} -\frac{1}{3}\, (1-4\la)\,
\pa \,\Phi^{(1),2}_{\frac{1}{2}}
\nonu \\
&
=&
\frac{i}{2}\,\Big(\,
Q^{\la,11}_{\frac{5}{2}}
+2i\sqrt{2}\,Q^{\la,21}_{\frac{5}{2}}
-2\,Q^{\la,22}_{\frac{5}{2}}
-2\,\bar{Q}^{\la,11}_{\frac{5}{2}}
-2i\sqrt{2}\,\bar{Q}^{\la,12}_{\frac{5}{2}}
+\bar{Q}^{\la,22}_{\frac{5}{2}}
\Big)\,,
\nonu\\
\tilde{\Phi}^{(1),3}_{\frac{3}{2}}
&
\equiv &
\Phi^{(1),3}_{\frac{3}{2}} -\frac{1}{3}\, (1-4\la)\,
\pa \,\Phi^{(1),3}_{\frac{1}{2}}
\nonu \\
&
=&
\frac{i}{2}\,\Big(
Q^{\la,11}_{\frac{5}{2}}
+i\sqrt{2}\,Q^{\la,12}_{\frac{5}{2}}
-2\,Q^{\la,22}_{\frac{5}{2}}
-2\,\bar{Q}^{\la,11}_{\frac{5}{2}}
-i\sqrt{2}\,\bar{Q}^{\la,21}_{\frac{5}{2}}
+\bar{Q}^{\la,22}_{\frac{5}{2}}
\Big)\,,
\nonu\\
\tilde{\Phi}^{(1),4}_{\frac{3}{2}}
&
\equiv &
\Phi^{(1),4}_{\frac{3}{2}} -\frac{1}{3}\, (1-4\la)\,
\pa \,\Phi^{(1),4}_{\frac{1}{2}}
\nonu \\
&
=&
\frac{1}{2}\,
\Big(
Q^{\la,11}_{\frac{5}{2}}
+2\,Q^{\la,22}_{\frac{5}{2}}
+2\,\bar{Q}^{\la,11}_{\frac{5}{2}}
+\bar{Q}^{\la,22}_{\frac{5}{2}}
\Big)\,.
\label{spin5half}
\eea
Note that by starting with (\ref{higherspin3half})
with minus signs in $Q^{\la, \bar{a} b}_{\frac{3}{2}}$
and increasing the weights by one, we reproduce the
above results (\ref{spin5half}).
For the case of $\la=0$ in \cite{AKK1910}, the
weight-$\frac{5}{2}$ operators are primary
but at nonzero $\la$, the above operators (\ref{spin5half})
are quasiprimary under the stress energy tensor (\ref{Lterm}),
although we are using the same notation
\footnote{The operators $\Phi^{(1),i}_{\frac{3}{2}}$,
  which are the component of ${\cal N}=4$ superfield
  later,
are not quasiprimary. See also Appendix $B$.}.
We expect that the half of other weight-$\frac{5}{2}$
operators will appear in the second ${\cal N}=4$ multiplet.

\subsubsection{ The
$\la$-dependent
  weight-$3$ quasiprimary operator}

Finally, by using the following
defining equation \cite{AK1509}
\bea
G^{i}(z) \, \Phi_{\frac{3}{2}}^{(1),j}(w)\Bigg|_{\frac{1}{(z-w)}} & = & 
 -  
 \Bigg[ \partial\Phi_{1}^{(1),ij}+ 
 \delta^{ij} \, \Phi_{2}^{(1)}\Bigg](w),
\label{Relfour}
 \eea
 we obtain the weight-$3$ quasiprimary
 operator, by taking two equal indices,
\bea
\tilde{\Phi}^{(1)}_{2}
&
\equiv &
\Phi^{(1)}_{2} -\frac{1}{3}\, (1-4\la)\,
\pa^2 \,\Phi^{(1)}_{0}
=
-2\,\Big(
W^{\la,11}_{\mathrm{B},3}
+W^{\la,22}_{\mathrm{B},3}
+W^{\la,11}_{\mathrm{F},3}
+W^{\la,22}_{\mathrm{F},3}
\Big).
\label{freespin3}
\eea
We observe that by increasing the weight by one  from
the stress energy tensor (\ref{Lterm}), the above
expression can be seen with the overall factor.
Compared to the $\la=0$ case in \cite{AKK1910}
where the corresponding operator is primary,
the above operator is quasiprimary
\footnote{The operator $\Phi^{(1)}_{2}$
is not quasiprimary and see also Appendix $B$.}.
The remaining seven other weight-$3$ operators will appear
in next ${\cal N}=4$ multiplets.
Six of them appear in the second ${\cal N}=4$ multiplet
and one of them appears in the third ${\cal N}=4$ multiplet.

\subsection{ The OPEs between the ${\cal N}=4$ stress energy tensor
  and the first ${\cal N}=4$ multiplet}

We calculate the OPEs between the
five kinds of operators in
the ${\cal N}=4$ stress energy tensor
and the lowest weight-$1$ operator
in  the first ${\cal N}=4$ multiplet.

\subsubsection{The OPE between the weight-$1$
  operator and the
  weight-$1$ operator}

From the explicit expressions in (\ref{U}) and (\ref{spinone}),
the following OPE can be obtained
\bea
U(z) \, \Phi_{0}^{(1) }(w) & = & 
\frac{1}{(z-w)^2}\Bigg[  N\Bigg] + \cdots.
\label{OPE1}
\eea
Compared to the standard result \cite{AK1509} which is trivial,
the above OPE has singular term on the right hand side.
This is due to the fact that this weight-$1$ operator has the
particular coefficients.

\subsubsection{The OPE between the weight-$\frac{1}{2}$ operators
  and the weight-$1$ operator}

Similarly, we obtain the following OPE
from (\ref{fourGa}) and the previous
weight-$1$ operator
\bea
\Gamma^{i}(z) \, \Phi_{0}^{(1)}(w) & = &
-\frac{1}{(z-w)}\Bigg[ \Ga^i\Bigg](w)+ \cdots.
\label{OPE2}
\eea
This implies that the weight-$1$ operator preserve
the structure of the weight-$\frac{1}{2}$ operator on the
left hand side and this is new, compared to the standard
result \cite{AK1509}.

\subsubsection{The OPE between the weight-$1$ operators
  and the weight-$1$ operator}

For the weight-$1$ operator (\ref{spin1}), we obtain the following
trivial result
\bea
T^{i j}(z) \, \Phi_{0}^{(1)}(w) & = &
+ \cdots.
\label{OPE3}
\eea

\subsubsection{The OPE between the weight-$\frac{3}{2}$ operators
  and the weight-$1$ operator}

By using (\ref{fourG}) and (\ref{spinone}), the
following OPE is satisfied 
\bea
G^{i}(z) \, \Phi_{0}^{(1)}(w) & = & 
-\frac{1}{(z-w)} \, \Phi_{\frac{1}{2}}^{(1),i}(w)
+\cdots.
\label{OPE4}
\eea
Under the action of the weight-$1$ operator,
the numerical coefficients appearing in the
weight-$\frac{3}{2}$ operators are shifted to the ones in
the weight-$\frac{3}{2}$ operators appearing on the right hand side
\footnote{We have the OPE
  \bea
  \Big(G^{i} - i \, (1-4\la)\, \pa \, \Ga^i \Big)(z) \,
  \Phi_{0}^{(1)}(w) & = & 
  -\frac{1}{(z-w)^2}\, i \, (1-4\la)\, \Ga^i(w)-
  \frac{1}{(z-w)} \, \Phi_{\frac{1}{2}}^{(1),i}(w)
+\cdots,
\label{Gitildespin1}
\eea
which will be used in the ${\cal N}=4$ superspace description.}.

\subsubsection{The OPE between the weight-$2$ operator
  and the weight-$1$ operator}

Finally, the last fundamental OPE
from the stress energy tensor (\ref{Lterm})
can be summarized by
\bea
L(z) \, \Phi_{0}^{(1)}(w) & = & 
\frac{1}{(z-w)^{2}} \, \Phi_{0}^{(1)}(w)+
\frac{1}{(z-w)} \, \partial\Phi_{0}^{(1)}(w)
+\cdots.
\label{OPE5}
\eea
This implies that the weight-$1$ operator is primary
\footnote{Similarly, the following OPE
  can be obtained
  \bea
  \Big( L + \frac{1}{2}\, (1-4\la)\, \pa \, U \Big)(z) \,
  \Phi_{0}^{(1)}(w) & = & 
-\frac{1}{(z-w)^3}\, N \, (1-4\la)+  
  \frac{1}{(z-w)^{2}} \, \Phi_{0}^{(1)}(w)+
\frac{1}{(z-w)} \, \partial\Phi_{0}^{(1)}(w)
\nonu \\
& + & \cdots,
\label{Ltildespin1}
\eea
which will be used in the ${\cal N}=4$ superspace description.}.

\subsubsection{The ${\cal N}=4$ supersymmetric OPE in
the ${\cal N}=4$ superspace}

As before, we write down each component operator
of the first ${\cal N}=4$ multiplet in the
${\cal N}=4$ superspace as follows \cite{Schoutens,BCG,AK1509}:
\bea
    {\bf \Phi^{(1)}} & = &
    \Phi_{0}^{(1)}+\theta^{i}\:\Phi_{\frac{1}{2}}^{(1),i}+
\theta^{4-ij}\:\Phi_{1}^{(1),ij}+\theta^{4-i}\:\Phi_{\frac{3}{2}}^{(1),i}
+\theta^{4-0}\:\Phi_{2}^{(1)}
\nonu \\
   & \equiv & \Bigg(
\Phi_{0}^{(1)},
\, \Phi_{\frac{1}{2}}^{(1),i},
\, \Phi_{1}^{(1),ij},
\, \Phi_{\frac{3}{2}}^{(1),i},
\, \Phi_{2}^{(1)} \Bigg), \qquad i,j = 1, \cdots, 4.
\label{Phiexp}
\eea
The precise relations between the components and its
superfields in (\ref{Phiexp}) can be described by
\cite{Schoutens,AK1509}, similar to
(\ref{comptosuper1}),
\bea
\Phi_{0}^{(1)}  & \leftrightarrow &  
    {\bf \Phi}^{(1)},
    \qquad
\Phi_{\frac{1}{2}}^{(1),i}  \leftrightarrow   D^i
    {\bf \Phi}^{(1)},  
\nonu \\
\Phi_{1}^{(1),ij}  & \leftrightarrow &  -\frac{1}{2!} \,
 \varepsilon^{ijkl} \, D^k D^l {\bf \Phi}^{(1)},
\qquad
\Phi_{\frac{3}{2}}^{(1),i}  \leftrightarrow  \frac{1}{3!} \,
\varepsilon^{ijkl} \, D^j D^k D^l {\bf \Phi}^{(1)},
\nonu \\
\Phi_{2}^{(1)}   & \leftrightarrow &  \frac{1}{ 4!} \,
\varepsilon^{ijkl} \, D^{i} D^j D^k D^l {\bf \Phi}^{(1)}.
\label{comptosuper2}
\eea
In other words,
by taking the fermionic coordinates on the right hand sides
to zero,
we obtain the corresponding operators on the left hand sides.

Therefore, the previous equations (\ref{OPE1}), (\ref{OPE2}),
(\ref{OPE3}), (\ref{OPE4}) (or (\ref{Gitildespin1}))
and (\ref{OPE5}) (or (\ref{Ltildespin1})), including
other various OPEs,
can be rewritten as 
\bea
{\bf J}(Z_{1})\,{\bf \Phi}^{(1)}(Z_{2})  & = & 
\Bigg[-\frac{\theta_{12}^{4-0}}{z_{12}^{3}}\,
2\,N\,(1-4\la) 
+\frac{\theta_{12}^{4-i}}{z_{12}^2}\,
 (1-4\la)\, D^i\, {\bf J}
(Z_{2})
-\frac{\theta_{12}^{i}}{z_{12}}\,
D^i \, {\bf J}
- \frac{1}{z_{12}} 
\, N \Bigg] \nonu \\
&+ & \frac{\theta_{12}^{4-0}}{z_{12}^{2}}\,
2\, {\bf \Phi}^{(1)}(Z_{2})
+\frac{\theta_{12}^{4-i}}{z_{12}}\,
D^i\, {\bf \Phi}^{(1)}(Z_{2})
+\frac{\theta_{12}^{4-0}}{z_{12}}\,
 2\, \pa \,  {\bf \Phi}^{(1)}(Z_{2})
 +\cdots.
 \label{JPhi}
\eea
This implies that the first ${\cal N}=4$ multiplet is not
a primary operator under the ${\cal N}=4$ supersymmetry
because there are the first four terms in (\ref{JPhi}).

In Appendix $B$, we present all the component OPEs explicitly.
As before, these can be checked by using the super derivatives
in \cite{AK1509}.

\subsection{ The OPEs between the first ${\cal N}=4$ multiplet and itself} 

We calculate the OPEs between the
five kinds of operators in
the   first ${\cal N}=4$ multiplet
and the lowest weight-$1$ operator
in that multiplet.

\subsubsection{The OPE between the weight-$1$ operator and itself}

By using the equations, (\ref{spinone}), (\ref{lowspincurrents}),
(\ref{VVQQla}) and (\ref{fundOPE}),
we obtain the following OPE
\bea
\Phi^{(1)}_0(z)\,\Phi^{(1)}_0(w)
&=&
\frac{1}{(z-w)^2}\,
\Bigg[2\,N(1-4\lambda)\Bigg]+\cdots.
\label{firstopein35}
\eea
Note the presence of the factor $(1-4\la)$ in the above.

\subsubsection{The OPE between the weight-$\frac{3}{2}$ operators
  and the weight-$1$ operator}

With the explicit expressions (\ref{higherspin3half}) and
the previous defining relations, we obtain the following
OPEs
\bea
\Phi^{(1),i}_{\frac{1}{2}}(z)\,\Phi^{(1)}_0(w)
&=&
-\frac{1}{(z-w)}\,
\Bigg[ G^i\Bigg](w)+\cdots.
\label{secondopein35}
\eea
This implies that the role of the
weight-$1$ operator
$\Phi^{(1)}_0(w)$ in this OPE changes the signs of
$\bar{Q}_{\frac{3}{2}}^{\la, \bar{a} b}$.
This leads to the first order pole on the
right hand side of the above OPE.

\subsubsection{The OPE between the weight-$2$ operators
  and the weight-$1$ operator}

With the help of (\ref{spintwo}),
we determine the following OPEs
\bea
\Phi^{(1),i j}_{1}(z)\,\Phi^{(1)}_0(w)
&=&
\frac{1}{(z-w)^2}\,\Bigg[ 2 i \, (1-4\la)\,T^{i j}+
  i \, \varepsilon^{i j k l}\, T^{k l} \Bigg](w)
\nonu \\
&+&  \frac{1}{(z-w)}\,\Bigg[ 2 i \, (1-4\la)\,\pa \, T^{i j}+
  i \, \varepsilon^{i j k l}\,\pa \,  T^{k l}  \Bigg](w)
m+\cdots.
\label{thirdopein35}
\eea
In this case, the role of
$\Phi^{(1)}_0(w)$ in this OPE decreases the weight of the
weight-$2$ operator on the left hand side by one and it turns out that
the second order pole of above OPE is a linear combination of
(\ref{spin1}). We have mentioned that there are some similarities
in the weight-$2$ operator $\Phi^{(1),i j}_{1}$ and the
weight-$1$ operator $T^{i j}$.
Furthermore, the relative coefficients $1$ and $1$
between the second and first order poles
on the right hand side can be understood
from the property of the standard conformal field theory
analysis based on the weights of  $\Phi^{(1),i j}_{1}$,
$\Phi^{(1)}_0$ and $T^{ij}$ which are primary under the stress energy
tensor.

\subsubsection{The OPE between the weight-$\frac{5}{2}$ operators
  and the weight-$1$ operator}

From the expressions of (\ref{spin5half}), we determine the
following OPEs
\bea
\Phi^{(1),i }_{\frac{3}{2}}(z)\,\Phi^{(1)}_0(w)
&=&
\frac{1}{(z-w)^3}\,\Bigg[ 16 i \, \la \, (1-2\la)\,\Ga^{i} \Bigg](w)
\nonu \\
& + & \frac{1}{(z-w)^2}\,\Bigg[ 32 i \, \la \, (1-2\la)\,
  \pa \, \Ga^{i}+3(1-4\la)\, G^i \Bigg](w)
\nonu \\
&+&
\frac{1}{(z-w)}\,\Bigg[ 24 i \, \la \, (1-2\la)\,
  \pa^2 \, \Ga^{i}+\frac{8}{3}(1-4\la)\, \pa \,
  G^i + \frac{1}{2}\, \Phi_{\frac{1}{2}}^{(2),i} \Bigg](w)
\nonu \\
& + & \cdots.
\label{fourthopein35}
\eea
Note that the weight-$\frac{5}{2}$ operator
on the left hand side is not a quasiprimary
operator, as mentioned before.
Therefore, the coefficients of the descendants in the above
OPE are not known in general. It turns out that
according to the realization of the $\beta \, \ga$
and $b \, c$ ghost system, we obtain the above
result. If we use the quasiprimary weight-$\frac{5}{2}$
operator $\tilde{\Phi}^{(1),i }_{\frac{3}{2}}$, then
the coefficient of the $G^i$ in the second order pole 
is given by $\frac{8}{3}(1-4\la)$ and others remain the same.
The contributions from the extra terms in
the  $\tilde{\Phi}^{(1),i }_{\frac{3}{2}}$
can be used from (\ref{secondopein35}).
The first order pole in the above does not change
when we use the weight-$\frac{5}{2}$ operator
$\Phi^{(1),i }_{\frac{3}{2}}$ or
$\tilde{\Phi}^{(1),i }_{\frac{3}{2}}$.

After subtracting the descendants in the first order pole,
we are left with a new quasiprimary operator which cannot be
written in terms of the previously known operators.
It turns out that there are 
\bea
\Phi^{(2),1}_{\frac{1}{2}}
&
=&
-2 \times \Bigg[\frac{1}{2}\,\Big(
Q^{\la,11}_{\frac{5}{2}}
+i\sqrt{2}\,Q^{\la,12}_{\frac{5}{2}}
+2i \sqrt{2}\,Q^{\la,21}_{\frac{5}{2}}
-2\,Q^{\la,22}_{\frac{5}{2}}
\nonu \\
& + & 2\,\bar{Q}^{\la,11}_{\frac{5}{2}}
+2i \sqrt{2}\, \bar{Q}^{\la,12}_{\frac{5}{2}}
+i\sqrt{2}\,\bar{Q}^{\la,21}_{\frac{5}{2}}
-\bar{Q}^{\la,22}_{\frac{5}{2}}
\Big)\Bigg]\,,
\nonu\\
\Phi^{(2),2}_{\frac{1}{2}}
&
=&
-2 \times \Bigg[-\frac{i}{2}\,\Big(
Q^{\la,11}_{\frac{5}{2}}
+2i\sqrt{2} \,Q^{\la,21}_{\frac{5}{2}}
-2 \,Q^{\la,22}_{\frac{5}{2}}
+2\, \bar{Q}^{\la,11}_{\frac{5}{2}}
+2i\sqrt{2} \, \bar{Q}^{\la,12}_{\frac{5}{2}}
-\bar{Q}^{\la,22}_{\frac{5}{2}}
\Big)\Bigg]\,,
\nonu\\
\Phi^{(2),3}_{\frac{1}{2}}
&
=&
-2 \times
\Bigg[
-\frac{i}{2}\,\Big(
Q^{\la,11}_{\frac{5}{2}}
+i\sqrt{2} \,Q^{\la,12}_{\frac{5}{2}}
-2\,Q^{\la,22}_{\frac{5}{2}}
+2\, \bar{Q}^{\la,11}_{\frac{5}{2}}
+i \sqrt{2} \, \bar{Q}^{\la,21}_{\frac{5}{2}}
-\bar{Q}^{\la,22}_{\frac{5}{2}}
\Big)\Bigg]\,,
\nonu\\
\Phi^{(2),4}_{\frac{1}{2}}
&
=&
-2 \times \Bigg[
-\frac{1}{2}\,Q^{\la,11}_{\frac{5}{2}}
-Q^{\la,22}_{\frac{5}{2}}
+\bar{Q}^{\la,11}_{\frac{5}{2}}
+\frac{1}{2}\,\bar{Q}^{\la,22}_{\frac{5}{2}} \Bigg]
\,.
\label{higherSpin5half}
\eea
We realize that this looks similar to the ones in
(\ref{higherspin3half}) in the sense that
we obtain the above expressions by increasing the
weight of the right hand sides of (\ref{higherspin3half})
by one and multiplying the overall factor $-2$.
At this moment, it is not clear how
the numerical factor $\frac{1}{2}$
in the coefficient $\frac{1}{2}$ of 
$\Phi_{\frac{1}{2}}^{(2),i}$ in the first order pole appears. 

\subsubsection{The OPE between the weight-$3$ operator
  and the weight-$1$ operator}

By using the expression in (\ref{freespin3}), we obtain
\bea
\Phi^{(1) }_{2}(z)\,\Phi^{(1)}_0(w)
&=&
\frac{1}{(z-w)^4}\,\Bigg[ 4 N \, (1-12\la + 24\la^2)\Bigg]+
\frac{1}{(z-w)^3}\,\Bigg[ 32 \, \la \, (1-2\la) U\Bigg](w)
\nonu \\
& + &
\frac{1}{(z-w)^2}\,\Bigg[ 48 \, \la \, (1-2\la) \, \pa \, U +
  2 \, \Big(\Phi^{(2) }_{0} - \frac{8}{3}\, (1-4\la) \, L \Big)
  \Bigg](w)
\nonu \\
& + &
\frac{1}{(z-w)}\,\Bigg[
   32 \, \la \, (1-2\la) \, \pa^2 \, U +
   2 \, 
   \Big(\pa \,\Phi^{(2) }_{0} - \frac{8}{3}\, (1-4\la) \,
   \pa \, L \Big)\Bigg](w)
\nonu \\
& + &
\cdots.
\label{fifthopein35}
\eea
Similarly, for the quasiprimary weight-$3$ operator
$\tilde{\Phi}^{(1) }_{2}(z)$, the corresponding OPE
has the fourth order term as $-16 \, N\, \la \, (1-2\la)$
where the relation (\ref{firstopein35}) is used
and the remaining singular terms remain the same as above.
Again, after subtracting the descendant in the second order pole,
there exists a new quasi primary operator of weight-$2$
which cannot be written in terms of previously known
operators.

It turns out that we obtain the lowest operator
in the second ${\cal N}=4$ multiplet shifted by the stress energy
tensor
\bea
\Phi^{(2) }_{0} - \frac{8}{3}\, (1-4\la) \, L
=-2 \times \Bigg[
4 \, \Big( (1-2\la)\, (W^{\la,11}_{\mathrm{F},2}+W^{\la,22}_{\mathrm{F},2})-
2\la \,(W^{\la,11}_{\mathrm{B},2}+W^{\la,22}_{\mathrm{B},2})\Big)
\Bigg]\,.
\label{secondlowestexp}
\eea
The structure of the right hand side  looks similar to the one in
(\ref{spinone}) and 
we obtain the above expression by increasing the
weight of the right hand side of (\ref{spinone})
by one and multiplying the overall factor $-2$.
Also it is not clear how
the numerical factor $2$
in front of 
the second terms in the second order pole appears
\footnote{We can write down the weight-$2$ operator
  from (\ref{secondlowestexp}) as 
\bea
\Phi^{(2) }_{0} 
= -\frac{16}{3}\, (1-\la)\, (W^{\la,11}_{\mathrm{F},2}+W^{\la,22}_{\mathrm{F},2})+
\frac{8}{3} \,(1+2\la)\,
(W^{\la,11}_{\mathrm{B},2}+W^{\la,22}_{\mathrm{B},2})\,.
\label{secondlowest}
\eea}. 

\subsubsection{The ${\cal N}=4$ supersymmetric OPE in
the ${\cal N}=4$ superspace}

Now we would like to construct
the single ${\cal N}=4$ supersymmetric OPE in
the ${\cal N}=4$ superspace based on the previous
component results.
The precise relations between the components and its
superfields can be summarized by (\ref{comptosuper1})
and (\ref{comptosuper2}).

Then we eventually determine the following
${\cal N}=4$ super OPE, after putting the above five fundamental OPEs
(\ref{firstopein35}), (\ref{secondopein35}), (\ref{thirdopein35}),
(\ref{fourthopein35}) and (\ref{fifthopein35}) into the
corresponding singular terms in the ${\cal N}=4$ superspace 
\bea
{\bf \Phi}^{(1)}(Z_{1})\,{\bf \Phi}^{(1)}(Z_{2})  & = & 
\frac{\theta_{12}^{4-0}}{z_{12}^{4}}\, 4N\, (1-12\la+24\la^2)
+\frac{\theta_{12}^{4-i}}{z_{12}^{3}}\,
\Bigg[ 16 \, \la \, (1-2\la)\, D^i \, {\bf J}\Bigg]
(Z_{2})
\nonu \\
& + & \frac{\theta_{12}^{4-0}}{z_{12}^{3}}\,
\Bigg[ 32 \, \la \, (1 - 2 \la) \, \pa \, {\bf J}\Bigg]
(Z_{2})
+
\frac{1}{z_{12}^{2}}\,
2 \, N\, (1 -4\la)
\nonu \\
& + &
\frac{\theta_{12}^{4-ij}}{z_{12}^{2}}\,
\Bigg[ 2\, (1-4\la)\, \frac{1}{2!}\, \varepsilon^{i j k l}\,
  D^k \, D^l \, {\bf J} +  \frac{1}{2!}\, \varepsilon^{i j k l}\,
  \varepsilon^{k l m n} \, D^m \, D^n \, {\bf J} \Bigg](Z_{2})
\nonu \\
& + &
\frac{\theta_{12}^{4-i}}{z_{12}^{2}}\,\Bigg[\,
  32 \, \la \, (1-2\la)\, \pa \, D^i \, {\bf J}
  \nonu \\
  & - & 3\, (1-4\la) \, \Big( \frac{1}{3!}\, \varepsilon^{i j k l}\,
  D^j \, D^k \, D^l \, {\bf J} -(1-4\la)\, \pa \, D^i \, {\bf J} \Big)
\,\Bigg](Z_{2})
\nonu\\
&
+&\frac{\theta_{12}^{4-0}}{z_{12}^{2}}\,\Bigg[\,
48 \, \la \, (1-2\la)\, \pa^2 \, {\bf J}
\nonu \\
&+ &
2\Big( {\bf \Phi}^{(2)} - \frac{8}{3}\, (1-4\la) \,
(\frac{1}{2}\, \frac{1}{4!} \, \varepsilon^{i j k l}\,
D^i \, D^j \, D^k \, D^l \,  {\bf J}
- \frac{1}{2}\,
  (1-4\la)\, \pa^2 \, {\bf J}
) \Big) \Bigg](Z_{2})
\nonu \\
&+ &  
\frac{\theta_{12}^{i}}{z_{12}}\,
\Bigg[ \frac{1}{3!}\, \varepsilon^{i j k l} \, D^j \, D^k \, D^l \,
  {\bf J} + (1-4\la)\, \pa \, D^i \, {\bf J}\Bigg](Z_{2})
\nonu \\
& + &
\frac{\theta_{12}^{4-ij}}{z_{12}}\,
\Bigg[ 2\, (1-4\la)\, \frac{1}{2!}\, \varepsilon^{i j k l}\,
  \pa \, D^k \, D^l \, {\bf J} +  \frac{1}{2!}\, \varepsilon^{i j k l}\,
  \varepsilon^{k l m n} \, \pa \, D^m \, D^n \, {\bf J}\Bigg](Z_{2})
\nonu \\
& + &
\frac{\theta_{12}^{4-i}}{z_{12}}\,\Bigg[\,
  24 \, \la \, (1-2\la)\, \pa^2 \, D^i \, {\bf J}
  +\frac{1}{2}\,
 D^i\, {\bf \Phi}^{(2)}
  \nonu \\
  & - & \frac{8}{3}\,
  (1-4\la) \, \Big( \frac{1}{3!}\, \varepsilon^{i j k l}\,
  \pa \, D^j \, D^k \, D^l \, {\bf J} -
  (1-4\la)\, \pa^2 \, D^i \, {\bf J} \Big) 
\,\Bigg](Z_{2})
\nonu \\ 
&
+&\frac{\theta_{12}^{4-0}}{z_{12}}\,\Bigg[\,
32 \, \la \, (1-2\la)\, \pa^3 \, {\bf J}
\nonu \\
&+ &
2\Big( \pa  {\bf \Phi}^{(2)} - \frac{8}{3} (1-4\la) 
(\frac{1}{2} \frac{1}{4!}  \varepsilon^{i j k l}
\pa  D^i  D^j  D^k  D^l  {\bf J}
- \frac{1}{2}
  (1-4\la) \pa^3  {\bf J}
) \Big) \Bigg](Z_{2})
\nonu \\
& + &  \cdots.
\label{super1super1}
\eea
The $\la$ dependence appears as $(1-4\la)$ or $\la\, (1-2\la)$
except the first central term of (\ref{super1super1}).
We have checked that the above OPE, except the two central
terms appearing in the first two lines of (\ref{super1super1}), 
is the same as the one \cite{AKK1910}
under the large $(N,k)$ limit. 

In Appendix $C$, we write down all the component OPEs explicitly
for convenience.
Equivalently, all the OPEs in (\ref{ope1})
can be obtained from (\ref{super1super1}) by acting the
various super derivatives on both sides and putting the fermionic
coordinates to zero.
In Appendix $D$,
the five fundamental OPEs in the ${\cal N}=4$ coset model
under the large $(N,k)$ limit
are given explicitly in (\ref{cosetresult}).
We observe that the previous OPEs
(\ref{firstopein35}), (\ref{secondopein35}), (\ref{thirdopein35}),
(\ref{fourthopein35}) and (\ref{fifthopein35})
are identical to the ones in (\ref{cosetresult}) together with
$\la =\frac{1}{2}\, \la_{co}$ in the footnote \ref{laco}.
Furthermore, other OPEs appearing in (\ref{ope1}) are
identified with the ones in \cite{AKK1910} we do not present
in this paper.

\subsection{ The OPEs between the other ${\cal N}=4$ multiplets} 

From the defining equation \cite{Schoutens,BCG,AK1509} of
\bea
G^{i}(z) \, \Phi_{0}^{(2)}(w)\Bigg|_{\frac{1}{(z-w)}} & = & 
- \Phi_{\frac{1}{2}}^{(2),i}(w),
\label{fireq}
\eea
which is obtained from the relation (\ref{Relone})
by changing the weight properly,
we can determine 
the quasiprimary operators of weight-$\frac{5}{2}$
appearing in (\ref{higherSpin5half}) by using (\ref{fireq}).

By using the following defining
equation \cite{AK1509}
\bea
G^{i}(z) \, \Phi_{\frac{1}{2}}^{(2),j}(w)\Bigg|_{\frac{1}{(z-w)}} & = & 
-
 \Bigg[ \delta^{ij}\, \partial\Phi_{0}^{(2)}-
   \frac{1}{2} \, \varepsilon^{i j k l} \,
   \Phi_{1}^{(2),k l} \Bigg](w),
\label{releq}
 \eea
coming from (\ref{Reltwo}),
we obtain the weight-$3$ quasiprimary operators, by taking
two different indices in (\ref{releq}), as follows:
\bea
\Phi^{(2),12}_{1}
&
=& -2 \times \Bigg[
2i\,W^{\la,11}_{\mathrm{B},3}
-\sqrt{2}\,W^{\la,12}_{\mathrm{B},3}
-2i\,\,W^{\la,22}_{\mathrm{B},3}
+2i\,W^{\la,11}_{\mathrm{F},3}
-2\sqrt{2}\,W^{\la,12}_{\mathrm{F},3}
-2i\,W^{\la,22}_{\mathrm{F},3} \Bigg]\,,
\nonu\\
\Phi^{(2),13}_{1}
&
=&
-2 \times \Bigg[
-2i\,W^{\la,11}_{\mathrm{B},3}
+4\sqrt{2}\,W^{\la,21}_{\mathrm{B},3}
+2i\,\,W^{\la,22}_{\mathrm{B},3}
-2i\,W^{\la,11}_{\mathrm{F},3}
+2\sqrt{2}\,W^{\la,21}_{\mathrm{F},3}
+2i\,W^{\la,22}_{\mathrm{F},3} \Bigg]\,,
\nonu\\
\Phi^{(2),14}_{1}
&
=&
-2\times \Bigg[
2\,W^{\la,11}_{\mathrm{B},3}
+i\sqrt{2}\,W^{\la,12}_{\mathrm{B},3}
+4i\sqrt{2}\,\,W^{\la,21}_{\mathrm{B},3}
-2\,W^{\la,22}_{\mathrm{B},3}
-2\,W^{\la,11}_{\mathrm{F},3}
-2i\sqrt{2}\,W^{\la,12}_{\mathrm{F},3}
\nonu \\
& - & 2i\sqrt{2}\,W^{\la,21}_{\mathrm{F},3}
+2\,W^{\la,22}_{\mathrm{F},3} \Bigg]\,,
\nonu\\
\Phi^{(2),23}_{1}
&
=& -2\times \Bigg[
-2\,W^{\la,11}_{\mathrm{B},3}
-i\sqrt{2}\,W^{\la,12}_{\mathrm{B},3}
-4i\sqrt{2}\,\,W^{\la,21}_{\mathrm{B},3}
+2\,W^{\la,22}_{\mathrm{B},3}
-2\,W^{\la,11}_{\mathrm{F},3}
-2i\sqrt{2}\,W^{\la,12}_{\mathrm{F},3}
\nonu \\
& - & 2i\sqrt{2}\,W^{\la,21}_{\mathrm{F},3}
+2\,W^{\la,22}_{\mathrm{F},3} \Bigg]\,,
\label{spinthree}
 \\
\Phi^{(2),24}_{1}
&
=& -2 \times \Bigg[
-2i\,W^{\la,11}_{\mathrm{B},3}
+4\sqrt{2}\,W^{\la,21}_{\mathrm{B},3}
+2i\,\,W^{\la,22}_{\mathrm{B},3}
+2i\,W^{\la,11}_{\mathrm{F},3}
-2\sqrt{2}\,W^{\la,21}_{\mathrm{F},3}
-2i\,W^{\la,22}_{\mathrm{F},3} \Bigg]\,,
\nonu\\
\Phi^{(2),34}_{1}
&
=& -2 \times \Bigg[
-2i\,W^{\la,11}_{\mathrm{B},3}
+\sqrt{2}\,W^{\la,12}_{\mathrm{B},3}
+2i\,\,W^{\la,22}_{\mathrm{B},3}
+2i\,W^{\la,11}_{\mathrm{F},3}
-2\sqrt{2}\,W^{\la,12}_{\mathrm{F},3}
-2i\,W^{\la,22}_{\mathrm{F},3} \Bigg]\,.
\nonu
\eea

From the defining equation \cite{AK1509} of
\bea
G^{i}(z) \, \Phi_{1}^{(2),jk}(w)\Bigg|_{\frac{1}{(z-w)}} & = & 
-  \Bigg[ 
\Big( \delta^{ij}\, \Phi_{\frac{3}{2}}^{(2),k}- \delta^{ik} \, 
\Phi_{\frac{3}{2}}^{(2),j}\Big)+
\varepsilon^{i j k l} \,
\partial\, \Phi_{\frac{1}{2}}^{(2),l} \Bigg](w),
\label{aboveeq}
\eea
which can be obtained from (\ref{Relthree}), 
we can determine the following weight-$\frac{7}{2}$
quasiprimary operators by taking the appropriate indices in (\ref{aboveeq})
\bea
\tilde{\Phi}^{(2),1}_{\frac{3}{2}}
&
\equiv &
\Phi^{(2),1}_{\frac{3}{2}} -\frac{1}{5}\, (1-4\la)\,
\pa \,\Phi^{(2),1}_{\frac{1}{2}}
\nonu \\
&=& -2 \times \Bigg[-\frac{1}{2}\,\Big(
Q^{\la,11}_{\frac{7}{2}}
+i\sqrt{2}\,Q^{\la,12}_{\frac{7}{2}}
+2i\sqrt{2}\,Q^{\la,21}_{\frac{7}{2}}
-2\,Q^{\la,22}_{\frac{7}{2}}
\nonu \\
& - & 2\,\bar{Q}^{\la,11}_{\frac{7}{2}}
-2i\sqrt{2}\,\bar{Q}^{\la,12}_{\frac{7}{2}}
-i\sqrt{2}\,\bar{Q}^{\la,21}_{\frac{7}{2}}
+\bar{Q}^{\la,22}_{\frac{7}{2}}
\Big)\Bigg]\,,
\nonu\\
\tilde{\Phi}^{(2),2}_{\frac{3}{2}}
&
\equiv &
\Phi^{(2),2}_{\frac{3}{2}} -\frac{1}{5}\, (1-4\la)\,
\pa \,\Phi^{(2),2}_{\frac{1}{2}}
\nonu \\
&
=& -2 \times \Bigg[
\frac{i}{2}\,\Big(\,
Q^{\la,11}_{\frac{7}{2}}
+2i\sqrt{2}\,Q^{\la,21}_{\frac{7}{2}}
-2\,Q^{\la,22}_{\frac{7}{2}}
-2\,\bar{Q}^{\la,11}_{\frac{7}{2}}
-2i\sqrt{2}\,\bar{Q}^{\la,12}_{\frac{7}{2}}
+\bar{Q}^{\la,22}_{\frac{7}{2}}
\Big) \Bigg]\,,
\nonu\\
\tilde{\Phi}^{(2),3}_{\frac{3}{2}}
&
\equiv &
\Phi^{(2),3}_{\frac{3}{2}} -\frac{1}{5}\, (1-4\la)\,
\pa \,\Phi^{(2),3}_{\frac{1}{2}}
\nonu \\
&
=& -2 \times \Bigg[
\frac{i}{2}\,\Big(
Q^{\la,11}_{\frac{7}{2}}
+i\sqrt{2}\,Q^{\la,12}_{\frac{7}{2}}
-2\,Q^{\la,22}_{\frac{7}{2}}
-2\,\bar{Q}^{\la,11}_{\frac{7}{2}}
-i\sqrt{2}\,\bar{Q}^{\la,21}_{\frac{7}{2}}
+\bar{Q}^{\la,22}_{\frac{7}{2}}
\Big)\Bigg]\,,
\nonu\\
\tilde{\Phi}^{(2),4}_{\frac{3}{2}}
&
\equiv &
\Phi^{(2),4}_{\frac{3}{2}} -\frac{1}{5}\, (1-4\la)\,
\pa \,\Phi^{(2),4}_{\frac{1}{2}}
\nonu \\
&
=& -2 \times \Bigg[
\frac{1}{2}\,
\Big(
Q^{\la,11}_{\frac{7}{2}}
+2\,Q^{\la,22}_{\frac{7}{2}}
+2\,\bar{Q}^{\la,11}_{\frac{7}{2}}
+\bar{Q}^{\la,22}_{\frac{7}{2}}
\Big)\Bigg]\,.
\label{spin7half}
\eea

Finally, by using the following
defining equation \cite{AK1509}
\bea
G^{i}(z) \, \Phi_{\frac{3}{2}}^{(2),j}(w)\Bigg|_{\frac{1}{(z-w)}} & = & 
 -  
 \Bigg[ \partial\Phi_{1}^{(2),ij}+ 
 \delta^{ij} \, \Phi_{2}^{(2)}\Bigg](w),
\label{lasteq1}
 \eea
which is obtained from (\ref{Relfour}),
we obtain the weight-$4$ quasiprimary
operator, by taking two equal indices in (\ref{lasteq1}),
\bea
\tilde{\Phi}^{(2)}_{2}
&
\equiv &
\Phi^{(2)}_{2} -\frac{1}{5} (1-4\la)
\pa^2 \Phi^{(2)}_{0}
=
-2 \times \Bigg[-2 \Big(
W^{\la,11}_{\mathrm{B},4}
+W^{\la,22}_{\mathrm{B},4}
+W^{\la,11}_{\mathrm{F},4}
+W^{\la,22}_{\mathrm{F},4}
\Big) \Bigg].
\label{freespin4}
\eea

Therefore, the second
${\cal N}=4$ multiplet by considering
the super weight-$2$ in (\ref{Phiexp}) is given by
(\ref{secondlowest}), (\ref{higherSpin5half}),
(\ref{spinthree}), (\ref{spin7half}) and (\ref{freespin4}).

We can further analyze the OPEs between
the next ${\cal N}=4$ multiplets.
By using the previous relations (\ref{spinone}),
(\ref{higherspin3half}), (\ref{spintwo}), (\ref{spin5half})
and (\ref{freespin3})
and the relation (\ref{secondlowest}),
the OPEs between the ${\cal N}=4$ stress energy tensor
and the second ${\cal N}=4$ multiplet 
are given in Appendix $E$ in ${\cal N}=4$ superspace explicitly.
The OPEs between the first and the second ${\cal N}=4$ multiplets
can be obtained and 
are given in Appendix $F$ in ${\cal N}=4$ superspace explicitly.
Moreover, the OPEs between the second ${\cal N}=4$ multiplet
and itself can be determined and they are given in Appendix $G$.
Then we can determine the third ${\cal N}=4$ multiplet in
(\ref{thirdmult}) and the fourth ${\cal N}=4$ multiplet (\ref{lasteq}).
In Appendices, $H$, $I$ and $J$,
the OPEs between the ${\cal N}=4$ stress energy tensor and
the third ${\cal N}=4$ multiplet,
the OPEs between the ${\cal N}=4$ stress energy tensor and
the fourth ${\cal N}=4$ multiplet,
and
the OPEs between the first ${\cal N}=4$ multiplet and
the third ${\cal N}=4$ multiplet are presented respectively.

\section{ Conclusions and outlook}

By using the ${\beta \, \ga}$
and $b \, c$ ghost systems explicitly,
we have constructed the generators in the ${\cal N}=4$
supersymmetric linear $W_{\infty}[\la]$ algebra: the ${\cal N}=4$
stress energy tensor, the first ${\cal N}=4$ multiplet and the
second ${\cal N}=4$ multiplet (and the third and fourth ${\cal N}=4$
multiplets).
Moreover, their algebras between these generators are determined
and in particular, the OPEs between the first ${\cal N}=4$ mutiplet
and itself are equivalent to the corresponding ones in the
${\cal N}=4$ coset model under the large $(N,k)$ limit.
Contrary to the findings in \cite{AK2009},
the modes of the currents in the present results
are not restricted to the wedges but can have any integers
or half integers because our construction is based on the OPEs
between the currents.

So far, we have considered the OPEs between the
${\cal N}=4$ stress energy tensor and the first ${\cal N}=4$ multiplet
(and other OPEs in Appendices
$E, \cdots, J$). Then it is natural to
ask what are the OPEs between the $h_1$-th ${\cal N}=4$ multiplet
and the $h_2$-th ${\cal N}=4$ multiplet for any weights $h_1$ and $h_2$.
In the analysis of (\ref{lasteq}), we can figure out
the explicit form for the five kinds of currents
for general weight-$h$. The lowest component can be obtained
easily up to the overall normalization. The remaining components
can be also determined with the weight dependent overall factors.
Then the question is how we can write down the OPEs between the
currents appearing in (\ref{WWQQnonzerola}) for each component in terms
of the $\la$ dependent structure constants introduced in
\cite{AK2009}.
It would be interesting to rewrite all the structure constants
obtained in this paper in terms of previously known ones
presented in \cite{Ahn2203}. This will give us some hints
to figure out their behaviors for generic weights \footnote{
  For example, the structure constant appearing in
  $W_{F,2}^{\la,\bar{a} b}\, \de_{b \bar{a}}(w)$
  of the OPE between $W_{F,4}^{\la,\bar{a} b}\, \de_{b \bar{a}}(z)$
  and
  $W_{F,4}^{\la,\bar{a} b}\, \de_{b \bar{a}}(w)$  is given by
  $\frac{2048}{5}\, (\la-1)\, (\la+1) \, (2\la-3)\, (2\la+3)$ around the
  equation $(3.18)$ in \cite{Ahn2203}.
  This $\la$ dependent function is related to
  $p_{F,4}^{4,4}(m,n,\la)$ appearing in that paper. 
  By realizing that we can extract
  $W_{F,4}^{\la,\bar{a} b}\, \de_{b \bar{a}}$ from the present context and
  we have $W_{F,4}^{\la,\bar{a} b}\, \de_{b \bar{a}}=
  \frac{1}{28}\, (3+2\la)\, \Phi^{(2)}_2-\frac{1}{384}\,
  \Phi^{(4)}_0 +\frac{1}{140}\, (3+2\la)\, (-1+4\la)\,
  \pa^2 \, \Phi^{(2)}_{0}$, we can check that
  the sixth order pole in the OPE between these three terms and itself
  reproduces the above structure constant.
Furthermore,
  by using the weight-$4$ current
  $W_{B,4}^{\la,\bar{a} b}\, \de_{b \bar{a}}=
  \frac{1}{14}\, (2-\la)\, \Phi^{(2)}_2+\frac{1}{384}\,
  \Phi^{(4)}_0 -\frac{1}{70}\, (-2+\la)\, (-1+4\la)\,
  \pa^2 \, \Phi^{(2)}_{0}$ and
the weight-$3$ current $W_{B,3}^{\la,\bar{a} b}\, \de_{b \bar{a}}=
  \frac{1}{10}\, (-3+2\la)\, \Phi^{(1)}_2-\frac{1}{48}\,
  \Phi^{(3)}_0 +\frac{1}{30}\, (-3+2\la)\, (-1+4\la)\,
  \pa^2 \, \Phi^{(1)}_{0}$, we obtain the sixth order pole in the
  OPE between them, which is
  equal to $512\, (\la-1)\,\la\,(2\la-3)\,(2\la+1)$.
  This structure constant appears in the equation $(B.2)$ of
  \cite{Ahn2203} and is related to $p_{B,4}^{4,3}(m,n,\la)$.
  In these examples, the bifundamental indices are contracted
  with each other. However, we should also obtain
  the OPEs between the currents with free bifundamental indices
  for generic weights.
\label{lastfootnote}}.

Because the second ${\cal N}=4$ multiplet from the free field
approach is not directly related to
the corresponding ${\cal N}=4$ multiplet from the
coset fields in \cite{AKK1910} (for example, the OPE between the
$\Phi_0^{(1)}$ and $\Phi_0^{(2)}$ in the former does not vanish
while that in the latter does vanish),
it would be interesting to
obtain the correct second ${\cal N}=4$ multiplet in the
coset model at finite $(N,k)$ as a first step. Note that
according to the free field approach in this paper,
the currents (\ref{secondlowest}), (\ref{higherSpin5half}),
(\ref{spinthree}), (\ref{spin7half}) and (\ref{freespin4})
are the quasiprimary operators
under the stress energy tensor.
We need to find out the correct basis where
the corresponding currents in the
${\cal N}=4$ coset model should reflect this quasiprimary
condition at least by calculating all the nonlinear terms.
After that, we also expect that under the large
$(N,k)$ limit, for example, the OPEs
between the first ${\cal N}=4$ multiplet and the second
${\cal N}=4$ multiplet in the coset model will produce
the ones on Appendix $F$.

In the context of
celestial holography \cite{PPR}, we have seen that
the wedge subalgebra of $w_{1+\infty}$ algebra \cite{Bakas} provides
the symmetries on the celestial sphere \cite{Strominger}.
See also \cite{GHPS}. Moreover, the analysis
for the ${\cal N}=1$
supersymmetric  $w_{1+\infty}$ algebra is obtained in
\cite{Ahn2111,Ahn2202}. In the present context,
the above $w_{1+\infty}$ algebra is related to the
OPEs between the currents $W_{F,h}^{\la, \bar{a} b}\, \de_{b \bar{a}}$
with the structure constants $p_{F,h}^{h_1,h_2}(m,n,\la)$ described in
the footnote \ref{lastfootnote}.
In the context of ${\cal N}=4$ supersymmetric
linear $W_{\infty}[\la]$ algebra,
the current of weight-$h$ are made of 1) the lowest current
in the $h$-th ${\cal N}=4$ multiplet,
2) the middle current in the $(h-1)$-th ${\cal N}=4$ multiplet
and 3) the highest current (and the lowest current
with two derivatives) in the $(h-2)$-th ${\cal N}=4$ multiplet.
It would be interesting
to observe whether the corresponding supersymmetric
Einstein-Yang-Mills theory at
nonzero deformation parameters
$\la$ (or $q$) reveals the OPEs we have obtained in this
paper or not.

In \cite{BVd1,BVd2}, the explicit representation of the corresponding
algebra is given by the differential operators in terms of
commuting parameters $\La^{\pm}(z)$ and anticommuting parameters
$\Theta^{\pm}(z)$ \footnote{We thank the referee for pointing
the questions raised in the remaining paragraphs out. }.
Under the symmetry generated by the currents in this paper,
the transformation of any fields (or operators) is given by the following
contour integrals over $z$ with
the OPEs between the currents and the fields (along the lines of
\cite{BPZ})
\bea
\delta_{\La^{\pm}_{\bar{a} b}} \, f(w) & = & \frac{1}{2\pi i} \, \oint_{C_w} \, dz \,
\La_{\bar{a} b}^{\pm}(z) \, V_{\la,\bar{a} b}^{(s)\pm}(z)\, f(w) \equiv
\Bigg[\frac{1}{2\pi i} \, \oint_{C_w} \, dz \,
\La_{\bar{a} b}^{\pm}(z) \, V_{\la,\bar{a} b}^{(s)\pm}(z),  f(w) \Bigg],
\nonu \\
\delta_{\Theta_{\bar{a} b}^{\pm}} \, f(w)  & = &
\frac{1}{2\pi i} \, \oint_{C_w} \, dz \,
\Theta_{\bar{a} b}^{\pm}(z) \, Q_{\la, _{\bar{a} b}}^{(s)\pm}(z)\, f(w)
\equiv \Bigg[ \frac{1}{2\pi i}  \oint_{C_w}  dz 
  \Theta_{\bar{a} b}^{\pm}(z)  Q_{\la, _{\bar{a} b}}^{(s)\pm}(z),
 f(w)
  \Bigg \},
\label{equationone}
\eea
where the contour $C_w$ surrounds the point $z$
and there are no summations over the indices $\bar{a}$ and $b$
on the right hand sides of (\ref{equationone}).
We can describe the corresponding (anti)commutator relations
between the `charges' and the fields.

By using the result of the following OPE
\bea
V_{\la,\bar{a} b}^{(s)+}(z)\, \beta^{\bar{m} c}(w) =
\delta_{c \bar{a}}\, \sum_{i=0}^{s-1}\, a^{i}(s,\la)\,
\Bigg(\pa_z^{s-1-i} \, \frac{1}{(z-w)} \Bigg)\,
\pa^i \, \beta^{\bar{m} b}(w)
+ \cdots,
\label{opeexp}
\eea
and substituting this (\ref{opeexp}) into the (\ref{equationone}),
we obtain
\bea
\delta_{\La^{+}_{\bar{a} b}} \, \beta^{\bar{m} c}(w)=
\Bigg(\delta_{c \bar{a}}\, \sum_{i=0}^{s-1}\, a^{i}(s,\la)\, (-1)^{s-1-i}\,
(s-1-i)! \, (\pa^{s-1-i}\,
\La^{(s)+}_{\bar{a} b}(w))\, \pa^i\Bigg) \, \beta^{\bar{m} c}(w).
\label{lineardiff}
\eea
This implies that we realize that there exist
the corresponding linear differential operators
appearing inside the bracket in (\ref{lineardiff}).
Therefore, the nontrivial differential
operators occur only when the first element $\bar{a}$ of
the currents $V_{\la,\bar{a} b}^{(s)+}(z)$ and
the second element $c$ of the operators $\beta^{\bar{m} c}(w)$
are equal to each other.

Similarly,
we act the above currents on the $b^{\bar{j} c}(w)$.
From the result of
\bea
V_{\la,\bar{a} b}^{(s)+}(z)\, b^{\bar{m} c}(w) =
\delta_{c \bar{a}}\, \sum_{i=0}^{s-1}\, a^{i}(s,\la+\frac{1}{2})\,
\Bigg(\pa_z^{s-1-i} \, \frac{1}{(z-w)} \Bigg)\,
\pa^i \, b^{\bar{m} b}(w)
+ \cdots,
\label{opeexp1}
\eea
we determine the following transformation with (\ref{opeexp1})
\bea
\delta_{\La^{+}_{\bar{a} b}} \, b^{\bar{j} c}(w)=
\Bigg(\delta_{c \bar{a}}\, \sum_{i=0}^{s-1}\, a^{i}(s,\la+\frac{1}{2})\,
(-1)^{s-1-i}\,
(s-1-i)! \, (\pa^{s-1-i}\,
\La^{(s)+}_{\bar{a} b}(w))\, \pa^i\Bigg) \, b^{\bar{j} c}(w).
\label{lineardiff1}
\eea
We observe that there exist
the corresponding linear differential operators
appearing inside the bracket in (\ref{lineardiff1}).

Due to the multiple derivatives of $\beta^{ \bar{j} b }(w)$
in the currents $V_{\la,\bar{a} b}^{(s)+}(z)$, the next OPE
is rather complicated and it turns out that
\bea
V_{\la,\bar{a} b}^{(s)+}(z)\, \ga^{j \bar{c}}(w) &=&
\delta_{b \bar{c}}\, \sum_{i=0}^{s-1}\, a^{i}(s,\la)\,
(-1)^s \, i! \,\sum_{t=0}^{i+1}\, (i+1-t)_{s-1-i} \, \frac{1}{t!}
\, \frac{1}{(z-w)^{s-t}}\,
\pa^t \, \ga^{j \bar{a} }(w)
\nonu \\
& + &  \cdots.
\label{opeexp2}
\eea
Note that there is a summation over $t$ and its maximum number
is given by $(i+1)$. There is also a summation over $i$.

Then we obtain the following result by using the relation (\ref{opeexp2}) 
\bea
\delta_{\La^{+}_{\bar{a} b}} \, \ga^{j \bar{c}}(w)=
\Bigg(\delta_{b \bar{c}}\, \sum_{i=0}^{s-1}\, a^{i}(s,\la)\,
(-1)^{s}\,
i! \,\sum_{t=0}^{i+1}\, (i+1-t)_{s-1-i} \, \frac{1}{t!}
\, (\pa^{s-1-t}\,
\La^{(s)+}_{\bar{a}b}(w))\, \pa^t\Bigg) \, \ga^{j \bar{c}}(w).
\label{lineardiff2}
\eea
There exist
the corresponding linear differential operators
with the  double summations
appearing inside the bracket in (\ref{lineardiff2}).

Again,
due to the multiple derivatives of $b^{ \bar{j} b }(w)$
in the currents $V_{\la,\bar{a} b}^{(s)+}(z)$, the next OPE
is complicated and it turns out that
\bea
V_{\la,\bar{a} b}^{(s)+}(z)\, c^{j \bar{c}}(w) &=&
\delta_{b \bar{c}}\, \sum_{i=0}^{s-1}\, a^{i}(s,\la+\frac{1}{2})\,
(-1)^s \, i! \,\sum_{t=0}^{i+1}\, (i+1-t)_{s-1-i} \, \frac{1}{t!}
\, \frac{1}{(z-w)^{s-t}}\,
\pa^t \, c^{j \bar{c} }(w)
\nonu \\
& + &  \cdots.
\label{opeexp3}
\eea
The corresponding transformation, with the help of
(\ref{opeexp3}), can be written as
\bea
\delta_{\La^{+}_{\bar{a} b}}  c^{j \bar{c}}(w)=
\Bigg(\delta_{b \bar{c}} \sum_{i=0}^{s-1} a^{i}(s,\la+\frac{1}{2})
(-1)^{s}
i! \sum_{t=0}^{i+1} (i+1-t)_{s-1-i}  \frac{1}{t!}
 (\pa^{s-1-t}
\La^{(s)+}_{\bar{a}b}(w)) \pa^t\Bigg)  c^{j \bar{c}}(w).
\label{lineardiff3}
\eea
The corresponding linear differential operators
with the  double summations
appearing inside the bracket in (\ref{lineardiff3}) occur.

Now we can consider the symmetry   generated by
fermionic currents and they can be described as follows:
\bea
\delta_{\Theta^{+}_{\bar{a} b}} \, \beta^{\bar{m} c}(w) & = &
-\Bigg(\delta_{c \bar{a}}\, \sum_{i=0}^{s-2}\, \beta^{i}(s,\la)\,
  (-1)^{s-2-i}\,
(s-2-i)! \, (\pa^{s-2-i}\,
\Theta^{(s)+}_{\bar{a} b}(w))\, \pa^i\Bigg) \, b^{\bar{m} c}(w),
\nonu \\
\delta_{\Theta^{+}_{\bar{a} b}} \, b^{\bar{j} c}(w) & = &
\Bigg(\delta_{c \bar{a}}\, \sum_{i=0}^{s-1}\, \alpha^{i}(s,\la)\,
(-1)^{s-1-i}\,
(s-1-i)! \, (\pa^{s-1-i}\,
\Theta^{(s)+}_{\bar{a} b}(w))\, \pa^i\Bigg) \, \beta^{\bar{j} c}(w),
\label{thetavar}
 \\
\delta_{\Theta^{+}_{\bar{a} b}} \, \ga^{j \bar{c}}(w) & = &
\Bigg(\delta_{b \bar{c}}\, \sum_{i=0}^{s-1}\, \alpha^{i}(s,\la)\,
(-1)^{s}\,
i! \,\sum_{t=0}^{i+1}\, (i+1-t)_{s-1-i} \, \frac{1}{t!}
\, (\pa^{s-1-t}\,
\Theta^{(s)+}_{\bar{a}b}(w))\, \pa^t\Bigg) \, c^{j \bar{c}}(w),
\nonu \\
\delta_{\Theta^{+}_{\bar{a} b}}  c^{j \bar{c}}(w) & = &
\Bigg(\delta_{b \bar{c}} \sum_{i=0}^{s-2} \beta^{i}(s,\la)
(-1)^{s-1}
i! \sum_{t=0}^{i+1} (i+1-t)_{s-2-i}  \frac{1}{t!}
 (\pa^{s-2-t}
\Theta^{(s)+}_{\bar{a}b}(w)) \pa^t\Bigg)  \ga^{j \bar{c}}(w).
\nonu
\eea
Therefore, we have the transformations of the $\beta \, \ga$
and $b \, c$ ghost systems under the bosonic and fermionic
currents, summarized by (\ref{lineardiff}), (\ref{lineardiff1}),
(\ref{lineardiff2}), (\ref{lineardiff3}) and (\ref{thetavar}).

Furthermore, for the remaining bosonic and fermionic currents,
we can calculate the corresponding OPEs with
 the $\beta \, \ga$
 and $b \, c$ ghost systems and we summarize the
 following results 
\bea
\delta_{\La^{-}_{\bar{a} b}}  \beta^{\bar{m} c}(w) & = &
-\Bigg(\delta_{c \bar{a}} \frac{(s-1+2\la)}{(2s-1)}
\sum_{i=0}^{s-1}\, a^{i}(s,\la) (-1)^{s-1-i}
(s-1-i)!  \nonu \\
& \times & (\pa^{s-1-i}
\La^{(s)-}_{\bar{a} b}(w)) \pa^i\Bigg)  \beta^{\bar{m} c}(w),
\nonu \\
\delta_{\La^{-}_{\bar{a} b}}  b^{\bar{j} c}(w) & = &
\Bigg(\delta_{c \bar{a}} \frac{(s-2\la)}{(2s-1)} 
\sum_{i=0}^{s-1} a^{i}(s,\la+\frac{1}{2})
(-1)^{s-1-i}
(s-1-i)!  \nonu \\
& \times & (\pa^{s-1-i}
\La^{(s)-}_{\bar{a} b}(w)) \pa^i\Bigg)  b^{\bar{j} c}(w), 
\nonu \\
\delta_{\La^{-}_{\bar{a} b}}  \ga^{j \bar{c}}(w) & = &
\Bigg(\delta_{b \bar{c}} \frac{(s-2\la)}{(2s-1)} 
\sum_{i=0}^{s-1} a^{i}(s,\la)
(-1)^{s}
i! \nonu \\
& \times & \sum_{t=0}^{i+1} (i+1-t)_{s-1-i}  \frac{1}{t!}
 (\pa^{s-1-t}
\La^{(s)-}_{\bar{a}b}(w)) \pa^t\Bigg)  \ga^{j \bar{c}}(w),
\nonu \\
\delta_{\La^{-}_{\bar{a} b}}  c^{j \bar{c}}(w) & = &
-\Bigg(\delta_{b \bar{c}}  \frac{(s-1+2\la)}{(2s-1)}
\sum_{i=0}^{s-1} a^{i}(s,\la+\frac{1}{2})
(-1)^{s}
i! \nonu \\
& \times & \sum_{t=0}^{i+1} (i+1-t)_{s-1-i}  \frac{1}{t!}
 (\pa^{s-1-t}
\La^{(s)-}_{\bar{a}b}(w)) \pa^t\Bigg) c^{j \bar{c}}(w),
\nonu \\
%
\delta_{\Theta^{-}_{\bar{a} b}} \, \beta^{\bar{m} c}(w) & = &
\Bigg(\delta_{c \bar{a}}\, \sum_{i=0}^{s-2}\, \beta^{i}(s,\la)\,
  (-1)^{s-2-i}\,
(s-2-i)! \, (\pa^{s-2-i}\,
\Theta^{(s)-}_{\bar{a} b}(w))\, \pa^i\Bigg) \, b^{\bar{m} c}(w),
\nonu \\
\delta_{\Theta^{-}_{\bar{a} b}} \, b^{\bar{j} c}(w) & = &
\Bigg(\delta_{c \bar{a}}\, \sum_{i=0}^{s-1}\, \alpha^{i}(s,\la)\,
(-1)^{s-1-i}\,
(s-1-i)! \, (\pa^{s-1-i}\,
\Theta^{(s)-}_{\bar{a} b}(w))\, \pa^i\Bigg) \, \beta^{\bar{j} c}(w),
\label{Lasteq1}
\\
\delta_{\Theta^{-}_{\bar{a} b}} \, \ga^{j \bar{c}}(w) & = &
\Bigg(\delta_{b \bar{c}}\, \sum_{i=0}^{s-1}\, \alpha^{i}(s,\la)\,
(-1)^{s}\,
i! \,\sum_{t=0}^{i+1}\, (i+1-t)_{s-1-i} \, \frac{1}{t!}
\, (\pa^{s-1-t}\,
\Theta^{(s)-}_{\bar{a}b}(w))\, \pa^t\Bigg) \, c^{j \bar{c}}(w),
\nonu \\
\delta_{\Theta^{-}_{\bar{a} b}}  c^{j \bar{c}}(w) & = &
-\Bigg(\delta_{b \bar{c}} \sum_{i=0}^{s-2} \beta^{i}(s,\la)
(-1)^{s-1}
i! \sum_{t=0}^{i+1} (i+1-t)_{s-2-i}  \frac{1}{t!}
 (\pa^{s-2-t}
\Theta^{(s)-}_{\bar{a}b}(w)) \pa^t\Bigg)  \ga^{j \bar{c}}(w).
\nonu
\eea
It is straightforward to obtain the
corresponding transformations of
 the $\beta \, \ga$
 and $b \, c$ ghost systems for the ${\cal N}=4$ currents
  studied in this paper,
 by taking the linear combinations between
 the above results summarized in
  (\ref{lineardiff}), (\ref{lineardiff1}),
 (\ref{lineardiff2}), (\ref{lineardiff3}), (\ref{thetavar})
 and (\ref{Lasteq1}).

\vspace{.7cm}

\centerline{\bf Acknowledgments}

We
would like to
thank M.H. Kim for the intensive discussions. 
This work was supported by
a National Research Foundation of Korea (NRF) grant
funded by the Korean government (MSIT)(No. 2020R1F1A1066893).

\newpage

\appendix

\renewcommand{\theequation}{\Alph{section}\mbox{.}\arabic{equation}}

\section{The OPEs between the ${\cal N}=4$
stress energy tensor and itself in
the component approach}

For the previous result (\ref{JJ}),
the component results can be summarized by
\bea
L(z) \, L(w) & = & 
\frac{1}{(z-w)^{4}}\Bigg[ \frac{1}{2} \, c\Bigg]+
\frac{1}{(z-w)^{2}}\Bigg[ 2 \, L\Bigg](w)+
\frac{1}{(z-w)} \Bigg[ \pa \, L\Bigg](w) + \cdots,
\nonu \\
L(z) \, G^{i}(w) & = & 
\frac{1}{(z-w)^{2}}\Bigg[ \frac{3}{2}\, G^{i}\Bigg](w)+
\frac{1}{(z-w)} \Bigg[
\partial \, G^{i}\Bigg](w) + \cdots, \nonu \\
L(z) \, T^{ij}(w) & = & \frac{1}{(z-w)^{2}} \Bigg[ T^{ij}\Bigg](w)+
\frac{1}{(z-w)} \Bigg[ \partial \, T^{ij}\Bigg](w) + \cdots,
\nonu \\
L(z) \, \Gamma^{i}(w) & = & 
\frac{1}{(z-w)^{2}}\Bigg[ \frac{1}{2} \, \Gamma^{i}\Bigg](w)+
\frac{1}{(z-w)} \Bigg[ \partial\, \Gamma^{i}\Bigg](w) +\cdots,\nonu \\
L(z) \, U(w) & = &{\tt -\frac{1}{(z-w)^3}\Bigg[ N \Bigg]}+
\frac{1}{(z-w)^{2}} \Bigg[ U\Bigg](w)+
\frac{1}{(z-w)} \Bigg[ \pa \, U\Bigg](w) +\cdots,
\nonu \\
G^{i}(z) \, G^{j}(w) & = & \frac{1}{(z-w)^{3}} \Bigg[
  \frac{2}{3} \, c \,
\delta^{ij} \Bigg]+
\frac{1}{(z-w)^{2}} \Bigg[ -2 \,i\,T^{ij} - i \,(1-4\la)
\, \varepsilon^{ijkl} \, T^{kl}\Bigg](w) 
\nonu \\
& + &
\frac{1}{(z-w)} \Bigg[
  2 \, \delta^{ij}\, L-i \, \partial T^{ij}-i\, \frac{1}{2}\,
  (1-4\la)\, \varepsilon^{ijkl}\,
\partial T^{kl}
\Bigg](w) + \cdots,
\nonu \\
G^{i}(z) \, T^{jk}(w) & = & 
-\frac{1}{(z-w)^{2}} \Bigg[ \varepsilon^{ijkl} \,
\Gamma^{l} + (1-4\la)\, (\delta^{ik}\Gamma^{j}-
\delta^{ij}\Gamma^{k})
\Bigg](w) \nonu \\
& - & \frac{1}{(z-w)} \, \Bigg[\varepsilon^{ijkl} \, \partial\Gamma^{l}
+   (1-4\la)\, (\delta^{ik} \pa \Gamma^{j}-
\delta^{ij} \pa \Gamma^{k}) 
+ i \, \delta^{ik} \, G^{j}-i \, \delta^{ij} \, G^{k} \Bigg](w) \nonu \\
& + & \cdots,
\nonu \\
G^{i}(z)\, \Gamma^{j}(w) & = &
{\tt -\frac{1}{(z-w)^2}\Bigg[ i\, N \, \de^{i j}\Bigg]}
+\frac{1}{(z-w)} \Bigg[
  -\frac{1}{2}\,
  \varepsilon^{ijkl} \, T^{kl}+  i\: \delta^{ij} \, U \Bigg](w) + \cdots,\nonu \\
G^{i}(z) \, U(w) & = & 
-\frac{1}{(z-w)^{2}} \Bigg[ i\, \Gamma^{i}\Bigg](w)-
\frac{1}{(z-w)} \Bigg[ i\, \partial \, \Gamma^{i}\Bigg](w) +\cdots,
\nonu \\
T^{ij}(z)\, T^{kl}(w) & = & 
\frac{1}{(z-w)^{2}} \Bigg[ \varepsilon^{ijkl}\, N \Bigg] \nonu \\
& - & 
\frac{1}{(z-w)} \, i \,
\Bigg[\delta^{ik} \, T^{jl}-\delta^{il}\, T^{jk}-\delta^{jk}\, T^{il}+
\delta^{jl}\, T^{ik} \Bigg](w) +\cdots,
\nonu \\
T^{ij}(z)\, \Gamma^{k}(w) & = & -
\frac{1}{(z-w)}\,
i\: \Bigg[\delta^{ik}\:\Gamma^{j}-\delta^{jk}\:\Gamma^{i} \Bigg](w)
+\cdots.
\label{16SCAOPEs}
\eea
Compared to the large ${\cal N}=4$  superconformal algebra
\cite{Schoutens,AK1509},
there are two additional central terms in the  fifth and
eighth of (\ref{16SCAOPEs}) \footnote{
  They are denoted by the typewriter fonts.}.
Moreover, there are trivial OPEs
$\Ga^i(z)\, \Ga^j(w)$ and $U(z) \, U(w)$.
Finally, the central term of the OPE
$T^{ij}(z) \, T^{kl}(w)$ which is proportional to
$(\de^{i k} \, \de^{j l}-\de^{i l}\,\de^{j k})$ does not appear.

\section{The OPEs between the ${\cal N}=4$ stress energy tensor
and the first ${\cal N}=4$ multiplet in the component approach}

We present the complete OPEs
corresponding to (\ref{JPhi}) as follows:
\bea
L(z) \, \Phi_{2}^{(1)}(w) & = &
{\tt -\frac{1}{(z-w)^5}\Bigg[ 96 \, N\, \la \, (1-2\la)\Bigg]}
\nonu \\
& + & \frac{1}{(z-w)^{4}}
\Bigg[ -6  \, (1-4\la) \,
\Phi_{0}^{(1)} {\tt - 96 \, \la \, (1-2\la)\, U} \Bigg](w)\nonu \\
& + & \frac{1}{(z-w)^{3}}
\Bigg[ 2\,(1-4\la)\,  \partial \, \Phi_{0}^{(1)}\Bigg](w) \nonu
\\
&+ & 
\frac{1}{(z-w)^{2}} \Bigg[ 3 \, \Phi_{2}^{(1)}\Bigg](w)+
\frac{1}{(z-w)} \Bigg[ \partial\Phi_{2}^{(1)}\Bigg](w)
+  \cdots \,,
\nonu \\
L(z) \, \Phi_{\frac{3}{2}}^{(1),i}(w) & = & 
{\tt -\frac{1}{(z-w)^4}\Bigg[ 24 \, i \,\la \, (1-2\la) \,
\Ga^i \Bigg](w)} 
+\frac{1}{(z-w)^{3}}\Bigg[(1-4\la)\, 
\Phi_{\frac{1}{2}}^{(1),i}\Bigg](w)\nonu \\
& + &
\frac{1}{(z-w)^{2}}\Bigg[ \frac{5}{2}\, \Phi_{\frac{3}{2}}^{(1),i}
  \Bigg](w)
+ 
\frac{1}{(z-w)}\Bigg[ \partial\, \Phi_{\frac{3}{2}}^{(1),i}
  \Bigg](w) +\cdots,
\nonu \\
L(z) \, \Phi_{1}^{(1),ij}(w) & = & 
\frac{1}{(z-w)^{2}}\Bigg[ 2 \, \Phi_{1}^{(1),ij}\Bigg](w)+
\frac{1}{(z-w)} \Bigg[ \partial\, \Phi_{1}^{(1),ij}\Bigg](w)
+\cdots,
\nonu \\
L(z) \, \Phi_{\frac{1}{2}}^{(1),i}(w) & = &
\frac{1}{(z-w)^{2}}\Bigg[ \frac{3}{2}\, \Phi_{\frac{1}{2}}^{(1),i}\Bigg](w)
+\frac{1}{(z-w)} \Bigg[ \partial\, \Phi_{\frac{1}{2}}^{(1),i}\Bigg](w)
+\cdots,
\nonu \\
L(z) \, \Phi_{0}^{(1)}(w) & = & 
\frac{1}{(z-w)^{2}} \Bigg[ \Phi_{0}^{(1)}\Bigg](w)+
\frac{1}{(z-w)^{1}} \Bigg[ \partial\Phi_{0}^{(1)}\Bigg](w)
+\cdots, \nonu \\
G^{i}(z) \, \Phi_{2}^{(1)}(w) & = &
{\tt \frac{1}{(z-w)^4}\Bigg[ 48 \, i \, \la \, (1-2\la)\,
\Ga^i\Bigg](w)}
\nonu \\
& + & \frac{1}{(z-w)^{3}} 
\Bigg[ {\tt -48 \, i \, \la \, (1-2\la)\,
\pa \, \Ga^i}- 6\, (1-4\la) \, \Phi_{\frac{1}{2}}^{(1),i}
  \Bigg](w)
\nonu \\
& + &
\frac{1}{(z-w)^{2}} \, \Bigg[-5 \, \Phi_{\frac{3}{2}}^{(1),i} 
+ (1-4\la)\,\partial\, \Phi_{\frac{1}{2}}^{(1),i}\Bigg](w)
 -  \frac{1}{(z-w)}\Bigg[ \partial\, \Phi_{\frac{3}{2}}^{(1),i}\Bigg](w)
 \nonu \\
 & + & \cdots,
\nonu \\
G^{i}(z) \, \Phi_{\frac{3}{2}}^{(1),j}(w) & = & 
{\tt \frac{1}{(z-w)^4}\Bigg[ 48 \, N \, \la \, (1-2\la) \, \de^{i j}
  \Bigg]}
\nonu \\
& + &
\frac{1}{(z-w)^{3}} \, \Bigg[
 {\tt  48 \, \la \, (1-2\la)\, \de^{i j} \, U
  - 8 \, i \, \la \, (1-2\la)\, \varepsilon^{i j k l}\,
  T^{k l}} \nonu \\
  & + &
 4\, (1-4\la)\, \delta^{ij}\, 
  \Phi_{0}^{(1)}
  \Bigg](w)\nonu \\
& - & \frac{1}{(z-w)^{2}} \, \Bigg[4 \, \Phi_{1}^{(1),ij}+
\frac{1}{2} \, (1-4\la) \, \varepsilon^{i j k l} \, \Phi_{1}^{(1),k l}+
(1-4\la) \, \delta^{ij}\,  \partial\, \Phi_{0}^{(1)} \Bigg](w)
\nonu \\
& - & 
\frac{1}{(z-w) } \, \Bigg[ \partial\, \Phi_{1}^{(1),ij}+ 
 \delta^{ij} \, \Phi_{2}^{(1)}\Bigg](w)
+\cdots,
\nonu \\
G^{i}(z) \, \Phi_{1}^{(1),jk}(w) & = & 
{\tt \frac{1}{(z-w)^3}\, \Bigg[16 \, i \, \la\,(1-2\la)
    \Big( \de^{i j}\, \Ga^k - \de^{i k}\, \Ga^j \Big) \Bigg](w)}
\nonu \\
&
+& \frac{1}{(z-w)^{2}} \Bigg[ -(1-4\la)\, \Big( \delta^{ij} \, 
\Phi_{\frac{1}{2}}^{(1),k}- \delta^{ik} \, 
\Phi_{\frac{1}{2}}^{(1),j}\Big)+3\,
 \varepsilon^{i j k l} \, 
\Phi_{\frac{1}{2}}^{(1),l}
\Bigg](w)\nonu \\
& + & \frac{1}{(z-w)}  \Bigg[- 
\Big( \delta^{ij}\, \Phi_{\frac{3}{2}}^{(1),k}- \delta^{ik} \, 
\Phi_{\frac{3}{2}}^{(1),j}\Big)+ 
 \varepsilon^{i j k l} \, \pa \,
\Phi_{\frac{1}{2}}^{(1),l} \Bigg](w)
+\cdots,
\nonu \\
G^{i}(z) \, \Phi_{\frac{1}{2}}^{(1),j}(w) & = & 
-\frac{1}{(z-w)^{2}} \Bigg[ 2\, \delta^{ij}\, \Phi_{0}^{(1)}
  \Bigg](w)-
\frac{1}{(z-w)}  \Bigg[ \delta^{ij}\, \partial\, \Phi_{0}^{(1)}-
\frac{1}{2}\, \varepsilon^{i j k l} \, \Phi_{1}^{(1),k l} \Bigg](w)
\nonu \\
& + & \cdots,
\nonu \\
G^{i}(z) \, \Phi_{0}^{(1)}(w) & = & 
-\frac{1}{(z-w)} \Bigg[ \Phi_{\frac{1}{2}}^{(1),i}\Bigg](w)
+\cdots, \nonu \\
T^{ij}(z) \, \Phi_{2}^{(1)}(w) & = &
{\tt \frac{1}{(z-w)^3} \Bigg[
  4\, (1-4\la)\, T^{i j}
  -2 \, \varepsilon^{i j k l}\, T^{k l}
  \Bigg](w)}\nonu \\
&+ &
\frac{1}{(z-w)^{2}}  \Bigg[
 {\tt -2\, (1-4\la)\, \pa \, T^{i j}
  +  \varepsilon^{i j k l}\, \pa \, T^{k l}}+
  4\,i\,
 \Phi_{1}^{(1),ij} \Bigg](w)
+\cdots,
\nonu \\
T^{ij}(z) \, \Phi_{\frac{3}{2}}^{(1),k}(w) & = & 
{\tt \frac{1}{(z-w)^3} \Bigg[
  -2 \, \Big(\de^{i k} \, \Ga^j -\de^{j k}\, \Ga^i \Big)
  + 2 \, (1-4\la)\, \varepsilon^{i j k l } \, \Ga^l
  \Bigg](w)}\nonu \\
& + &
\frac{1}{(z-w)^{2}} \, \Bigg[
  {\tt (3-8\la+16\la^2) \,
  \Big(\de^{i k} \, \pa \,\Ga^j -\de^{j k}\, \pa \, \Ga^i \Big)
  -3 \, (1-4\la)\, \varepsilon^{i j k l } \, \pa \, \Ga^l}
  \nonu \\
  & - &  3\, i\, \varepsilon^{i j k l} 
  \, \Phi_{\frac{1}{2}}^{(1),l}
  {\tt +i \,(1-4\la)\,
  \Big(\de^{i k} \, G^j -\de^{j k}\,  G^i \Big)
  -i \, \varepsilon^{i j k l }  \, G^l}
  \Bigg](w)
\nonu \\
& - & \frac{1}{(z-w)} \Bigg[ i\,
  \delta^{ik}\, \Phi_{\frac{3}{2}}^{(1),j}
 -i\, \delta^{jk}\, \Phi_{\frac{3}{2}}^{(1),i}\Bigg](w)
+\cdots,
\nonu \\
T^{ij}(z) \, \Phi_{1}^{(1),kl}(w) & = & 
{\tt \frac{1}{(z-w)^3}\Bigg[ 2 \,N\, i \de^{i k}\,
  \de^{j l}- 2 \, N \, i \, \de^{i l}\,
  \de^{j k} - 2 \,N \, i \, (1-4\la)\, \varepsilon^{i j k l}
  \Bigg]}
\nonu \\
& + &
\frac{1}{(z-w)^{2}} \Bigg[
{\tt   -2 \, i \, (1-4\la)\, \varepsilon^{i j k l}\, U
  -2 \, i \Big(\de^{i l} \, \de^{j k} \,  -
\de^{i k} \, \de^{j l} \
\Big)\, U } \nonu \\
&  - & {\tt
(1-4\la)\, \Big( \de^{i k}\, T^{j l}-
\de^{i l}\, T^{j k}-
\de^{j k}\, T^{i l}+
 \de^{j l}\, T^{i k}\Big)}
 \nonu \\
 &  + & {\tt 
 \frac{1}{2}  \Big( \de^{i k}\,
\varepsilon^{j l m n}
-
\de^{i l}\, \varepsilon^{j k m n}-
\de^{j k}\, \varepsilon^{i l m n}+
\de^{j l}\, \varepsilon^{i k m n}\Big)\,
T^{m n} }
 +2\, i \, 
\varepsilon^{ijkl} \, \Phi_{0}^{(1)}\Bigg](w)
\nonu \\
& - &
\frac{1}{(z-w)}  \Bigg[  
  i\, \delta^{ik}\, \Phi_{1}^{(1),jl} -  i\, \delta^{il}\, \Phi_{1}^{(1),jk} -
  i\, \delta^{jk}\, \Phi_{1}^{(1),il} +  i\,
  \delta^{jl}\, \Phi_{1}^{(1),ik}\Bigg](w)
+\cdots,
\nonu \\
T^{ij}(z)\;\Phi_{\frac{1}{2}}^{(1),k}(w)
& = &
{\tt \frac{1}{(z-w)^2} \Bigg[(1-4\la)\, \Big( \de^{i k} \,
  \Ga^j - \de^{j k}\, \Ga^i \Big) -
  \varepsilon^{i j k l} \, \Ga^l\Bigg](w)}
\nonu \\
& + &
\frac{1}{(z-w)}
\Bigg[
  -i\, \delta^{ik}\,\Phi_{\frac{1}{2}}^{(1),j}+
  i\, \delta^{jk}\,\Phi_{\frac{1}{2}}^{(1),i}
\Bigg](w)
+\cdots,\nonu \\
U(z) \, \Phi_{2}^{(1)}(w) & = & 
{\tt \frac{1}{(z-w)^4} \Bigg[
6 N\, (1-4\la)\Bigg]}+
\frac{1}{(z-w)^{3}}  \Bigg[ -4\, \Phi_{0}^{(1)}
{\tt +8 \, (1-4\la)\, U}
\Bigg](w)\nonu \\
& + &
\frac{1}{(z-w)^{2}}  \Bigg[ 2\, \partial\, \Phi_{0}^{(1)}
 {\tt -4 \, (1-4\la)\, \pa \, U -8 \, L} \Bigg](w)
+\cdots,
\nonu \\
U(z) \, \Phi_{\frac{3}{2}}^{(1),i}(w) & = & 
{\tt \frac{1}{(z-w)^3}\Bigg[ 2 \, i \, (1-4\la)\, \Ga^i\Bigg](w)}
\nonu \\
& + & \frac{1}{(z-w)^{2}} 
\Bigg[ {\tt -3 \, i \, (1-4\la)\, \pa \, \Ga^i}+
  \Phi_{\frac{1}{2}}^{(1),i} {\tt + 3 \, G^i} \Bigg](w)
+\cdots,
\nonu \\
U(z) \, \Phi_{1}^{(1),i j}(w) & = & 
{\tt \frac{1}{(z-w)^2}\Bigg[ 2 \, i \, T^{i j}\Bigg](w)}+ \cdots,
\nonu \\
U(z) \, \Phi_{\frac{1}{2}}^{(1),i }(w) & = & 
{\tt -\frac{1}{(z-w)^2}\Bigg[  i \, \Ga^{i}\Bigg](w)}+ \cdots,
\nonu \\
U(z) \, \Phi_{0}^{(1) }(w) & = & 
{\tt \frac{1}{(z-w)^2}\Bigg[  N\Bigg]} + \cdots,
\nonu \\
\Gamma^{i}(z) \, \Phi_{2}^{(1)}(w) & = & 
{\tt \frac{1}{(z-w)^3}\Bigg[ 2\, (1-4\la)\, \Ga^i\Bigg](w)}
\nonu \\
& + & \frac{1}{(z-w)^{2}} 
\Bigg[
 {\tt -6\, (1-4\la)\, \pa \, \Ga^i}-
  3 \, i\, \Phi_{\frac{1}{2}}^{(1),i} {\tt -3 \, i\, G^i} \Bigg](w)
\nonu \\
& + & \frac{1}{(z-w)} \Bigg[
{\tt   (1-4\la)\, \pa^2 \, \Ga^i}+
   i\, \pa \, \Phi_{\frac{1}{2}}^{(1),i} {\tt +  i\, \pa \,G^i}
\Bigg](w) +\cdots,
\nonu \\
\Gamma^{i}(z) \, \Phi_{\frac{3}{2}}^{(1),j}(w) & = & 
      {\tt -\frac{1}{(z-w)^3}\Bigg[ 2\, N\, i \, (1-4\la) \,
          \de^{i j} \Bigg]}
+  \frac{1}{(z-w)^{2}} \, \Bigg[
  {\tt   2\, T^{i j}} \nonu \\
  & + &
  2\, i\, \delta^{ij}\, \Phi_{0}^{(1)}
  {\tt - 3 \, i\, (1-4\la) \, \de^{i j}\, U
  -\frac{1}{2}\, (1-4\la)\, \varepsilon^{i j k l} \, T^{k l}}
  \Bigg](w)
\nonu \\
& + & \frac{1}{(z-w)} \, \Bigg[
 {\tt -\pa \, T^{i j}}
  -i\, \delta^{ij}\, \pa \, \Phi_{0}^{(1)}
 {\tt + i\, (1-4\la) \, \de^{i j}\, \pa \, U}
  -\frac{1}{2}\, i\, \varepsilon^{i j k l}  \, \Phi_1^{(1), k l}
  \nonu \\
  &+ & {\tt 
2 \, i \, \de^{i j} \, L } \Bigg](w)
+\cdots \,,
\nonu \\
\Gamma^{i}(z) \, \Phi_{1}^{(1),jk}(w) & = & 
{\tt \frac{1}{(z-w)^2}\,
\Bigg[ -(1-4\la)\, \Big( \de^{i k}\, \Ga^j -
  \de^{i j}\, \Ga^k \Big)+ \varepsilon^{i j k l} \, \Ga^l
  \Bigg](w)}\nonu \\
&+ & \frac{1}{(z-w)} \, \Bigg[
 {\tt (1-4\la)\, \Big( \de^{i k}\, \pa \, \Ga^j -
  \de^{i j}\, \pa \, \Ga^k \Big)- \varepsilon^{i j k l} \, \pa \, \Ga^l}-
  i \, \delta^{ij} \, \Phi_{\frac{1}{2}}^{(1),k}
  \nonu \\
  & + & i \, \delta^{ik} \, \Phi_{\frac{1}{2}}^{(1),j}
 {\tt  +  i \, \Big( \de^{i k} \, G^j -
  \de^{i j}\, G^k \Big)}
  \Bigg](w) +\cdots,
\nonu \\
\Gamma^{i}(z) \, \Phi_{\frac{1}{2}}^{(1),j}(w) & = & 
{\tt -\frac{1}{(z-w)^2}\Bigg[ N\, i\, \de^{i j} \Bigg]
+\frac{1}{(z-w)}\, \Bigg[ -i \, \de^{i j} \, U -\frac{1}{2}
  \, \varepsilon^{i j k l}\, T^{k l} \Bigg](w)} +\cdots,
\nonu \\
\Gamma^{i}(z) \, \Phi_{0}^{(1)}(w) & = &
{\tt -\frac{1}{(z-w)}\Bigg[ \Ga^i\Bigg](w)}+ \cdots.
\label{jphicomp}
\eea
Compared to the ${\cal N}=4$ primary condition \cite{AK1509}
of the first ${\cal N}=4$ multiplet,
there are additional terms in (\ref{jphicomp}):
either central terms or the field contents of
the ${\cal N}=4$ stress energy tensor
\footnote{They have the typewriter fonts.}.
In other words,
the first ${\cal N}=4$ multiplet is not ${\cal N}=4$ primary.

\section{
The OPEs between the first ${\cal N}=4$ multiplet
and itself in the component approach }

As before, we
list all the component results corresponding to
(\ref{super1super1}) as follows:
\bea
\Phi^{(1)}_0(z)\,\Phi^{(1)}_0(w)
&=&
\frac{1}{(z-w)^2}\Bigg[
2\,N\, (1-4\lambda)\Bigg]+\cdots
\,,
\nonu\\
\Phi^{(1)}_0(z)\,\Phi^{(1),i}_{\frac{1}{2}}(w)
&=&
\frac{1}{(z-w)}\Bigg[
G^i\Bigg](w)+\cdots
\,,
\nonu\\
\Phi^{(1)}_0(z)\,\Phi^{(1),ij}_{1}(w)
&=&
\frac{1}{(z-w)^2}
\Bigg[2 \, i \, (1-4\lambda)\, T^{ij}+i \, \varepsilon^{i j k l} \,
  T^{k l} 
  \Bigg](w)+\cdots
\,,
\nonu\\
\Phi^{(1)}_0(z)\,\Phi^{(1),i}_{\frac{3}{2}}(w)
&=&
-\frac{1}{(z-w)^3}\,
\Bigg[ 16 \, i \, \lambda\, (1-2\lambda)\, \Gamma^i \Bigg](w)
\nonu \\
& + & \frac{1}{(z-w)^2}
\Bigg[
  3\,(1-4\lambda)\,G^i+16 \, i \, \lambda\,
  (1-2\lambda)\,\partial \, \Gamma^i
\Bigg](w)
\nonu\\
&
+&\frac{1}{(z-w)}\Bigg[
 \frac{1}{3}\,(1-4\lambda)\,\pa \, G^i 
  -\frac{1}{2}\,\Phi^{(2),i}_{\frac{1}{2}}\Bigg](w)+\cdots
\,,
\nonu\\
\Phi^{(1)}_0(z)\,\Phi^{(1)}_{2}(w)
&=&
\frac{1}{(z-w)^4}\Bigg[ 4 \, N\,(1-12\la +24\la^2)\Bigg]-
\frac{1}{(z-w)^3}
\Bigg[32\,\lambda \, (1-2\lambda)\,U\Bigg](w)
\nonu \\
& + & \frac{1}{(z-w)^2}
\Bigg[
2\,\Big( \Phi^{(2)}_0 -\frac{8}{3}\, (1-4\la)\, L\Big)
+16\, \lambda\, (1-2\lambda)\,\partial \, U
\Bigg](w)+\cdots
\,,
\nonu \\
\Phi^{(1),i}_{\frac{1}{2}}(z)\,\Phi^{(1),j}_{\frac{1}{2}}(w)
&=&
-\frac{1}{(z-w)^3}\Bigg[
4\,N \, (1-4\lambda) \, \delta^{ij}\Bigg]
\nonu \\
& + & \frac{1}{(z-w)^2}\,
\Bigg[ 2 \,i \,T^{ij}+ i \, (1-4\lambda)\,
  \varepsilon^{i j k l} \, T^{k l} \,
  \Bigg](w)
\nonu\\
&
+ &\frac{1}{(z-w)}\,\Bigg[
-2\,\delta^{ij}\,L
+i\,\partial \, T^{ij}+\frac{1}{2}\, i\,(1-4\lambda)\,
\varepsilon^{i j k l} \, \pa \,
T^{k l}
\,\Bigg](w)+\cdots
\,,
\nonu\\
\Phi^{(1),i}_{\frac{1}{2}}(z)\,\Phi^{(1),jk}_{1}(w)
&=&
\frac{1}{(z-w)^3}\,\Bigg[ 16 \, i\,\lambda \, (1-2\lambda)\,
\Big(\delta^{ij}\,\Gamma^k-\delta^{ik}\,\Gamma^j\Big)\Bigg](w)
\nonu\\
&+&\frac{1}{(z-w)^2}
\Bigg[
-(1-4\lambda)\, \Big(\delta^{ij}\,G^k-\delta^{ik}\,G^j\Big)
+3\,\varepsilon^{ijkl}\,G^l
\,\Bigg](w)
\nonu\\
&+&\frac{1}{(z-w)}
\Bigg[
\delta^{ij}\, \Big(
\frac{1}{2}\,\Phi^{(2),k}_{\frac{1}{2}}-\frac{1}{3}\, (1-4\lambda)\,
\partial \, G^k \Big) 
\nonu \\
& - &
\delta^{ik}\, \Big(
\frac{1}{2}\,\Phi^{(2),j}_{\frac{1}{2}}-\frac{1}{3}\, (1-4\lambda)\,
\partial \, G^j
\Big) 
+
  \varepsilon^{ijkl}\,\partial \, G^l \Bigg](w)+\cdots \,,
\nonu\\
\Phi^{(1),i}_{\frac{1}{2}}(z)\,\Phi^{(1),j}_{\frac{3}{2}}(w)
&=&
-\frac{1}{(z-w)^4}\Bigg[ 4 \, N\,(1-12\la+24\la^2) \, \de^{ij} \Bigg]
\nonu \\
&+ & \frac{1}{(z-w)^3}
\Bigg[
16\, \delta^{ij}\,\lambda \, (1-2\lambda)\,U
-
4 i\, (1-4\lambda)\,T^{ij} \nonu \\
& - & 2 \, i \, (1+4\la-8\la^2) \,\varepsilon^{i j k l}
\, T^{k l}
\,\Bigg](w)
\nonu\\
&
+&\frac{1}{(z-w)^2}\Bigg[
  i\, (1-4\lambda)\,\pa \, T^{ij} +
  \frac{1}{2} \, i \, (1-4\la)^2 \,\varepsilon^{i j k l}
\, \pa \, T^{k l}
  \nonu \\
 & - & 2\,\delta^{ij}\, \Phi^{(2)}_{0} -
  \frac{2}{3}\, (1-4\la) \, L\Bigg](w)
\nonu \\
& + & \frac{1}{(z-w)}
\Bigg[
  \frac{1}{3}\, i\, (1-4\lambda)\,\pa^2 \,
  T^{ij} +  \frac{1}{6} \, i \, (1-4\la)^2 \,
  \varepsilon^{i j k l}
\, \pa^2 \, T^{k l}
\nonu \\
& - & \frac{2}{3}\, (1-4\la) \, \de^{i j}\, \pa\, L
-\frac{1}{2} \, \delta^{ij}\,\partial \, \Phi^{(2)}_{0}
+\frac{1}{4}\,\varepsilon^{i j k l} \,
\Phi^{(2),k l}_{1} \Bigg](w)+\cdots,
\nonu\\
\Phi^{(1),i}_{\frac{1}{2}}(z)\,\Phi^{(1)}_{2}(w)
&=&
\frac{1}{(z-w)^4}\Bigg[48 \, i\,\lambda\,
  (1-2\lambda)\,\Gamma^i\Bigg](w)
\nonu\\
&
+&\frac{1}{(z-w)^3}\Bigg[
-6\, (1-4\lambda)\,G^i-48\, i\,\lambda\, (1-2\lambda)\,\partial \Gamma^i
\Bigg](w)
\nonu \\
& + & \frac{1}{(z-w)^2}
\Bigg[
-\frac{2}{3}\, (1-4\lambda)\,\pa \, G^i
  +\frac{5}{2}\,\Phi^{(2),i}_{\frac{1}{2}}\Bigg](w)
\nonu\\
&
+&\frac{1}{(z-w)}\Bigg[
-\frac{1}{3}\, (1-4\lambda)\,\pa^2 \, G^i+
  \frac{1}{2}\,\partial\, \Phi^{(2),i}_{\frac{1}{2}}\Bigg](w)+
\cdots \,,
\nonu \\
\Phi^{(1),ij}_{1}(z)\,\Phi^{(1),kl}_{1}(w)
&=&
\frac{1}{(z-w)^4}\,\Bigg[\,
-12\,N \, (1-4\lambda)\,
\Big(\delta^{ik}\delta^{jl}-\delta^{il}\delta^{jk}\Big)
\nonu \\
& + & 4\, N \, (1-12\la+24\la^2)\,\varepsilon^{ijkl}
\Bigg]
\nonu\\
&
+&\frac{1}{(z-w)^3} \Bigg[ 
  -4 i\, (1+4\la-8\la^2)\nonu \\
  &\times & 
\Big( -\delta^{ik}  
\, T^{jl}
+\delta^{i l}  
\, T^{jk}
+\delta^{j k}  
\, T^{il}-
\delta^{jl}  
\, T^{ik}
\Big)
-  2 \, i \, (1-4\la)\nonu \\
&\times & \Big(
 -\delta^{ik}  \, \varepsilon^{j l m n}
+\delta^{i l} \, \varepsilon^{j k m n}  
+\delta^{j k} \, \varepsilon^{i l m n}  
-
\delta^{jl} \, \varepsilon^{i k m n}  
\, \Big) \, T^{m n}
\,\Bigg](w)
\nonu\\
&
+&\frac{1}{(z-w)^2} \Bigg[
  -2 i\, (1+4\la-8\la^2)\nonu \\
  &\times & 
\Big( -\delta^{ik}  
\, \pa \, T^{jl}
+\delta^{i l}  
\, \pa \, T^{jk}
+\delta^{j k}  
\, \pa \, T^{il}-
\delta^{jl}  
\, \pa \, T^{ik}
\Big)
-   i \, (1-4\la)\nonu \\
&\times & \Big(
 -\delta^{ik}  \, \varepsilon^{j l m n}
+\delta^{i l} \, \varepsilon^{j k m n}  
+\delta^{j k} \, \varepsilon^{i l m n}  
-
\delta^{jl} \, \varepsilon^{i k m n}  
\, \Big) \, \pa \, T^{m n}
\nonu \\
&-&
8 \, \Big( \delta^{ik}\delta^{jl}-\delta^{il}\delta^{jk} \Big)\,L
+2\, \varepsilon^{ijkl}\,
\Phi^{(2)}_{0} +\frac{8}{3}\, (1-4\la)\,
\varepsilon^{ijkl}\,
L
\,\Bigg](w)
\nonu\\
&
+&\frac{1}{(z-w)}\Bigg[
  -\frac{2}{3}\, i\, (1+4\la-8\la^2)\nonu \\
  &\times & 
\Big( -\delta^{ik}  
\, \pa^2 \, T^{jl}
+\delta^{i l}  
\, \pa^2 \, T^{jk}
+\delta^{j k}  
\, \pa^2 \, T^{il}-
\delta^{jl}  
\, \pa^2 \, T^{ik}
\Big)
-   i \, \frac{1}{3}\, (1-4\la)\nonu \\
& \times & \Big(
 -\delta^{ik}  \, \varepsilon^{j l m n}
+\delta^{i l} \, \varepsilon^{j k m n}  
+\delta^{j k} \, \varepsilon^{i l m n}  
-
\delta^{jl} \, \varepsilon^{i k m n}  
\, \Big) \, \pa^2 \, T^{m n}
\nonu \\
&-&
4 \, \Big( \delta^{ik}\delta^{jl}-\delta^{il}\delta^{jk}
\Big)\, \pa \, L
+ \varepsilon^{ijkl}\,
\pa \, \Phi^{(2)}_{0} +\frac{4}{3}\, (1-4\la)\,
\varepsilon^{ijkl}\,
\pa \, L
\nonu\\
&
+ & \frac{1}{2}\, \Big( -\delta^{ik}   
\, \Phi^{(2),jl}_{1} +
\delta^{il}   
\, \Phi^{(2),jk}_{1}
+\delta^{jk}   
\, \Phi^{(2),il}_{1}-
\delta^{jl}   
\, \Phi^{(2),ik}_{1}
\,\Big)
\,\Bigg](w)+  \cdots \,,
\nonu \\
\Phi^{(1),ij}_{1}(z)\,\Phi^{(1),k}_{\frac{3}{2}}(w)
&=&
-\frac{1}{(z-w)^4}\Bigg[
  48 \, i\, \la\, (1-2\la)\,
  \varepsilon^{ijkl}\,\Gamma^l
\,\Bigg](w)
\nonu\\
&
+&\frac{1}{(z-w)^3}
\Bigg[
2\, (5+8\la-16\la^2)\,\Big(\delta^{ik}\,G^j-\delta^{jk}\,G^i\Big)
+
  2 \, (1-4\la)\,
  \varepsilon^{ijkl}\,G^l
  \nonu \\
  &- & 16 \, i \,\la\, (1-2\la)\,(1-4\la)\,
  \Big(\delta^{ik}\, \pa \, \Ga^j-
  \delta^{jk}\,\pa \, \Ga^i\Big)
\,\Bigg](w)
\nonu\\
&
+&\frac{1}{(z-w)^2}
\Bigg[
  (3-4\la)\, (1+4\la)\,
  \Big(\delta^{ik}\,\pa \, G^j-\delta^{jk}\,\pa \, G^i\Big)
\nonu \\
&+&
  \frac{5}{3}\, (1-4\la)\,
  \varepsilon^{ijkl}\,\pa \, G^l
  \nonu \\
  & - & 8 i \,\la\, (1-2\la)\,(1-4\la)\,
  \Big(\delta^{ik}\, \pa^2 \, \Ga^j-
  \delta^{jk}\,\pa^2 \, \Ga^i\Big)
- 
\frac{5}{2} \, \varepsilon^{ijkl}\,\Phi^{(2),l}_{\frac{1}{2}}
\,\Bigg](w)
\nonu\\
&
+&\frac{1}{(z-w)}
\Bigg[
  \frac{2}{3}\,(1+4\la-8\la^2)\,
  \Big(\delta^{ik}\,\pa^2 \, G^j-\delta^{jk}\,\pa^2 \, G^i\Big)
\nonu \\
&+&
  \frac{2}{3}\, (1-4\la)\,
  \varepsilon^{ijkl}\,\pa^2 \, G^l
  \nonu \\
  & - & \frac{8}{3} \,
  i \,\la \, (1-2\la)\,(1-4\la)\,
  \Big(\delta^{ik}\, \pa^3 \, \Ga^j-
  \delta^{jk}\,\pa^3 \, \Ga^i\Big)
\nonu\\
&
- & 
 \varepsilon^{ijkl}\,\pa \, \Phi^{(2),l}_{\frac{1}{2}}
-\frac{1}{2} \,
\Big(\delta^{ik}\, \Phi^{(2),j}_{\frac{3}{2}}-
\delta^{jk} \, \Phi^{(2),i}_{\frac{3}{2}}\Big)
\,\Bigg](w)+\cdots\,,
\nonu\\
\Phi^{(1),ij}_{1}(z)\,\Phi^{(1)}_{2}(w)
&=&
\frac{1}{(z-w)^4}\Bigg[
-12 \, i \,(1+8\la-16\la^2)\, T^{ij}
-6 \, i \, (1-4\lambda)\,\varepsilon^{i j k l}\, T^{k l}
\Bigg](w)
\nonu\\
&
+ & \frac{1}{(z-w)^3}
\Bigg[
4 \, i \,(1-4\la)^2\, \pa \, T^{ij}
+2 \, i \, (1-4\lambda)\,\varepsilon^{i j k l}\, \pa \, T^{k l}
\Bigg](w)
\nonu \\
&
+ & \frac{1}{(z-w)^2}
\Bigg[
2 \, i \,(1-4\la)^2\, \pa^2 \, T^{ij}
+i \, (1-4\lambda)\,\varepsilon^{i j k l}\, \pa^2 \, T^{k l}
\nonu \\
& + & 3 \, \Phi^{(2), i j}_{1}
\Bigg](w)
\nonu \\
& + & \frac{1}{(z-w)}
\Bigg[
\frac{2}{3} \, i \,(1-4\la)^2\, \pa^3 \, T^{ij}
+\frac{1}{3}\, i \, (1-4\lambda)\,\varepsilon^{i j k l}\, \pa^3 \, T^{k l}
\nonu \\
& + &  \pa \, \Phi^{(2), i j}_{1}
\Bigg](w)+\cdots\,,
\nonu \\
\Phi^{(1),i}_{\frac{3}{2}}(z)\,\Phi^{(1),j}_{\frac{3}{2}}(w)
&=&
\frac{1}{(z-w)^5}\Bigg[ 48\,N\, (1-4\lambda)\, \delta^{ij}\Bigg]
\nonu\\
&
+&\frac{1}{(z-w)^4}\Bigg[
-12 \, i \, (1+8\la-16\la^2)\,T^{ij}
-  6 \, i \, (1-4\lambda)\,\varepsilon^{i j k l} \, T^{k l}
\,\Bigg](w)
\nonu\\
&
+& \frac{1}{(z-w)^3}\Bigg[
-6 \, i \, (1+8\la-16\la^2)\,\pa \, T^{ij}
-  3 \, i \, (1-4\lambda)\,\varepsilon^{i j k l} \, \pa \, T^{k l}
\nonu \\
&+& 4\, (9+8 \la -16\la^2) \,\de^{i j} \, L
+ 16\, \la \, (1-2\la)\, (1-4\la)\, \de^{i j}\, \pa \, U
\Bigg](w)
\nonu\\
&
+& \frac{1}{(z-w)^2}\Bigg[
-i \, (1+24\la-48\la^2)\,\pa^2 \, T^{ij}
-  \frac{1}{2} \,i \,
(1-4\lambda)\,\varepsilon^{i j k l} \, \pa^2 \, T^{k l}
\nonu \\
&+& 2\, (9+8 \la -16\la^2) \,\pa \, L
+ 8\, \la \, (1-2\la)\, (1-4\la)\, \de^{i j}\, \pa^2 \, U
\nonu \\
&
+ & 3\,\Phi^{(2),ij}_{1}
\,\Bigg](w)
\nonu\\
&
+&\frac{1}{(z-w)}\Bigg[
-8 \, i \, \la\, (1-2\la)\,\pa^3 \, T^{ij}
\nonu \\
&+& \frac{8}{3}\, (2-2\la)\, (1+2\la) \,\pa^2 \, L
+ \frac{8}{3}\, \la \, (1-2\la)\,
(1-4\la)\, \de^{i j}\, \pa^3 \, U
\nonu\\
&
+& \frac{3}{2}\,\partial \, \Phi^{(2),ij}_{1}
+ \frac{1}{2}\, \de^{i j}\, \Phi^{(2)}_{2}
\,\Bigg](w)
+  \cdots \,,
\nonu\\
\Phi^{(1),i}_{\frac{3}{2}}(z)\,\Phi^{(1)}_{2}(w)
&=&
\frac{1}{(z-w)^4}\Bigg[
  -6\,(7+24\la-48\la^2)\,G^i \nonu \\
  & + &
144 \, i \la \, (1-2\la)\, (1-4\la)\, \pa \, \Ga^i
  \Bigg](w)
\nonu\\
&
+&\frac{1}{(z-w)^3}\Bigg[
  -4\, (3+16\la-32\la^2)\,\pa \, G^i \nonu \\
  & + &
64 \, i \la \, (1-2\la)\, (1-4\la)\, \pa^2 \, \Ga^i
  \,\Bigg](w)
\nonu\\
&
+&\frac{1}{(z-w)^2}\Bigg[
  -\frac{5}{3}\, (1+16\la-32\la^2)\,\pa^2 \, G^i
  \nonu \\
  & + &
\frac{80}{3} \, i \la \, (1-2\la)\, (1-4\la)\, \pa^3 \, \Ga^i
+  
\frac{7}{2}\,\Phi^{(2),i}_{\frac{3}{2}}
\Bigg](w)
\nonu\\
&
+&\frac{1}{(z-w)}
\Bigg[
 -8 \, \la \, (1-2\la)\,\pa^3 \, G^i +
8 \, i \, \la \, (1-2\la)\, (1-4\la)\, \pa^4 \, \Ga^i
\nonu \\
&+&  
\frac{3}{2}\,\pa \, \Phi^{(2),i}_{\frac{3}{2}}
\Bigg](w)
+  \cdots \,,
\nonu \\
\Phi^{(1)}_{2}(z)\,\Phi^{(1)}_{2}(w)
&=&
\frac{1}{(z-w)^6}\Bigg[
240\,N \, (1-4\lambda)\Bigg]
\nonu\\
&
+&\frac{1}{(z-w)^4}
\Bigg[
  192\,(1+2\lambda-4\la^2)\,L+
 192 \, \la \, (1-2\la)\, (1-4\la)\, \pa \, U
\,\Bigg](w)
\nonu\\
&
+&\frac{1}{(z-w)^3}
\Bigg[
  96 \,(1+2\lambda-4\la^2)\,\pa \, L+
 96 \, \la \, (1-2\la)\, (1-4\la)\, \pa^2 \, U
\,\Bigg](w)
\nonu\\
&
+&\frac{1}{(z-w)^2}
\Bigg[
  \frac{16}{3} \,(5+14\lambda-28\la^2)\,\pa^2 \, L+
  \frac{112}{3} \, \la \, (1-2\la)\, (1-4\la)\, \pa^3 \, U
  \nonu \\
&+& 
4\,\Phi^{(2)}_{2} \Bigg](w)
\nonu\\
&
+&\frac{1}{(z-w)}
\Bigg[
  \frac{16}{3} \,(1+4\lambda-8\la^2)\,\pa^3 \, L+
  \frac{32}{3}\, \la \, (1-2\la)\, (1-4\la)\, \pa^4 \, U
  \nonu \\
&+& 
2\, \pa \, \Phi^{(2)}_{2} \Bigg](w)
+  \cdots \,.
\label{ope1}
\eea
Note that on the right hand sides of
(\ref{ope1}), the field contents of
the ${\cal N}=4$ stress energy tensor and the second
${\cal N}=4$ multiplet appear which is manifest in
(\ref{super1super1}).

\section{ The ${\cal N}=4$ coset model results under the
large $(N,k)$ limit}

We rewrite the previous results
under the large $(N,k)$ limit
in \cite{AKK1910} as
\bea
\Phi^{(1)}_0(z)\,\Phi^{(1)}_0(w)
&=&
\frac{1}{(z-w)^2}\,
2\,N(1-\lambda_{co})+\cdots,
\nonu \\
\Phi^{(1),i}_{\frac{1}{2}}(z)\,\Phi^{(1)}_0(w)
&=&
-\frac{1}{(z-w)}\,
G^i(w)+\cdots,
\nonu \\
\Phi^{(1),i j}_{1}(z)\,\Phi^{(1)}_0(w)
&=&
\frac{1}{(z-w)^2}\,\Bigg[ 2 i \, (1-2\la_{co})\,T^{i j}+
  i \, \varepsilon^{i j k l}\, T^{k l} \Bigg](w)
\nonu \\
&+&  \frac{1}{(z-w)}\,\Bigg[ 2 i \, (1-2\la_{co})\,\pa \, T^{i j}+
  i \, \varepsilon^{i j k l}\,\pa \,  T^{k l}  \Bigg](w)
+\cdots,
\nonu \\
\Phi^{(1),i }_{\frac{3}{2}}(z)\,\Phi^{(1)}_0(w)
&=&
\frac{1}{(z-w)^3}\,\Bigg[ 8 i \, \la_{co} \, (1-\la_{co})\,\Ga^{i} \Bigg](w)
\nonu \\
& + & \frac{1}{(z-w)^2}\,\Bigg[ 16 i \, \la_{co} \, (1-\la_{co})\,
  \pa \, \Ga^{i}+3(1-2\la_{co})\, G^i \Bigg](w)
\nonu \\
&+&
\frac{1}{(z-w)}\,\Bigg[ 12 i \, \la_{co} \, (1-\la_{co})\,
  \pa^2 \, \Ga^{i}+\frac{8}{3}(1-2\la_{co})\, \pa \,
  G^i + \frac{1}{2}\, \Phi_{\frac{1}{2}}^{(2),i} \Bigg](w)
\nonu \\
& + & \cdots,
\nonu \\
\Phi^{(1) }_{2}(z)\,\Phi^{(1)}_0(w)
&=&
\frac{1}{(z-w)^4}\,\Bigg[ 4 N \, (1-\la_{co})( 1- 2\la_{co})\Bigg]+
\frac{1}{(z-w)^3}\,\Bigg[ 16 \, \la_{co} \, (1-\la_{co}) U\Bigg](w)
\nonu \\
& + &
\frac{1}{(z-w)^2}\,\Bigg[ 24 \, \la_{co} \, (1-\la_{co}) \, \pa \, U +
  2 \, \Big(\Phi^{(2) }_{0} - \frac{8}{3}\, (1-2\la_{co}) \, L \Big)
  \Bigg](w)
\nonu \\
& + &
\frac{1}{(z-w)}\,\Bigg[
   16 \, \la_{co} \, (1-\la_{co}) \, \pa^2 \, U +
   2 \, 
   \Big(\pa \,\Phi^{(2) }_{0} - \frac{8}{3}\, (1-2\la_{co}) \,
   \pa \, L \Big)\Bigg](w)
\nonu \\
& + &
\cdots.
\label{cosetresult}
\eea
It is straightforward to
express the OPE like as in (\ref{super1super1}) in ${\cal N}=4$
superspace. 

\section{
  The OPEs between the ${\cal N}=4$ stress energy tensor
  and the second ${\cal N}=4$ multiplet }

We present
the super  OPE between the ${\cal N}=4$ stress energy tensor 
and the second ${\cal N}=4$ multiplet
as follows:
\bea
{\bf J}(Z_{1})\,{\bf \Phi}^{(2)}(Z_{2})  & = & 
-\frac{\theta_{12}^{4-0}}{z_{12}^{4}}\,
8\,N\, (1-4\la+8\la^2)
-\frac{1}{z_{12}^2} 
\, \frac{4}{3}\, N\, (1-4\la)
\nonu \\
& + & \frac{\theta_{12}^{4-i}}{z_{12}^3}\,
 \frac{16}{3} (1-2\la+4\la^2)\, D^i\, {\bf J}
(Z_{2})-\frac{\theta_{12}^{i}}{z_{12}^2}\,
\frac{8}{3} \,(1-4\la)\, D^i \, {\bf J}(Z_2)
\nonu \\
& + & \frac{\theta_{12}^{4-0}}{z_{12}^{3}}\,
\Bigg[ 8\, (1-4\la) \, {\bf \Phi}^{(1)}
  +\frac{16}{3}\, (1-4\la)^2 \, \pa \, {\bf J}\Bigg](Z_2)
\nonu \\
& + & 
 \frac{1}{z_{12}}\Bigg[ 4 \, {\bf \Phi}^{(1)}-
  \frac{8}{3}\, (1-4\la)\, \pa {\bf J}
  \Bigg](Z_2)
 \nonu \\
& + & \frac{\theta_{12}^{4-ij}}{z_{12}^2}\Bigg[
  -\frac{4}{3}\, (1-4\la)\, \frac{1}{2!} \, \varepsilon^{i j k l}
  D^k\, D^l \, {\bf J}
  -2 \, \varepsilon^{i j k l} \, \frac{1}{2} \, \varepsilon^{ k l m n}
  \, D^m \, D^n \, {\bf J}
  \Bigg](Z_2)
\nonu \\
 & + & \frac{\theta_{12}^{4-i}}{z_{12}^2}\Bigg[
  -\frac{8}{3} (1-4\la)^2\, \pa \, D^i\, {\bf J}
  +2\, (1-4\la)\, D^i \, {\bf \Phi}^{(1)}
  \nonu \\
  & + & 2\, (1-4\la)\, \Big(-\frac{1}{3!}\, \varepsilon^{i j k l}\,
  D^j \, D^k \, D^l \, {\bf J} -(1-4\la)\, \pa \, D^i \, {\bf J}
  \Big) 
\Bigg] (Z_{2})
\nonu \\
& + & 
 \frac{\theta_{12}^{i}}{z_{12}}\Bigg[
  \frac{8}{3} \,(1-4\la)\, \pa \, D^i \, {\bf J}
  + 2\, D^i \, {\bf \Phi}^{(1)}
  \nonu \\
  & + & 2 \, \Big( -\frac{1}{3!}\, \varepsilon^{i j k l}\,
  D^j \, D^k \, D^l \, {\bf J} -(1-4\la)\, \pa \, D^i \,
  {\bf J} \Big)
  \Bigg](Z_2)
\nonu \\
\nonu \\
&+ & \frac{\theta_{12}^{4-0}}{z_{12}^{2}}\,
4\, {\bf \Phi}^{(2)}(Z_{2})
+\frac{\theta_{12}^{4-i}}{z_{12}}\,
D^i\, {\bf \Phi}^{(2)}(Z_{2})
+\frac{\theta_{12}^{4-0}}{z_{12}}\,
 2\, \pa \,  {\bf \Phi}^{(2)}(Z_{2})
 +\cdots.
 \label{JPhitwo}
\eea
Compared to the ${\cal N}=4$ primary
condition \cite{AK1509} for the second ${\cal N}=4$ multiplet,
there are additional terms except the last line of
(\ref{JPhitwo}). The ${\cal N}=4$ stress energy tensor
and the first ${\cal N}=4$ multiplet including their
descendants appear in these extra terms.

\section{
The OPEs between the first ${\cal N}=4$ multiplet
and the second ${\cal N}=4$ multiplet }

We describe
the super  OPE between the first ${\cal N}=4$ multiplet
and the second ${\cal N}=4$ multiplet
as follows:
\bea
&& {\bf \Phi}^{(1)}(Z_{1})\,{\bf \Phi}^{(2)}(Z_{2})  =
-\frac{\theta_{12}^{4-0}}{z_{12}^5}\,\Bigg[
  128 \, N \, \la \, (1-2\la)\, (1-4\la)\Bigg]
\nonu \\
&& +  \frac{\theta_{12}^{4-0}}{z_{12}^4}\,\Bigg[
  -32 \, (-1-4\la+8\la^2)\, {\bf \Phi}^{(1)}
  -64 \, \la \, (1-2\la)\, (1-4\la)\, \pa \, {\bf J}
  \Bigg]
\nonu \\
&& +  \frac{\theta_{12}^{4-i}}{z_{12}^4}\,
\Bigg[\, 32 \, \la \, (1-2\la)\, (1-4\la)\, D^i \, {\bf J} 
\,\Bigg]
(Z_2)\nonu \\
&& + 
\frac{\theta_{12}^{4-i}}{z_{12}^3}\,
\Bigg[\, \frac{64}{3}  \, (1-\la)\, (1+2\la)\, D^i \, {\bf \Phi}^{(1)} 
  \,\Bigg](Z_2)
\nonu \\
&& +  \frac{\theta_{12}^{i}}{z_{12}^3}\,
\Bigg[\, -32  \, \la \,  (1-2\la)\, D^i \, {\bf J} 
  \,\Bigg](Z_2)
\nonu\\
&&
+ \frac{\theta_{12}^{4-0}}{z_{12}^3}\,
\Bigg[\,
   \frac{128}{3} \, (1-\la)\, (1+2\la)\, \pa \, {\bf \Phi}^{(1)}
\,\Bigg](Z_2) 
+\frac{1}{z_{12}^3}\, \Bigg[\,
  -32 \, N\, \la \, (1-2\la)
  \,\Bigg]
\nonu \\
&& +
\frac{1}{z_{12}^2}\, \Bigg[\,
  -\frac{16}{3}\, (1-4\la)\, {\bf \Phi}^{(1)} -
  32 \,\la \, (1-2\la)\, \pa {\bf J}
  \,\Bigg](Z_2)
\nonu \\
&& +  \frac{\theta_{12}^{4-ij}}{z_{12}^2}\, \Bigg[\, 
  -\frac{8}{3}\, (1-4\la)\, \frac{1}{2}\, \varepsilon^{i j k l}\,
  D^k \, D^l \, {\bf \Phi}^{(1)} -  4
  \, \varepsilon^{i j k l } \, \frac{1}{2}\,
  \varepsilon^{ k l m n }\,
  D^m \, D^n \, {\bf \Phi}^{(1)} 
  \,\Bigg](Z_2)
\nonu\\
&& 
+  \frac{\theta_{12}^{4-i}}{z_{12}^2}\,\Bigg[\, 
  -\frac{2}{3}\, (-23-8\la+16\la^2)\, \pa \, D^i \, {\bf \Phi}^{(1)}
  \nonu \\
&  & -  \frac{10}{3}\, (1-4\la)\, \frac{1}{3!}\, \varepsilon^{i j k l}\,
  D^j \, D^k \, D^l \, {\bf \Phi}^{(1)}
  \,\Bigg](Z_2)
\nonu\\
& &
+ \frac{\theta_{12}^{4-0}}{z_{12}^2}\,
\Bigg[\,
 -\frac{8}{15} \,(-41-32\la+64\la^2)\,  \pa^2 \, {\bf \Phi}^{(1)}
 \nonu \\
 && -  \frac{8}{5}\, (1-4\la)\, \frac{1}{4!} \, \varepsilon^{i j k l}
 \, D^i \, D^j \, D^k \,D^l \, {\bf \Phi}^{(1)}
 + {\bf \Phi}^{(3)} 
 \,\Bigg](Z_2)
\nonu\\
& &
+\frac{\theta_{12}^{i}}{z_{12}}\, \Bigg[\, 
  -\frac{2}{3} \, (1-4\la)\, \pa  \, D^i \, {\bf \Phi}^{(1)}
  + 2\, \frac{1}{3!} \, \varepsilon^{i j k l}
 \, D^j \, D^k \,D^l \, {\bf \Phi}^{(1)}
  \,\Bigg](Z_2)
\nonu \\
&& +  \frac{\theta_{12}^{4-ij}}{z_{12}}\,\Bigg[\, 
 -\frac{4}{3}\, (1-4\la)\, \frac{1}{2}\, \varepsilon^{i j k l}\,
 \pa \, D^k \, D^l \, {\bf \Phi}^{(1)}  -  2
  \, \varepsilon^{i j k l } \, \frac{1}{2}\,
  \varepsilon^{ k l m n }\, \pa \, 
  D^m \, D^n \, {\bf \Phi}^{(1)} 
  \,\Bigg](Z_2)
\nonu \\
&& +  \frac{\theta_{12}^{4-i}}{z_{12}}\,\Bigg[\, 
\frac{8}{15}\, (-11-2\la+4\la^2)\, \pa^2 \, D^i \, {\bf \Phi}^{(1)}
\nonu \\
&& -  \frac{8}{5}\, (1-4\la)\, \frac{1}{3!}\, \varepsilon^{i j k l}\,
\pa \,   D^j \, D^k \, D^l \, {\bf \Phi}^{(1)}
+\frac{1}{6}\, D^i \, {\bf \Phi}^{(3)}
\,\Bigg](Z_2)
\nonu\\
& &
+\frac{\theta_{12}^{4-0}}{z_{12}}\,
\Bigg[\,
  -\frac{16}{15} \,(-7-4\la+8\la^2)\,  \pa^3 \, {\bf \Phi}^{(1)}
 \nonu \\
 && -  \frac{16}{5}\, (1-4\la)\, \frac{1}{4!} \, \varepsilon^{i j k l}
 \, \pa \, D^i \, D^j \, D^k \,D^l \, {\bf \Phi}^{(1)}
 + \frac{2}{3} \, \pa \, {\bf \Phi}^{(3)} 
\,\Bigg](Z_2)
+\cdots.
\label{singleOPE}
\eea
On the right hand sides of (\ref{singleOPE}),
there are the ${\cal N}=4$ stress energy tensor,
the first ${\cal N}=4$ multiplet, the third ${\cal N}=4$ multiplet
as well as their descendants.

The third ${\cal N}=4$ multiplet can be summarized by
\bea
\Phi^{(3) }_{0} & = &
-\frac{48}{5} \, (-3+2\la)\,
(W^{\la,11}_{\mathrm{F},3}+W^{\la,22}_{\mathrm{F},3})-
\frac{96}{5}\,(1+\la)\,
(W^{\la,11}_{\mathrm{B},3}+W^{\la,22}_{\mathrm{B},3})\,,
\nonu \\
\Phi^{(3),1}_{\frac{1}{2}}
&
=&
(-6)(-2) \times \Bigg[ \frac{1}{2}\,\Big(
Q^{\la,11}_{\frac{7}{2}}
+i\sqrt{2}\,Q^{\la,12}_{\frac{7}{2}}
+2i \sqrt{2}\,Q^{\la,21}_{\frac{7}{2}}
-2\,Q^{\la,22}_{\frac{7}{2}}
\nonu \\
& + & 2\,\bar{Q}^{\la,11}_{\frac{7}{2}}
+2i \sqrt{2}\,\bar{Q}^{\la,12}_{\frac{7}{2}}
+i\sqrt{2}\,\bar{Q}^{\la,21}_{\frac{7}{2}}
-\bar{Q}^{\la,22}_{\frac{7}{2}}
\Big) \Bigg]\,,
\nonu\\
\Phi^{(3),2}_{\frac{1}{2}}
&
=&
(-6)(-2) \times \Bigg[ -\frac{i}{2}\,\Big(
Q^{\la,11}_{\frac{7}{2}}
+2i\sqrt{2} \,Q^{\la,21}_{\frac{7}{2}}
-2 \,Q^{\la,22}_{\frac{7}{2}}
+2\, \bar{Q}^{\la,11}_{\frac{7}{2}}
+2i\sqrt{2} \, \bar{Q}^{\la,12}_{\frac{7}{2}}
\nonu \\
& - & \bar{Q}^{\la,22}_{\frac{7}{2}}
\Big)\Bigg]\,,
\nonu\\
\Phi^{(3),3}_{\frac{1}{2}}
&
=& (-6)(-2) \times \Bigg[
-\frac{i}{2}\,\Big(
Q^{\la,11}_{\frac{7}{2}}
+i\sqrt{2} \,Q^{\la,12}_{\frac{7}{2}}
-2\,Q^{\la,22}_{\frac{7}{2}}
+2\, \bar{Q}^{\la,11}_{\frac{7}{2}}
+i \sqrt{2} \, \bar{Q}^{\la,21}_{\frac{7}{2}}
-\bar{Q}^{\la,22}_{\frac{7}{2}}
\Big)\Bigg]\,,
\nonu\\
\Phi^{(3),4}_{\frac{1}{2}}
&
=& (-6)(-2) \times \Bigg[
-\frac{1}{2}\,Q^{\la,11}_{\frac{7}{2}}
-Q^{\la,22}_{\frac{7}{2}}
+\bar{Q}^{\la,11}_{\frac{7}{2}}
+\frac{1}{2}\,\bar{Q}^{\la,22}_{\frac{7}{2}} \Bigg]
\,,
\nonu \\
\Phi^{(3),12}_{1}
&
=& (-6)(-2) \times \Bigg[
2i\,W^{\la,11}_{\mathrm{B},4}
-\sqrt{2}\,W^{\la,12}_{\mathrm{B},4}
-2i\,\,W^{\la,22}_{\mathrm{B},4}
+2i\,W^{\la,11}_{\mathrm{F},4}
-2\sqrt{2}\,W^{\la,12}_{\mathrm{F},4}
\nonu \\
& - & 2i\,W^{\la,22}_{\mathrm{F},4} \Bigg]\,,
\nonu\\
\Phi^{(3),13}_{1}
&
=&
(-6)(-2) \times \Bigg[
-2i\,W^{\la,11}_{\mathrm{B},4}
+4\sqrt{2}\,W^{\la,21}_{\mathrm{B},4}
+2i\,\,W^{\la,22}_{\mathrm{B},4}
-2i\,W^{\la,11}_{\mathrm{F},4}
+2\sqrt{2}\,W^{\la,21}_{\mathrm{F},4}
\nonu \\
& + & 2i\,W^{\la,22}_{\mathrm{F},4} \Bigg]\,,
\nonu\\
\Phi^{(3),14}_{1}
&
=&
(-6)(-2)\times \Bigg[
2\,W^{\la,11}_{\mathrm{B},4}
+i\sqrt{2}\,W^{\la,12}_{\mathrm{B},4}
+4i\sqrt{2}\,\,W^{\la,21}_{\mathrm{B},4}
-2\,W^{\la,22}_{\mathrm{B},4}
-2\,W^{\la,11}_{\mathrm{F},3}
\nonu \\
& - & 2i\sqrt{2}\,W^{\la,12}_{\mathrm{F},4}
-  2i\sqrt{2}\,W^{\la,21}_{\mathrm{F},4}
+2\,W^{\la,22}_{\mathrm{F},4} \Bigg]\,,
\nonu\\
\Phi^{(3),23}_{1}
&
=& (-6)(-2)\times \Bigg[
-2\,W^{\la,11}_{\mathrm{B},4}
-i\sqrt{2}\,W^{\la,12}_{\mathrm{B},4}
-4i\sqrt{2}\,\,W^{\la,21}_{\mathrm{B},4}
+2\,W^{\la,22}_{\mathrm{B},4}
-2\,W^{\la,11}_{\mathrm{F},4}
\nonu \\
& - & 2i\sqrt{2}\,W^{\la,12}_{\mathrm{F},4}
-  2i\sqrt{2}\,W^{\la,21}_{\mathrm{F},4}
+2\,W^{\la,22}_{\mathrm{F},4} \Bigg]\,,
\nonu  \\
\Phi^{(3),24}_{1}
&
=& (-6)(-2) \times \Bigg[
-2i\,W^{\la,11}_{\mathrm{B},4}
+4\sqrt{2}\,W^{\la,21}_{\mathrm{B},4}
+2i\,\,W^{\la,22}_{\mathrm{B},4}
+2i\,W^{\la,11}_{\mathrm{F},4}
-2\sqrt{2}\,W^{\la,21}_{\mathrm{F},4}
\nonu \\
& - & 2i\,W^{\la,22}_{\mathrm{F},4} \Bigg]\,,
\nonu\\
\Phi^{(3),34}_{1}
&
=& (-6)(-2) \times \Bigg[
-2i\,W^{\la,11}_{\mathrm{B},4}
+\sqrt{2}\,W^{\la,12}_{\mathrm{B},4}
+2i\,\,W^{\la,22}_{\mathrm{B},4}
+2i\,W^{\la,11}_{\mathrm{F},4}
-2\sqrt{2}\,W^{\la,12}_{\mathrm{F},4}
\nonu \\
& - & 2i\,W^{\la,22}_{\mathrm{F},4} \Bigg]\,,
\nonu \\
\tilde{\Phi}^{(3),1}_{\frac{3}{2}}
&
\equiv &
\Phi^{(3),1}_{\frac{3}{2}} -\frac{1}{7}\, (1-4\la)\,
\pa \,\Phi^{(3),1}_{\frac{1}{2}}
\nonu \\
&=& (-6)(-2) \times \Bigg[-\frac{1}{2}\,\Big(
Q^{\la,11}_{\frac{9}{2}}
+i\sqrt{2}\,Q^{\la,12}_{\frac{9}{2}}
+2i\sqrt{2}\,Q^{\la,21}_{\frac{9}{2}}
-2\,Q^{\la,22}_{\frac{9}{2}}
\nonu \\
& - & 2\,\bar{Q}^{\la,11}_{\frac{9}{2}}
-2i\sqrt{2}\,\bar{Q}^{\la,12}_{\frac{9}{2}}
-i\sqrt{2}\,\bar{Q}^{\la,21}_{\frac{9}{2}}
+\bar{Q}^{\la,22}_{\frac{9}{2}}
\Big)\Bigg]\,,
\nonu\\
\tilde{\Phi}^{(3),2}_{\frac{3}{2}}
&
\equiv &
\Phi^{(3),2}_{\frac{3}{2}} -\frac{1}{7}\, (1-4\la)\,
\pa \,\Phi^{(3),2}_{\frac{1}{2}}
\nonu \\
&
=& (-6)(-2) \times \Bigg[
\frac{i}{2}\,\Big(\,
Q^{\la,11}_{\frac{9}{2}}
+2i\sqrt{2}\,Q^{\la,21}_{\frac{9}{2}}
-2\,Q^{\la,22}_{\frac{9}{2}}
-2\,\bar{Q}^{\la,11}_{\frac{9}{2}}
-2i\sqrt{2}\,\bar{Q}^{\la,12}_{\frac{9}{2}}
+\bar{Q}^{\la,22}_{\frac{9}{2}}
\Big) \Bigg]\,,
\nonu\\
\tilde{\Phi}^{(3),3}_{\frac{3}{2}}
&
\equiv &
\Phi^{(3),3}_{\frac{3}{2}} -\frac{1}{7}\, (1-4\la)\,
\pa \,\Phi^{(3),3}_{\frac{1}{2}}
\nonu \\
&
=& (-6)(-2) \times \Bigg[
\frac{i}{2}\,\Big(
Q^{\la,11}_{\frac{9}{2}}
+i\sqrt{2}\,Q^{\la,12}_{\frac{9}{2}}
-2\,Q^{\la,22}_{\frac{9}{2}}
-2\,\bar{Q}^{\la,11}_{\frac{9}{2}}
-i\sqrt{2}\,\bar{Q}^{\la,21}_{\frac{9}{2}}
+\bar{Q}^{\la,22}_{\frac{9}{2}}
\Big)\Bigg]\,,
\nonu\\
\tilde{\Phi}^{(3),4}_{\frac{3}{2}}
&
\equiv &
\Phi^{(3),4}_{\frac{3}{2}} -\frac{1}{7}\, (1-4\la)\,
\pa \,\Phi^{(3),4}_{\frac{1}{2}}
\nonu \\
&
=& (-6)(-2) \times \Bigg[
\frac{1}{2}\,
\Big(
Q^{\la,11}_{\frac{9}{2}}
+2\,Q^{\la,22}_{\frac{9}{2}}
+2\,\bar{Q}^{\la,11}_{\frac{9}{2}}
+\bar{Q}^{\la,22}_{\frac{9}{2}}
\Big)\Bigg]\,,
\nonu \\
\tilde{\Phi}^{(3)}_{2}
&
\equiv &
\Phi^{(3)}_{2} -\frac{1}{7} (1-4\la)
\pa^2 \Phi^{(3)}_{0}
\nonu \\
& = &
(-6)(-2) \times \Bigg[-2 \Big(
W^{\la,11}_{\mathrm{B},5}
+W^{\la,22}_{\mathrm{B},5}
+W^{\la,11}_{\mathrm{F},5}
+W^{\la,22}_{\mathrm{F},5}
\Big) \Bigg].
\label{thirdmult}
\eea
All of these in (\ref{thirdmult})
are quasiprimary under the stress energy tensor (\ref{Lterm}).

\section{
The OPEs between 
the second ${\cal N}=4$ multiplet and itself }

We summarize
the super  OPE between 
the second ${\cal N}=4$ multiplet and itself
as follows:
\bea
&& {\bf \Phi}^{(2)}(Z_{1})\,{\bf \Phi}^{(2)}(Z_{2}) =
\frac{1}{z_{12}^{4}}\, \Bigg[ \frac{128}{3}\, N\, (1-4\la)\, (1+2\la-
  4\la^2) \Bigg]
\nonu \\
&& +\frac{\theta_{12}^{4-0}}{z_{12}^{6}}\,
\Bigg[\, 
-\frac{512}{3}\, N\, (-1+10\la-80\la^3+80\la^4)
  \,\Bigg]
\nonu \\
&& +\frac{\theta_{12}^{4-i}}{z_{12}^{5}}\,
\Bigg[\, 
  \frac{2048}{3}\, \la \, (1-\la)\, (1-2\la)\, (1+2\la)\, D^i
  {\bf J}
  \,\Bigg](Z_{2})
\nonu \\
&& +\frac{\theta_{12}^{4-0}}{z_{12}^{5}}\,
\Bigg[\, 
  \frac{4096}{3}\, \la \, (1-\la)\, (1-2\la)\, (1+2\la)\,
  \pa \, {\bf J}
  \,\Bigg](Z_{2})
\nonu \\
&& +
\frac{\theta_{12}^{4-ij}}{z_{12}^{4}}\,
\Bigg[\,
  -\frac{256}{3}\, (1-\la)\, (1+2\la)\, (1-4\la)\,
  \frac{1}{2!}\, \varepsilon^{i j k l}\, D^k \, D^l \, {\bf J}
  \nonu \\
  && - \frac{128}{3} \, (1-\la)\, (1+2\la)\,
  \varepsilon^{i j k l} \,
   \frac{1}{2!}\, \varepsilon^{ k l m n}\, D^m \, D^n \, {\bf J}
  \,\Bigg](Z_{2})
\nonu \\
&& +\frac{\theta_{12}^{4-i}}{z_{12}^{4}}\,\Bigg[\,
 \frac{4096}{3}\, \la \, (1-\la)\, (1-2\la)\, (1+2\la)\, \pa \, D^i
      {\bf J}
      \nonu \\
      && + 128\, (1-\la)\, (1+2\la) \, (1-4\la)\, \Big(
      -\frac{1}{3!}\, \varepsilon^{i j k l}\,
      D^j \, D^k \, D^l {\bf J} -(1-4\la)\, \pa \, D^i \, {\bf J}\Big)
  \,\Bigg](Z_{2})
 \nonu\\
&&
+\frac{\theta_{12}^{4-0}}{z_{12}^{4}}\,\Bigg[\,
2048\, \la \, (1-\la)\, (1-2\la)\, (1+2\la)\,
  \pa^2 \, {\bf J}
  \nonu \\
  && -\frac{512}{3} \, (1-\la)\, (1+2\la)\, (1-4\la)\,
  \Big( \frac{1}{2 \cdot 4!}\, \varepsilon^{i j k l}\,
  D^i\, D^j \, D^k \, D^l {\bf J} -
  \frac{1}{2} \, (1-4\la)\, \pa^2  \, {\bf J}
  \Big)
  \nonu \\
  && -32\, (-7-4\la+8\la^2)\, {\bf \Phi}^{(2)}
  \,\Bigg](Z_{2})
\nonu \\
&& +
\frac{\theta_{12}^{i}}{z_{12}^{3}}\,\Bigg[\,
  -\frac{128}{3}\, (1-\la)\, (1+2\la)\,
  \Big( -\frac{1}{3!}\, \varepsilon^{i j k l} \,
  D^j \, D^k \, D^l \, {\bf J}-(1-4\la)\, \pa \, D^i \,
  {\bf J}\Big)
  \,
\Bigg]
(Z_{2})
\nonu \\
&& +
\frac{\theta_{12}^{4-ij}}{z_{12}^{3}}\,\Bigg[\,
-\frac{256}{3}\, (1-\la)\, (1+2\la)\, (1-4\la)\,
  \frac{1}{2!}\, \varepsilon^{i j k l}\, \pa \, D^k \, D^l \, {\bf J}
  \nonu \\
  && - \frac{128}{3} \, (1-\la)\, (1+2\la)\,
  \varepsilon^{i j k l} \,
   \frac{1}{2!}\, \varepsilon^{ k l m n}\, \pa \, D^m \, D^n \, {\bf J}
  \,\Bigg](Z_{2})
\nonu \\
&& +\frac{\theta_{12}^{4-i}}{z_{12}^{3}}\,\Bigg[\,
 1024\, \la \, (1-\la)\, (1-2\la)\, (1+2\la)\, \pa^2 \, D^i
      {\bf J}
      \nonu \\
      && + \frac{1024}{9}\, (1-\la)\, (1+2\la) \, (1-4\la)\, \Big(
      -\frac{1}{3!}\, \varepsilon^{i j k l}\,
      \pa \,
      D^j \, D^k \, D^l {\bf J} -(1-4\la)\, \pa^2 \, D^i \, {\bf J}\Big)
      \nonu \\
      && -\frac{16}{3}\, (-11-2\la+4\la^2)\, D^i \, {\bf \Phi}^{(2)}
      \,\Bigg](Z_{2})
\nonu\\
&&
+\frac{\theta_{12}^{4-0}}{z_{12}^{3}}\,\Bigg[\,
\frac{4096}{3}\, \la \, (1-\la)\, (1-2\la)\, (1+2\la)\,
  \pa^3 \, {\bf J}
  \nonu \\
  && +\frac{1792}{9} \, (1-\la)\, (1+2\la)\, (1-4\la)\,
  \Big( \frac{1}{2 \cdot 4!}\, \varepsilon^{i j k l}\,
 \pa\,  D^i\, D^j \, D^k \, D^l {\bf J} -
  \frac{1}{2} \, (1-4\la)\, \pa^3  \, {\bf J}
  \Big)
  \nonu \\
  && -\frac{16}{3}\, (-43-16\la+32\la^2)\, \pa \, {\bf \Phi}^{(2)}
  \,\Bigg](Z_{2})
\nonu \\
&& +  \frac{1}{z_{12}^{2}}\,
\Bigg[\,
  -\frac{16}{3}\, (1-4\la)\, {\bf \Phi}^{(2)}
  \nonu \\
  && +  \frac{256}{9}\, (1-\la)\, (1+2\la)\, \Big(
  \frac{1}{2 \cdot 4!}\, \varepsilon^{i jk l}\, D^i \, D^j \, D^k \,
  D^l \, {\bf J}-\frac{1}{2}\, (1-4\la)\, \pa^2 \, {\bf J}
  \Big)
  \,\Bigg](Z_{2})
\nonu \\
&&
+
\frac{\theta_{12}^{i}}{z_{12}^{2}}\,\Bigg[\,
-\frac{256}{9}\, (1-\la)\, (1+2\la)\,
  \Big( -\frac{1}{3!}\, \varepsilon^{i j k l} \,
 \pa\,  D^j \, D^k \, D^l \, {\bf J}-(1-4\la)\, \pa^2 \, D^i \,
        {\bf J}\Big)
        \nonu \\
        && -\frac{8}{3}\, (1-4\la)\,
   D^i \, {\bf \Phi}^{(2)}
  \,\Bigg](Z_{2})
\nonu \\
&& +
\frac{\theta_{12}^{4-ij}}{z_{12}^{2}}\,\Bigg[\,
\frac{128}{3}\, (1-\la)\, (1+2\la)\, (1-4\la)\,
  \frac{1}{2!}\, \varepsilon^{i j k l}\, \pa^2 \, D^k \, D^l \, {\bf J}
  \nonu \\
  && - \frac{64}{3} \, (1-\la)\, (1+2\la)\,
  \varepsilon^{i j k l} \,
   \frac{1}{2!}\, \varepsilon^{ k l m n}\, \pa^2 \, D^m \, D^n \, {\bf J}
   -4 \, (1-4\la)\,
   \frac{1}{2!}\, \varepsilon^{ i j k l }\,  D^k \, D^l \, {\bf \Phi}^{(2)}
   \nonu \\
   && - 6\, 
   \varepsilon^{i j k l}\,
   \frac{1}{2!}\, \varepsilon^{ k l m n}\,
   D^m \, D^n \, {\bf \Phi}^{(2)}
   \,\Bigg](Z_{2})
\nonu\\
&&
+
\frac{\theta_{12}^{4-i}}{z_{12}^{2}}\,\Bigg[\,
 \frac{4096}{9}\, \la \, (1-\la)\, (1-2\la)\, (1+2\la)\, \pa^3 \, D^i
      {\bf J}
      \nonu \\
      && + \frac{160}{3}\, (1-\la)\, (1+2\la) \, (1-4\la)\, \Big(
      -\frac{1}{3!}\, \varepsilon^{i j k l}\,
      \pa^2 \, D^j \, D^k \, D^l {\bf J} -
      (1-4\la)\, \pa^3 \, D^i \, {\bf J}\Big)
      \nonu \\
      && -\frac{2}{3}\, (-71-8\la+16\la^2)\, \pa \, D^i \, {\bf \Phi}^{(2)}
      - \frac{14}{3} \, (1-4\la)\,
      \frac{1}{3!}\, \varepsilon^{i j k l}\,
       D^j \, D^k \, D^l {\bf \Phi}^{(2)}
      \,\Bigg](Z_{2})
\nonu\\
&&
+\frac{\theta_{12}^{4-0}}{z_{12}^{2}}\,\Bigg[\,
\frac{5120}{9}\, \la \, (1-\la)\, (1-2\la)\, (1+2\la)\,
  \pa^4 \, {\bf J}
  \nonu \\
  && -\frac{512}{5} \, (1-\la)\, (1+2\la)\, (1-4\la)\,
  \Big( \frac{1}{2 \cdot 4!}\, \varepsilon^{i j k l}\,
 \pa^2 \,  D^i\, D^j \, D^k \, D^l {\bf J} -
  \frac{1}{2} \, (1-4\la)\, \pa^4  \, {\bf J}
  \Big)
  \nonu \\
  && -\frac{64}{21}\, (-38-11\la+22\la^2)\, \pa^2 \, {\bf \Phi}^{(2)}
  -\frac{64}{21}\, (1-4\la)\,
   \frac{1}{ 4!}\, \varepsilon^{i j k l}\,
  \,  D^i\, D^j \, D^k \, D^l {\bf \Phi}^{(2)}
  \nonu \\
  && +  \frac{1}{3}\,
{\bf \Phi}^{(4)}
  \,\Bigg](Z_{2})
\nonu\\
&&
+\frac{1}{z_{12}}\,
\Bigg[\,
 -\frac{8}{3}\, (1-4\la)\, \pa \, {\bf \Phi}^{(2)}
  \nonu \\
  && +  \frac{128}{9}\, (1-\la)\, (1+2\la)\, \Big(
  \frac{1}{2 \cdot 4!}\, \varepsilon^{i jk l}\,
  \pa \, D^i \, D^j \, D^k \,
  D^l \, {\bf J}-\frac{1}{2}\, (1-4\la)\, \pa^3 \, {\bf J}
  \Big)
  \,\Bigg](Z_{2})
\nonu \\
&& +
\frac{\theta_{12}^{i}}{z_{12}}\,\Bigg[\,
-\frac{32}{3}\, (1-\la)\, (1+2\la)\,
  \Big( -\frac{1}{3!}\, \varepsilon^{i j k l} \,
 \pa^2 \,  D^j \, D^k \, D^l \, {\bf J}-(1-4\la)\, \pa^3 \, D^i \,
        {\bf J}\Big)
        \nonu \\
        && -2 \, (1-4\la) \,
        \pa \,  D^i \, {\bf \Phi}^{(2)}
        +2 \,  \frac{1}{3!}\, \varepsilon^{i j k l} \,
           D^j \, D^k \, D^l \, {\bf \Phi}^{(2)}
  \,\Bigg](Z_{2})
\nonu\\
&&
+
\frac{\theta_{12}^{4-ij}}{z_{12}}\,\Bigg[\,
-\frac{128}{9}\, (1-\la)\, (1+2\la)\, (1-4\la)\,
  \frac{1}{2!}\, \varepsilon^{i j k l}\, \pa^3 \, D^k \, D^l \, {\bf J}
  \nonu \\
  && - \frac{64}{9} \, (1-\la)\, (1+2\la)\,
  \varepsilon^{i j k l} \,
   \frac{1}{2!}\, \varepsilon^{ k l m n}\, \pa^3 \, D^m \, D^n \, {\bf J}
   +\frac{8}{3} \, (1-4\la)\,
   \frac{1}{2!}\, \varepsilon^{ i j k l }\,  \pa \,
   D^k \, D^l \, {\bf \Phi}^{(2)}
   \nonu \\
   && + 4\, 
   \varepsilon^{i j k l}\,
   \frac{1}{2!}\, \varepsilon^{ k l m n}\,
   \pa \, D^m \, D^n \, {\bf \Phi}^{(2)}
  \,\Bigg](Z_{2})
\nonu \\ 
&&
+\frac{\theta_{12}^{4-i}}{z_{12}}\,\Bigg[\,
\frac{1280}{9}\, \la \, (1-\la)\, (1-2\la)\, (1+2\la)\, \pa^4 \, D^i
      {\bf J}
      \nonu \\
      &&  \frac{256}{15}\, (1-\la)\, (1+2\la) \, (1-4\la)\, \Big(
      -\frac{1}{3!}\, \varepsilon^{i j k l}\,
      \pa^3 \, D^j \, D^k \, D^l {\bf J} -
      (1-4\la)\, \pa^4 \, D^i \, {\bf J}\Big)
      \nonu \\
      && -\frac{32}{21}\, (-13-\la+2\la^2)\, \pa^2 \, D^i \, {\bf \Phi}^{(2)}
      - \frac{64}{21} \, (1-4\la)\,
      \frac{1}{3!}\, \varepsilon^{i j k l}\,
      \pa \, D^j \, D^k \, D^l \, {\bf \Phi}^{(2)}
      \nonu \\
      && + \frac{1}{24}\,   D^i \, {\bf \Phi}^{(4)}
  \,\Bigg](Z_{2})
\nonu \\ 
&&
+\frac{\theta_{12}^{4-0}}{z_{12}}\,\Bigg[\,
\frac{512}{3}\, \la \, (1-\la)\, (1-2\la)\, (1+2\la)\,
  \pa^5 \, {\bf J}
  \nonu \\
  && -\frac{512}{15} \, (1-\la)\, (1+2\la)\, (1-4\la)\,
  \Big( \frac{1}{2 \cdot 4!}\, \varepsilon^{i j k l}\,
 \pa^3 \,  D^i\, D^j \, D^k \, D^l {\bf J} -
  \frac{1}{2} \, (1-4\la)\, \pa^5 \, {\bf J}
  \Big)
  \nonu \\
  && -\frac{67}{7}\, (-17-4\la+8\la^2)\, \pa^3 \, {\bf \Phi}^{(2)}
  -\frac{16}{7}\, (1-4\la)\,
   \frac{1}{ 4!}\, \varepsilon^{i j k l}\,
  \,  \pa \, D^i\, D^j \, D^k \, D^l {\bf \Phi}^{(2)}
  \nonu \\
  && + \frac{1}{4}\,\pa \,
{\bf \Phi}^{(4)}
  \,
\Bigg](Z_{2})
+\cdots.
\label{superopeappF}
\eea
There exist the ${\cal N}=4$ stress energy tensor,
the second ${\cal N}=4$ multiplet,
the fourth ${\cal N}=4$ multiplet,
as well as their descendants on the right hand
sides of (\ref{superopeappF}).

The fourth ${\cal N}=4$ multiplet can be summarized by
\bea
\Phi^{(4) }_{0} & = & \frac{768}{7} \, (-2+\la)\,
(W^{\la,11}_{\mathrm{F},4}+W^{\la,22}_{\mathrm{F},4})+
\frac{384}{7}\,(3+2\la)\,
(W^{\la,11}_{\mathrm{B},4}+W^{\la,22}_{\mathrm{B},4})\,,
\nonu \\
\Phi^{(4),1}_{\frac{1}{2}}
&
=&
(-8)(-6)(-2) \times \Bigg[ \frac{1}{2}\,\Big(
Q^{\la,11}_{\frac{9}{2}}
+i\sqrt{2}\,Q^{\la,12}_{\frac{9}{2}}
+2i \sqrt{2}\,Q^{\la,21}_{\frac{9}{2}}
-2\,Q^{\la,22}_{\frac{9}{2}}
\nonu \\
& + & 2\,\bar{Q}^{\la,11}_{\frac{9}{2}}
+2i \sqrt{2}\,\bar{Q}^{\la,12}_{\frac{9}{2}}
+i\sqrt{2}\,\bar{Q}^{\la,21}_{\frac{9}{2}}
-\bar{Q}^{\la,22}_{\frac{9}{2}}
\Big) \Bigg]\,,
\nonu\\
\Phi^{(4),2}_{\frac{1}{2}}
&
=&
(-8)(-6)(-2) \times \Bigg[ -\frac{i}{2}\,\Big(
Q^{\la,11}_{\frac{9}{2}}
+2i\sqrt{2} \,Q^{\la,21}_{\frac{9}{2}}
-2 \,Q^{\la,22}_{\frac{9}{2}}
+2\, \bar{Q}^{\la,11}_{\frac{9}{2}}
+2i\sqrt{2} \, \bar{Q}^{\la,12}_{\frac{9}{2}}
\nonu \\
& - & \bar{Q}^{\la,22}_{\frac{9}{2}}
\Big)\Bigg]\,,
\nonu\\
\Phi^{(4),3}_{\frac{1}{2}}
&
=& (-8)(-6)(-2) \times \Bigg[
-\frac{i}{2}\,\Big(
Q^{\la,11}_{\frac{9}{2}}
+i\sqrt{2} \,Q^{\la,12}_{\frac{9}{2}}
-2\,Q^{\la,22}_{\frac{9}{2}}
+2\, \bar{Q}^{\la,11}_{\frac{9}{2}}
+i \sqrt{2} \, \bar{Q}^{\la,21}_{\frac{9}{2}}
\nonu \\
& - & \bar{Q}^{\la,22}_{\frac{9}{2}}
\Big)\Bigg]\,,
\nonu\\
\Phi^{(4),4}_{\frac{1}{2}}
&
=& (-8)(-6)(-2) \times \Bigg[
-\frac{1}{2}\,Q^{\la,11}_{\frac{9}{2}}
-Q^{\la,22}_{\frac{9}{2}}
+\bar{Q}^{\la,11}_{\frac{9}{2}}
+\frac{1}{2}\,\bar{Q}^{\la,22}_{\frac{9}{2}} \Bigg]
\,,
\nonu \\
\Phi^{(4),12}_{1}
&
=& (-8)(-6)(-2) \times \Bigg[
2i\,W^{\la,11}_{\mathrm{B},5}
-\sqrt{2}\,W^{\la,12}_{\mathrm{B},5}
-2i\,\,W^{\la,22}_{\mathrm{B},5}
+2i\,W^{\la,11}_{\mathrm{F},5}
-2\sqrt{2}\,W^{\la,12}_{\mathrm{F},5}
\nonu \\
& - & 2i\,W^{\la,22}_{\mathrm{F},5} \Bigg]\,,
\nonu\\
\Phi^{(4),13}_{1}
&
=&
(-8)(-6)(-2) \times \Bigg[
-2i\,W^{\la,11}_{\mathrm{B},5}
+4\sqrt{2}\,W^{\la,21}_{\mathrm{B},5}
+2i\,\,W^{\la,22}_{\mathrm{B},5}
-2i\,W^{\la,11}_{\mathrm{F},5}
+2\sqrt{2}\,W^{\la,21}_{\mathrm{F},5}
\nonu \\
& + & 2i\,W^{\la,22}_{\mathrm{F},5} \Bigg]\,,
\nonu\\
\Phi^{(4),14}_{1}
&
=&
(-8)(-6)(-2)\times \Bigg[
2\,W^{\la,11}_{\mathrm{B},5}
+i\sqrt{2}\,W^{\la,12}_{\mathrm{B},5}
+4i\sqrt{2}\,\,W^{\la,21}_{\mathrm{B},5}
-2\,W^{\la,22}_{\mathrm{B},5}
-2\,W^{\la,11}_{\mathrm{F},5}
\nonu \\
& - & 2i\sqrt{2}\,W^{\la,12}_{\mathrm{F},5}
-  2i\sqrt{2}\,W^{\la,21}_{\mathrm{F},5}
+2\,W^{\la,22}_{\mathrm{F},5} \Bigg]\,,
\nonu\\
\Phi^{(4),23}_{1}
&
=& (-8)(-6)(-2)\times \Bigg[
-2\,W^{\la,11}_{\mathrm{B},5}
-i\sqrt{2}\,W^{\la,12}_{\mathrm{B},5}
-4i\sqrt{2}\,\,W^{\la,21}_{\mathrm{B},5}
+2\,W^{\la,22}_{\mathrm{B},5}
-2\,W^{\la,11}_{\mathrm{F},5}
\nonu \\
& - & 2i\sqrt{2}\,W^{\la,12}_{\mathrm{F},5}
-  2i\sqrt{2}\,W^{\la,21}_{\mathrm{F},5}
+2\,W^{\la,22}_{\mathrm{F},5} \Bigg]\,,
\nonu  \\
\Phi^{(4),24}_{1}
&
=& (-8)(-6)(-2) \times \Bigg[
-2i\,W^{\la,11}_{\mathrm{B},5}
+4\sqrt{2}\,W^{\la,21}_{\mathrm{B},5}
+2i\,\,W^{\la,22}_{\mathrm{B},5}
+2i\,W^{\la,11}_{\mathrm{F},5}
-2\sqrt{2}\,W^{\la,21}_{\mathrm{F},5}
\nonu \\
& - & 2i\,W^{\la,22}_{\mathrm{F},5} \Bigg]\,,
\nonu\\
\Phi^{(4),34}_{1}
&
=& (-8)(-6)(-2) \times \Bigg[
-2i\,W^{\la,11}_{\mathrm{B},5}
+\sqrt{2}\,W^{\la,12}_{\mathrm{B},5}
+2i\,\,W^{\la,22}_{\mathrm{B},5}
+2i\,W^{\la,11}_{\mathrm{F},5}
-2\sqrt{2}\,W^{\la,12}_{\mathrm{F},5}
\nonu \\
& - & 2i\,W^{\la,22}_{\mathrm{F},5} \Bigg]\,,
\nonu \\
\tilde{\Phi}^{(4),1}_{\frac{3}{2}}
&
\equiv &
\Phi^{(4),1}_{\frac{3}{2}} -\frac{1}{9}\, (1-4\la)\,
\pa \,\Phi^{(4),1}_{\frac{1}{2}}
\nonu \\
&=& (-8)(-6)(-2) \times \Bigg[-\frac{1}{2}\,\Big(
Q^{\la,11}_{\frac{11}{2}}
+i\sqrt{2}\,Q^{\la,12}_{\frac{11}{2}}
+2i\sqrt{2}\,Q^{\la,21}_{\frac{11}{2}}
-2\,Q^{\la,22}_{\frac{11}{2}}
\nonu \\
& - & 2\,\bar{Q}^{\la,11}_{\frac{11}{2}}
-2i\sqrt{2}\,\bar{Q}^{\la,12}_{\frac{11}{2}}
-i\sqrt{2}\,\bar{Q}^{\la,21}_{\frac{11}{2}}
+\bar{Q}^{\la,22}_{\frac{11}{2}}
\Big)\Bigg]\,,
\nonu\\
\tilde{\Phi}^{(4),2}_{\frac{3}{2}}
&
\equiv &
\Phi^{(4),2}_{\frac{3}{2}} -\frac{1}{9}\, (1-4\la)\,
\pa \,\Phi^{(4),2}_{\frac{1}{2}}
\nonu \\
&
=& (-8)(-6)(-2) \times \Bigg[
\frac{i}{2}\,\Big(\,
Q^{\la,11}_{\frac{11}{2}}
+2i\sqrt{2}\,Q^{\la,21}_{\frac{11}{2}}
-2\,Q^{\la,22}_{\frac{11}{2}}
-2\,\bar{Q}^{\la,11}_{\frac{11}{2}}
-2i\sqrt{2}\,\bar{Q}^{\la,12}_{\frac{11}{2}}
\nonu \\
& + & \bar{Q}^{\la,22}_{\frac{11}{2}}
\Big) \Bigg]\,,
\nonu\\
\tilde{\Phi}^{(4),3}_{\frac{3}{2}}
&
\equiv &
\Phi^{(4),3}_{\frac{3}{2}} -\frac{1}{9}\, (1-4\la)\,
\pa \,\Phi^{(4),3}_{\frac{1}{2}}
\nonu \\
&
=& (-8)(-6)(-2) \times \Bigg[
\frac{i}{2}\,\Big(
Q^{\la,11}_{\frac{11}{2}}
+i\sqrt{2}\,Q^{\la,12}_{\frac{11}{2}}
-2\,Q^{\la,22}_{\frac{11}{2}}
-2\,\bar{Q}^{\la,11}_{\frac{11}{2}}
-i\sqrt{2}\,\bar{Q}^{\la,21}_{\frac{11}{2}}
+\bar{Q}^{\la,22}_{\frac{11}{2}}
\Big)\Bigg]\,,
\nonu\\
\tilde{\Phi}^{(4),4}_{\frac{3}{2}}
&
\equiv &
\Phi^{(4),4}_{\frac{3}{2}} -\frac{1}{9}\, (1-4\la)\,
\pa \,\Phi^{(4),4}_{\frac{1}{2}}
\nonu \\
&
=& (-8)(-6)(-2) \times \Bigg[
\frac{1}{2}\,
\Big(
Q^{\la,11}_{\frac{11}{2}}
+2\,Q^{\la,22}_{\frac{11}{2}}
+2\,\bar{Q}^{\la,11}_{\frac{11}{2}}
+\bar{Q}^{\la,22}_{\frac{11}{2}}
\Big)\Bigg]\,,
\nonu \\
\tilde{\Phi}^{(4)}_{2}
&
\equiv &
\Phi^{(4)}_{2} -\frac{1}{9} (1-4\la)
\pa^2 \Phi^{(4)}_{0}
\nonu \\
& = &
(-8)(-6)(-2) \times \Bigg[-2 \Big(
W^{\la,11}_{\mathrm{B},6}
+W^{\la,22}_{\mathrm{B},6}
+W^{\la,11}_{\mathrm{F},6}
+W^{\la,22}_{\mathrm{F},6}
\Big) \Bigg].
\label{lasteq}
\eea
All of these are quasiprimary under the stress energy tensor
(\ref{Lterm}).
The $\la$-dependence in the weight-$4$ operator in
(\ref{lasteq}) can be obtained
from the factor $(4-2\la)$ appearing in $W_{B,4}^{\la, \bar{a} a}$
and the factor $(3+2\la)$ appearing in $W_{F,4}^{\la, \bar{a} a}$
respectively.

This implies that for the $h$-th ${\cal N}=4$ multiplet,
the $\la$-dependence in the weight-$h$ operator
can be obtained from the factor $(h-2\la)$
appearing in $W_{F,h}^{\la, \bar{a} a}$
and the factor $(h-1+2\la)$
appearing in $W_{B,h}^{\la, \bar{a} a}$ respsectively.
For the four weight-$(h+\frac{1}{2})$ operators,
we simply take $(-2h)\, \cdots \, (-8)\, (-6) \, (-2)$
multiplied by the quantities inside the brackets after we
replace
$\frac{9}{2}$ with $(h+\frac{1}{2})$.
For the six weight-$(h+1)$ operators,
we simply take $(-2h)\, \cdots \, (-8)\, (-6) \, (-2)$
multiplied by the quantities inside the brackets after we
replace
$5$ with $(h+1)$. Similarly,
for the four weight-$(h+\frac{3}{2})$ operators,
we simply take $(-2h)\, \cdots \, (-8)\, (-6) \, (-2)$
multiplied by the quantities inside the brackets after we
replace
$\frac{11}{2}$ with $(h+\frac{3}{2})$.
For the weight-$(h+2)$ operator,
we simply take $(-2h)\, \cdots \, (-8)\, (-6) \, (-2)$
multiplied by the quantity inside the bracket after we
replace
$6$ with $(h+2)$. 
The  $W_{F,h}^{\la, \bar{a} b}$ and the
$W_{B,h}^{\la, \bar{a} b}$ can be written in terms of
$\Phi_0^{(h)}$, $\Phi^{(h-1), i j}_1$ and $\Phi_2^{(h-2)}$.
Similarly, the  $Q_{h+\frac{1}{2}}^{\la, \bar{a} b}$ and the
$\bar{Q}_{h+\frac{1}{2}}^{\la, a \bar{b}}$ can be written in terms of
$\Phi_{\frac{1}{2}}^{(h),i}$ and $\Phi_{\frac{3}{2}}^{(h-1), i }$.

\section{
The OPEs between 
the ${\cal N}=4$ stress energy tensor  and the third ${\cal N}=4$
multiplet}

The OPEs between the ${\cal N}=4$ stress energy tensor
and the third ${\cal N}=4$ multiplet can be described by
\bea
&& {\bf J}(Z_{1})\,{\bf \Phi}^{(3)}(Z_{2})  = 
-\frac{\theta_{12}^{4-0}}{z_{12}^5}\,\Bigg[
-\frac{1536}{5} \, N\, \la \, (1-2\la)\, (1-4\la)
\Bigg]
\nonu \\
&& +  \frac{1}{z_{12}^3}\, \Bigg[\,
-\frac{192}{15}\,N\, (1-3\la + 6\la^2)  \,\Bigg]
\nonu \\
&& +  \frac{\theta_{12}^{4-i}}{z_{12}^4}\,
\Bigg[\,  
\frac{1536}{5} \, \la \, (1-2\la)\, (1-4\la)\, D^i \, {\bf J}
  \,\Bigg]
(Z_2)
\nonu \\
&& +  \frac{\theta_{12}^{i}}{z_{12}^3}\,
\Bigg[\, 
-\frac{192}{5}\, (1-3\la+6\la^2)\, D^i \, {\bf J}
  \,\Bigg](Z_2)
\nonu \\
&& + 
\frac{\theta_{12}^{4-ij}}{z_{12}^3}\,
\Bigg[\,
  \frac{192}{5}\, (1+2\la-4\la^2)\, \frac{1}{2!} \,
  \varepsilon^{ i j k l }\, D^k \, D^l\, {\bf J}
  \nonu  \\
  && -  \frac{96}{5}\, (1-4\la)\, \varepsilon^{i j k l}\,
   \frac{1}{2!}\, \varepsilon^{  k l m n }\, D^m \, D^n\, {\bf J}
  \,\Bigg](Z_2)  
\nonu \\
&& +  \frac{\theta_{12}^{4-0}}{z_{12}^4}\,\Bigg[
\frac{1152}{5}\, (1+2\la-4\la^2 )\, {\bf \Phi}^{(1)}
-\frac{1536}{5}\, \la \, (1-2\la)\, (1-4\la)\, \pa \, {\bf J}
\Bigg](Z_2)
\nonu \\
&& +
\frac{1}{z_{12}^2}\, \Bigg[\,
  \frac{96}{5}\, (1-4\la)\, {\bf \Phi}^{(1)}
  - \frac{192}{5}\, (1-3\la+6\la^2)\, \pa \, {\bf J}
  \,\Bigg](Z_2)
\nonu \\
&& + 
\frac{\theta_{12}^{4-i}}{z_{12}^3}\,
\Bigg[\, 
  -\frac{1536}{5} \, \la \, (1-2\la)\, (1-4\la)\, \pa \,
  D^i \, {\bf J} +
  \frac{576}{5} \, (1+2\la-4\la^2)\, D^i\, {\bf \Phi}^{(1)}
  \nonu \\
  & & + 192 \, \la \, (1-2\la)\, \Big(
 -\frac{1}{3!}\, \varepsilon^{i j k l}\, D^j \, D^k \, D^l
{\bf J} -(1-4\la)\, \pa \, D^i \, {\bf J}
  \Big)
  \Bigg](Z_2)
  \nonu \\
&& +  \frac{\theta_{12}^{i}}{z_{12}^2}\,
\Bigg[\, 
\frac{384}{5}\, (1-3\la+6\la^2)\, \pa \, D^i \, {\bf J}
-\frac{96}{5} \, (1-4\la)\, D^i \, {\bf \Phi}^{(1)}
\nonu \\
&& - \frac{96}{5} \, (1-4\la)\,
\Big( -\frac{1}{3!}\, \varepsilon^{i j k l}\, D^j \, D^k \, D^l
{\bf J} -(1-4\la)\, \pa \, D^i \, {\bf J} \Big)
\,\Bigg](Z_2)
\nonu\\
& & +
\frac{1}{z_{12}}\, \Bigg[\,
  -\frac{96}{5}\, (1-4\la)\, \pa \, {\bf \Phi}^{(1)}
  + \frac{192}{5}\, (1-3\la+6\la^2)\, \pa^2 \, {\bf J}
  \nonu \\
  & & +  \frac{64}{5}\, (1-4\la)\, \Big( \frac{1}{2\cdot 4!}
  \, \varepsilon^{i j k l}\, D^i \, D^j \, D^k \, D^l \, {\bf J}
  -\frac{1}{2}\, (1-4\la)\, \pa^2 \, {\bf J}\Big)
   + 24 \, {\bf \Phi}^{(2)}
  \,\Bigg](Z_2)
\nonu \\
&& +  \frac{\theta_{12}^{4-ij}}{z_{12}^2}\, \Bigg[\, 
-\frac{96}{5}\, (1+2\la-4\la^2)\, \frac{1}{2!} \,
  \varepsilon^{ i j k l }\, \pa \, D^k \, D^l\, {\bf J}
  \nonu \\
  && + \frac{48}{5}\, (1-4\la)\, \varepsilon^{i j k l}\,
\frac{1}{2!}\, \varepsilon^{  k l m n }\, \pa \, D^m \, D^n\, {\bf J}
   -\frac{48}{5}\, (1-4\la)\,
    \frac{1}{2!} \,
  \varepsilon^{ i j k l } \, D^k \, D^l\, {\bf \Phi}^{(1)}
  \nonu \\
  && -  24 \, \varepsilon^{i j k l}\,
 \frac{1}{2!} \,
  \varepsilon^{  k l m n } \, D^m \, D^n\, {\bf \Phi}^{(1)}
  \,\Bigg](Z_2)
\nonu\\
&&
+\frac{\theta_{12}^{i}}{z_{12}}\, \Bigg[\, 
-\frac{96}{5}\, (1-3\la+6\la^2)\, \pa^2 \, D^i \, {\bf J}
+\frac{12}{5} \, (1-4\la)\, \pa \, D^i \, {\bf \Phi}^{(1)}
\nonu \\
&& + \frac{32}{5} \, (1-4\la)\,
\Big( -\frac{1}{3!}\, \varepsilon^{i j k l}\, \pa \, D^j \, D^k \, D^l
 {\bf J} -(1-4\la)\, \pa^2 \, D^i \, {\bf J} \Big)
\nonu \\
&& + 
12 \, \frac{1}{3!}\, \varepsilon^{i j k l} \, D^j \, D^k \, D^l\,
{\bf \Phi}^{(1)} - 6 \, D^i \, {\bf \Phi}^{(2)}
\,\Bigg](Z_2)
\nonu \\
&&
+ \frac{\theta_{12}^{4-0}}{z_{12}^2}\,
\Bigg[\,
6 \, {\bf \Phi}^{(3)}
\,\Bigg](Z_2)
+  \frac{\theta_{12}^{4-i}}{z_{12}}\,\Bigg[\, 
D^i \, {\bf \Phi}^{(3)}
  \,\Bigg](Z_2)
+\frac{\theta_{12}^{4-0}}{z_{12}}\,
\Bigg[\, 2 \, \pa \, {\bf \Phi}^{(3)}
\,\Bigg](Z_2) +  \cdots.
\label{JPhi3}
\eea
Except the last three terms
for the ${\cal N}=4$ primary condition \cite{AK1509},
the additional terms consisting of the ${\cal N}=4$
stress energy tensor, the first and the second
${\cal N}=4$ multiplets (and their descendants)
appear in (\ref{JPhi3}).

\section{
The OPEs between 
the ${\cal N}=4$ stress energy tensor  and the fourth ${\cal N}=4$
multiplet}

The OPEs between the operators in (\ref{BigJ})
and the operators in (\ref{lasteq}) can be
summarized by
\bea
&& {\bf J}(Z_{1})\,{\bf \Phi}^{(4)}(Z_{2}) =
\frac{1}{z_{12}^{4}}\, \Bigg[
  - \frac{3072}{35}\, N \, (1-4\la)\,
  (3-2\la+4\la^2)
    \Bigg]
\nonu \\
&& +\frac{\theta_{12}^{4-0}}{z_{12}^{6}}\,
\Bigg[\, 
-\frac{147456}{7} \, N\, \la \, (1-2\la)\, (1-\la +2\la^2)
  \,\Bigg]
\nonu \\
&& +\frac{\theta_{12}^{4-i}}{z_{12}^{5}}\,
\Bigg[\, 
  \frac{147456}{7} \, \la \, (1-2\la)\, (1-\la +2\la^2)\,
D^i \, {\bf J}  
  \,\Bigg](Z_{2})
\nonu \\
&& +\frac{\theta_{12}^{i}}{z_{12}^{4}}\,\Bigg[\,
  -\frac{12288}{35}\, (1-4\la)\, (3-2\la+4\la^2)\,
  D^i \, {\bf J}
  \,\Bigg](Z_{2})
\nonu \\
&& +\frac{\theta_{12}^{4-0}}{z_{12}^{5}}\,
\Bigg[\, 
  \frac{49152}{7}\, \la \, (1-2\la)\, (1-4\la)\, {\bf \Phi}^{(1)}
  -\frac{294912}{7}\, \la \, (1-2\la)\, (1-\la+2\la^2)\,
  \pa \, {\bf J}
  \,\Bigg](Z_{2})
\nonu \\
&& +
\frac{\theta_{12}^{4-ij}}{z_{12}^{4}}\,
\Bigg[\,
  \frac{18432}{35}\, (1-\la)\, (1+2\la)\, (1-4\la)\,
  \frac{1}{2!}\, \varepsilon^{i j k l} \, D^k \, D^l \, {\bf J}
  \nonu \\
  && -\frac{9216}{7}\, (1-\la+2\la^2)\,
  \varepsilon^{i j k l} \,
 \frac{1}{2!}\, \varepsilon^{ k l m n} \, D^m \, D^n \, {\bf J}
  \,\Bigg](Z_{2})
\nonu \\
&&
+\frac{1}{z_{12}^{3}}\,\Bigg[\,
  \frac{6144}{7}\, (1-\la+2\la^2)\,
  {\bf \Phi}^{(1)}
  - \frac{12288}{35}\, (1-4\la)\,
  (3-2\la +4\la^2)\, \pa \, {\bf J}
  \,\Bigg](Z_{2})
\nonu \\
&& +\frac{\theta_{12}^{4-i}}{z_{12}^{4}}\,\Bigg[\,
-\frac{294912}{7} \, \la \, (1-2\la)\, (1-\la +2\la^2)\,\pa\,
D^i \, {\bf J}
+\frac{24576}{7}\, \la \, (1-2\la)\, (1-4\la)\,
D^i \, {\bf \Phi}^{(1)}
\nonu \\
&& -\frac{36864}{35}\, (1-4\la)\, (2-3\la +6\la^2)\,
\Big(
-\frac{1}{3!}\, \varepsilon^{i j k l}\,
       D^j \, D^k \, D^l {\bf J} -
      (1-4\la) \, \pa\, D^i \, {\bf J}
\Big)
  \,\Bigg](Z_{2})
 \nonu\\
&& +\frac{\theta_{12}^{i}}{z_{12}^{3}}\,\Bigg[\,
 \frac{36864}{35}\, (1-4\la)\, (3-2\la+4\la^2)\,
 \pa \, D^i \, {\bf J}
 - \frac{9216}{7}\, (1-\la +2\la^2)\, D^i \, {\bf \Phi}^{(1)}
 \nonu \\
 && - \frac{9216}{7}\, (1-\la +2\la^2)\, \Big(
-\frac{1}{3!}\, \varepsilon^{i j k l}\,
       D^j \, D^k \, D^l {\bf J} -
      (1-4\la) \, D^i \, {\bf J}
 \Big)
  \,\Bigg](Z_{2})
\nonu\\
 &&
+\frac{\theta_{12}^{4-0}}{z_{12}^{4}}\,\Bigg[\,
- \frac{24576}{7}\, \la \, (1-2\la)\, (1-4\la)\, \pa \, {\bf \Phi}^{(1)}
  +\frac{147456}{7}\, \la \, (1-2\la)\, (1-\la+2\la^2)\,
  \pa^2 \, {\bf J}\nonu \\
  && -\frac{9216}{7}\, (-4 -3\la +6\la^2)\, {\bf \Phi}^{(2)}
  \nonu \\
  && -\frac{24576}{35} \, (1-4\la)\, (2-3\la+6\la^2)\,
  \Big(
\frac{1}{2 \cdot 4!}\, \varepsilon^{i j k l}\,
   D^i\, D^j \, D^k \, D^l {\bf J} -
  \frac{1}{2} \, (1-4\la)\, \pa^2 \, {\bf J}
  \Big)
  \,\Bigg](Z_{2}
\nonu \\
&& +
\frac{\theta_{12}^{4-ij}}{z_{12}^{3}}\,\Bigg[\,
-\frac{18432}{35}\, (1-\la)\, (1+2\la)\, (1-4\la)\,
  \frac{1}{2!}\, \varepsilon^{i j k l} \, \pa \, D^k \, D^l \, {\bf J}
  \nonu \\
  && + \frac{9216}{7}\, (1-\la+2\la^2)\,
  \varepsilon^{i j k l} \,
 \frac{1}{2!}\, \varepsilon^{ k l m n} \, \pa \, D^m \, D^n \, {\bf J}
 \nonu \\
 && -\frac{1536}{7}\, (-3-4\la+8\la^2)
 \frac{1}{2!}\, \varepsilon^{i j k l } \,  D^k \, D^l \,
      {\bf \Phi}^{(1)}
      -\frac{2304}{7}\, (1-4\la)\,
      \varepsilon^{i j k l }\,
 \frac{1}{2!}\, \varepsilon^{k l m n } \,  D^m \, D^n \,
      {\bf \Phi}^{(1)}     
 \,\Bigg](Z_{2})
\nonu \\
&& +  \frac{1}{z_{12}^{2}}\,
\Bigg[\,
- \frac{9216}{7}\, (1-\la+2\la^2)\,
  \pa \, {\bf \Phi}^{(1)}
  + \frac{18432}{35}\, (1-4\la)\,
  (3-2\la +4\la^2)\, \pa^2 \, {\bf J}
  \nonu \\
  && + \frac{6144}{7}\, (1-\la +2\la^2)
  \, \Big(
 \frac{1}{2 \cdot 4!}\, \varepsilon^{i j k l}\,
   D^i\, D^j \, D^k \, D^l {\bf J} -
  \frac{1}{2} \, (1-4\la)\, \pa^2 \, {\bf J}
  \Big) \nonu \\
  && + \frac{1152}{7}\, (1-4\la)\, {\bf \Phi}^{(2)}
  \,\Bigg](Z_{2})
\nonu \\
&& +\frac{\theta_{12}^{4-i}}{z_{12}^{3}}\,\Bigg[\,
\frac{73728}{7} \, \la \, (1-2\la)\, (1-\la +2\la^2)\,\pa^2\,
D^i \, {\bf J}
-\frac{4608}{7}\, \la \, (1-2\la)\, (1-4\la)\,
\pa \, D^i \, {\bf \Phi}^{(1)}
\nonu \\
&& +\frac{12288}{35}\, (1-4\la)\, (2-3\la +6\la^2)\,
\Big(
-\frac{1}{3!}\, \varepsilon^{i j k l}\,
       \pa \, D^j \, D^k \, D^l {\bf J} -
      (1-4\la) \, \pa^2 \, D^i \, {\bf J}
\Big)
\nonu \\
&& - 1536\, \la \, (1-2\la)\,
\frac{1}{3!}\, \varepsilon^{i j k l}\,
       \pa \, D^j \, D^k \, D^l \, {\bf \Phi}^{(1)}
       - \frac{2304}{7}\, (-4-3\la+6\la^2)\,
D^i \, {\bf \Phi}^{(2)}       
       \,\Bigg](Z_{2})
\nonu\\
&&
+
\frac{\theta_{12}^{i}}{z_{12}^{2}}\,\Bigg[\,
 -\frac{18432}{35}\, (1-4\la)\, (3-2\la+4\la^2)\,
 \pa^2 \, D^i \, {\bf J}
 + \frac{384}{7}\, (15-8\la +16\la^2)\, \pa \, D^i \, {\bf \Phi}^{(1)}
 \nonu \\
 && + \frac{6144}{7}\, (1-\la +2\la^2)\, \Big(
-\frac{1}{3!}\, \varepsilon^{i j k l}\,
      \pa \, D^j \, D^k \, D^l {\bf J} -
      (1-4\la)\, \pa^2 \, D^i \, {\bf J}
      \Big)\nonu \\
      && +\frac{1152}{7}\, (1-4\la)\,
   \frac{1}{3!}\, \varepsilon^{i j k l}\,
    D^j \, D^k \, D^l\, {\bf \Phi}^{(1)}
   -\frac{576}{7}\, (1-4\la)\, D^i\, {\bf \Phi}^{(2)}
 \,\Bigg](Z_{2})
\nonu \\
&& +
\frac{\theta_{12}^{4-ij}}{z_{12}^{2}}\,\Bigg[\,
\frac{3072}{35}\, (1-\la)\, (1+2\la)\, (1-4\la)\,
  \frac{1}{2!}\, \varepsilon^{i j k l} \, \pa^2 \, D^k \, D^l \, {\bf J}
  \nonu \\
  && - \frac{1536}{7}\, (1-\la+2\la^2)\,
  \varepsilon^{i j k l} \,
 \frac{1}{2!}\, \varepsilon^{ k l m n} \, \pa^2 \, D^m \, D^n \, {\bf J}
 \nonu \\
 && +\frac{384}{7}\, (-3-4\la+8\la^2)
 \frac{1}{2!}\, \varepsilon^{i j k l } \,  \pa \, D^k \, D^l \,
      {\bf \Phi}^{(1)}
      +\frac{576}{7}\, (1-4\la)\,
      \varepsilon^{i j k l }\,
 \frac{1}{2!}\, \varepsilon^{k l m n } \,  \pa \, D^m \, D^n \,
      {\bf \Phi}^{(1)}     
      \nonu \\
      &&
      - \frac{288}{7} \,(1-4\la)\,
      \frac{1}{2!}\, \varepsilon^{i j k l }  \, D^k \, D^l \,
           {\bf \Phi}^{(2)}
           - 144\,
           \varepsilon^{i j k l }\,
  \frac{1}{2!}\, \varepsilon^{k l m n }  \, D^m \, D^n \,
           {\bf \Phi}^{(2)}         
           \,\Bigg](Z_{2})
\nonu\\
&&
+\frac{1}{z_{12}}\,
\Bigg[\,
 \frac{384}{35}\, (39-32\la+64\la^2)\,
  \pa^2 \, {\bf \Phi}^{(1)}
  - \frac{6144}{35}\, (1-4\la)\,
  (3-2\la +4\la^2)\, \pa^3 \, {\bf J}
  \nonu \\
  && - \frac{3072}{7}\, (1-\la +2\la^2)
  \, \Big(
 \frac{1}{2 \cdot 4!}\, \varepsilon^{i j k l}\,
   \pa \, D^i\, D^j \, D^k \, D^l \, {\bf J} -
  \frac{1}{2} \, (1-4\la)\, \pa^3 \, {\bf J}
  \Big)\nonu \\
  && - \frac{576}{7}\, (1-4\la)\, \pa\,
      {\bf \Phi}^{(2)}
       +48 \, {\bf \Phi}^{(3)}
      +\frac{1152}{35}\, (1-4\la)\,
  \frac{1}{ 4!}\, \varepsilon^{i j k l}\,
    D^i\, D^j \, D^k \, D^l    {\bf \Phi}^{(1)}
  \,\Bigg](Z_{2})
\nonu \\
&& +
\frac{\theta_{12}^{i}}{z_{12}}\,\Bigg[\,
\frac{2048}{35}\, (1-4\la)\, (3-2\la+4\la^2)\,
 \pa^3 \, D^i \, {\bf J}
 - \frac{384}{35}\, (9-2\la +4\la^2)\, \pa^2 \, D^i \, {\bf \Phi}^{(1)}
 \nonu \\
 && - \frac{768}{7}\, (1-\la +2\la^2)\, \Big(
-\frac{1}{3!}\, \varepsilon^{i j k l}\,
      \pa^2 \, D^j \, D^k \, D^l {\bf J} -
      (1-4\la)\, \pa^3 \, D^i \, {\bf J}
      \Big)\nonu \\
      && -\frac{1152}{35}\, (1-4\la)\,
   \frac{1}{3!}\, \varepsilon^{i j k l}\,
   \pa \, D^j \, D^k \, D^l\, {\bf \Phi}^{(1)}
   +\frac{48}{7}\, (1-4\la)\, \pa \, D^i\, {\bf \Phi}^{(2)}
   \nonu \\
   &&-  8 \,D^i \, {\bf \Phi}^{(3)} +48\,
    \frac{1}{3!}\, \varepsilon^{i j k l}\,
    D^j \, D^k \, D^l\, {\bf \Phi}^{(2)}
   \,\Bigg](Z_{2})
\nonu\\
&&
+\frac{\theta_{12}^{4-0}}{z_{12}^{2}}\,\Bigg[\,
8 \, {\bf \Phi}^{(4)}
  \,\Bigg](Z_{2})
+\frac{\theta_{12}^{4-i}}{z_{12}}\,\Bigg[\,
D^i \, {\bf \Phi}^{(4)}
  \,\Bigg](Z_{2})
+\frac{\theta_{12}^{4-0}}{z_{12}}\,\Bigg[\,
2 \, \pa \, {\bf \Phi}^{(4)}
  \,
\Bigg](Z_{2})
+\cdots.
\label{JPhi4}
\eea
Except the last three terms
for the ${\cal N}=4$ primary condition \cite{AK1509},
the additional terms consisting of the ${\cal N}=4$
stress energy tensor, the first, the second and the third
${\cal N}=4$ multiplets (and their descendants)
appear in (\ref{JPhi4}).
We expect that the OPEs between the ${\cal N}=4$ stress energy tensor
and the $h$-th ${\cal N}=4$ multiplet contain
the first, the second, $\cdots$ the $h$-th ${\cal N}=4$
multiplets.

\section{
The OPEs between 
the first ${\cal N}=4$ multiplet and the third ${\cal N}=4$
multiplet}

The OPEs between the operators in (\ref{Phiexp})
and the operators in (\ref{thirdmult}) can be
summarized by
\bea
&& {\bf \Phi}^{(1)}(Z_{1})\,{\bf \Phi}^{(3)}(Z_{2}) =
\frac{1}{z_{12}^{4}}\, \Bigg[
  -\frac{768}{5}\, N\, \la \, (1-2\la)\, (1-4\la) \Bigg]
\nonu \\
&& +\frac{\theta_{12}^{4-0}}{z_{12}^{6}}\,
\Bigg[\, 
3072\, N\, \la \, (-1+4\la-8\la^2 +8 \la^3)
  \,\Bigg]
\nonu \\
&& +\frac{\theta_{12}^{4-i}}{z_{12}^{5}}\,
\Bigg[\, 
  \frac{6144}{5}\, \la\, (1-2\la)\, (2-\la+2\la^2)\,
  D^i \, {\bf J}
  \,\Bigg](Z_{2})
\nonu \\
&& +
\frac{\theta_{12}^{i}}{z_{12}^{4}}\,\Bigg[\,
-\frac{1536}{5} \, \la \, (1-2\la)\, (1-4\la)\, D^i \, {\bf J}
  \,
\Bigg]
(Z_{2})
\nonu \\
&& +\frac{\theta_{12}^{4-0}}{z_{12}^{5}}\,
\Bigg[\,
  1536\, \la \, (1-2\la)\, (1-4\la)\,
  {\bf \Phi}^{(1)}
  +\frac{6144}{5}\, \la \, (1-2\la)\, (1-4\la)^2 \,
  \pa \, {\bf J}
  \,\Bigg](Z_{2})
\nonu \\
&& +
\frac{1}{z_{12}^{3}}\,\Bigg[\,
  384 \, \la \, (1-2\la)\, {\bf \Phi}^{(1)} -
  \frac{1536}{5}\, \la \, (1-2\la)\, (1-4\la)\, \pa \,
   {\bf J} \,
\Bigg]
(Z_{2})
\nonu \\
&& +
\frac{\theta_{12}^{4-ij}}{z_{12}^{4}}\,
\Bigg[\,
 - \frac{384}{5}\, (1-4\la)\, (3+2\la-4\la^2)\,
  \frac{1}{2!}\, \varepsilon^{i j k l}\, D^k \, D^l \,
       {\bf J}  
       \nonu \\
       && -\frac{576}{5}\, (1+2\la-4\la^2)\,
       \varepsilon^{i j k l}\,
  \frac{1}{2!}\, \varepsilon^{k l m n }\, D^m \, D^n \,
       {\bf J}       
       \,\Bigg](Z_{2})
\nonu \\
&& +\frac{\theta_{12}^{4-i}}{z_{12}^{4}}\,\Bigg[\,
 -\frac{1536}{5}\, \la\, (1-2\la)\, (1-4\la)^2\,
 \pa \, D^i \, {\bf J}
 + 192 \, \la \, (1-2\la)\, (1-4\la)\,
 D^i \, {\bf \Phi}^{(1)}
 \nonu \\
 && + \frac{576}{5}\, (1-4\la)\,(1+2\la-4\la^2)\,
 \Big(
-\frac{1}{3!}\, \varepsilon^{i j k l}\,
      D^j \, D^k \, D^l {\bf J} -(1-4\la)\, \pa \, D^i \, {\bf J} 
 \Big)
  \,\Bigg](Z_{2})
 \nonu\\
&& +
\frac{\theta_{12}^{i}}{z_{12}^{3}}\,\Bigg[\,
  \frac{1536}{5} \, \la \, (1-2\la)\, (1-4\la)\, \pa \,
  D^i \, {\bf J}
  - 192 \, \la \, (1-2\la)\, D^i \, {\bf \Phi}^{(1)}
  \nonu \\
  && -\frac{576}{5}\, (1+2\la-4\la^2)\,
  \Big(
-\frac{1}{3!}\, \varepsilon^{i j k l}\,
      D^j \, D^k \, D^l {\bf J} -(1-4\la)\, \pa \, D^i \, {\bf J}
  \Big)
 \,
\Bigg]
(Z_{2})
\nonu \\
 &&
+\frac{\theta_{12}^{4-0}}{z_{12}^{4}}\,\Bigg[\,
 -384 \, \la \, (1-2\la)\, (1-4\la)\,
  \pa \, {\bf \Phi}^{(1)}
  +\frac{1536}{5}\, \la \, (1-2\la)\, (1-4\la)^2 \,
  \pa^2 \, {\bf J}
  \nonu \\
  && +\frac{768}{5}\, (1-4\la)\, (1+2\la-4\la^2)\,
  \Big(
\frac{1}{2 \cdot 4!}\, \varepsilon^{i j k l}\,
  D^i\, D^j \, D^k \, D^l {\bf J} -
  \frac{1}{2} \, (1-4\la)\, \pa^2  \, {\bf J}
  \Big) \nonu \\
  && +\frac{288}{5}\, (11+12\la-24\la^2)\, {\bf \Phi}^{(2)}
  \,\Bigg](Z_{2})
\nonu \\
&& +  \frac{1}{z_{12}^{2}}\,
\Bigg[\,
 -192 \, \la \, (1-2\la)\, \pa \, {\bf \Phi}^{(1)} +
  \frac{768}{5}\, \la \, (1-2\la)\, (1-4\la)\, \pa^2 \,
   {\bf J}
-\frac{144}{5}\, (1-4\la)\, {\bf \Phi}^{(2)} 
\nonu \\
&& -  \frac{384}{5}\, (-1-2\la+4\la^2)\, \Big(
\frac{1}{2 \cdot 4!}\, \varepsilon^{i j k l}\,
  D^i\, D^j \, D^k \, D^l {\bf J} -
  \frac{1}{2} \, (1-4\la)\, \pa^2  \, {\bf J}
\Big)
\,\Bigg](Z_{2})
\nonu \\
&& +\frac{\theta_{12}^{4-i}}{z_{12}^{3}}\,\Bigg[\,
\frac{288}{5}\, (1+\la)\, (3-2\la)\,  D^i \, {\bf \Phi}^{(2)}
  \,\Bigg](Z_{2})
\nonu\\
&& +
\frac{\theta_{12}^{4-ij}}{z_{12}^{2}}\,\Bigg[\,
-\frac{36}{5}\, (1-4\la)\,
 \frac{1}{2!}\, \varepsilon^{i j k l}\, D^k \, D^l \, {\bf \Phi}^{(2)}
- 18\,
 \varepsilon^{i j k l}\,
 \frac{1}{2!}\, \varepsilon^{ k l m n}\, D^m \, D^n \, {\bf \Phi}^{(2)} 
 \,\Bigg](Z_{2})
\nonu\\
&&
+
\frac{\theta_{12}^{4-0}}{z_{12}^{3}}\,\Bigg[\,
  \frac{576}{5}\, (1+\la)\, (3-2\la)\,
  \pa \, {\bf \Phi}^{(2)}
  \,\Bigg](Z_{2})
\nonu\\
&&
+
\frac{\theta_{12}^{4-i}}{z_{12}^{2}}\,\Bigg[\,
  \frac{6}{5}\, (59+8\la -16\la^2)\,
  \pa\, D^i \, {\bf \Phi}^{(2)}
  -\frac{42}{5}\, (1-4\la)\,
\frac{1}{3!}\, \varepsilon^{i j k l}\,
      D^j \, D^k \, D^l \, {\bf \Phi}^{(2)}  
  \,\Bigg](Z_{2})
\nonu\\
&& +
\frac{\theta_{12}^{i}}{z_{12}}\,\Bigg[\,
-\frac{6}{5}\, (1-4\la)\, \pa \, D^i \, {\bf \Phi}^{(2)}
+ 6 \,
\frac{1}{3!}\, \varepsilon^{i j k l}\,
      D^j \, D^k \, D^l \, {\bf \Phi}^{(2)}
\,\Bigg](Z_{2})
\nonu\\
&&
+\frac{\theta_{12}^{4-0}}{z_{12}^{2}}\,\Bigg[\,
  \frac{96}{35}\, (38+11\la-22\la^2)\, \pa^2 \,
       {\bf \Phi}^{(2)}
       - \frac{96}{35}\, (1-4\la)\,
 \frac{1}{ 4!}\, \varepsilon^{i j k l}\,
 D^i\, D^j \, D^k \, D^l {\bf \Phi}^{(2)}
 + {\bf \Phi}^{(4)}
   \,\Bigg](Z_{2})
\nonu\\
&&
+
\frac{\theta_{12}^{4-ij}}{z_{12}}\,\Bigg[\,
-\frac{12}{5}\, (1-4\la)\,
\frac{1}{2!}\, \varepsilon^{i j k l}\, \pa\,
D^k \, D^l \, {\bf \Phi}^{(2)}
- 6\,
 \varepsilon^{i j k l}\,
 \frac{1}{2!}\, \varepsilon^{ k l m n}\, \pa \,
 D^m \, D^n \, {\bf \Phi}^{(2)} 
  \,\Bigg](Z_{2})
\nonu \\ 
&&
+\frac{\theta_{12}^{4-i}}{z_{12}}\,\Bigg[\,
 \frac{48}{35}\, (13+\la -2\la^2)\,
  \pa^2 \, D^i \, {\bf \Phi}^{(2)}
  -\frac{96}{35}\, (1-4\la)\,
\frac{1}{3!}\, \varepsilon^{i j k l}\,
      \pa \, D^j \, D^k \, D^l \, {\bf \Phi}^{(2)}  
      \nonu \\
      && +\frac{1}{8} \,
D^i \, {\bf \Phi}^{(4)}
      \,\Bigg](Z_{2})
\nonu \\ 
&&
+\frac{\theta_{12}^{4-0}}{z_{12}}\,\Bigg[\,
  \frac{48}{35}\, (17+4\la-8\la^2)\, \pa^3 \,
       {\bf \Phi}^{(2)}
       - \frac{48}{35}\, (1-4\la)\,
 \frac{1}{ 4!}\, \varepsilon^{i j k l}\,
 \pa \, D^i\, D^j \, D^k \, D^l {\bf \Phi}^{(2)}
 \nonu \\
 && + \frac{1}{2} \, \pa \, {\bf \Phi}^{(4)}
  \,
\Bigg](Z_{2})
+\cdots.
\label{lastlast}
\eea
There are the ${\cal N}=4$ stress energy tensor,
the second and the fourth ${\cal N}=4$ multiplets
(and their descendants) in (\ref{lastlast}).

\end{document}